%% file: main.tex
\documentclass[ALICE,manyauthors]{cernphprep}

\usepackage[comma,square,numbers,sort&compress]{natbib}
\usepackage{hyperref}
\usepackage{lineno}
\usepackage{color}
\usepackage{units}
\usepackage{ulem}
\usepackage[T1]{fontenc}
\usepackage{orcidlink}
\input{commands.tex}
\begin{document}%
%%%%%%%%%%%%%%%  Title page %%%%%%%%%%%%%%%%%%
\begin{titlepage}
\PHyear{2022}
\PHnumber{238}      % required, will be obtained from PH
\PHdate{03 November}  % required, will be obtained from PH
%

%%% Put your own title + short title here:
\title{Measurement of electrons from beauty-hadron decays in pp and Pb--Pb collisions at $\sqrt{\boldmath{s}_{\rm{\bf{NN}}}}$ = ${\rm \bf 5.02~TeV}$}
\ShortTitle{Measurement of electrons from beauty-hadron decays in Pb--Pb collisions}   % appears on right page headers

%%% Do not change the next lines
\Collaboration{ALICE Collaboration\thanks{See Appendix~\ref{app:collab} for the list of collaboration members}}
\ShortAuthor{ALICE Collaboration} % appears on left page headers, do not change

\begin{abstract}
The production of electrons from beauty-hadron decays was measured at midrapidity in proton--proton (pp) and central Pb--Pb collisions at center-of-mass energy per nucleon--nucleon pair \comPbPb, using the ALICE detector at the LHC. The cross section measured in pp collisions in the transverse momentum interval \ptrange{2}{8} was compared with models based on perturbative quantum chromodynamics calculations. The yield in the 10\% most central Pb--Pb collisions, measured in the interval \ptrange{2}{26}, was used to compute the nuclear modification factor $\raa$, extrapolating the pp reference cross section to $\pt$~larger than 8~\GeVc. The measured $\raa$ shows significant suppression of the yield of electrons from beauty-hadron decays at high \pt~and does not show a significant dependence on \pt~above $8~\GeVc$ within uncertainties.
The results are described by several theoretical models based on different implementations of the interaction of heavy quarks with a quark--gluon plasma, which  predict a smaller energy loss for beauty quarks compared to light and charm quarks. 

\end{abstract}
\end{titlepage}
\setcounter{page}{2}

\input{Introduction}

\input{ALICEDetector}

\input{Methodology}

\input{Results}
\input{Summary}

%%%%% acknowledgements
\newenvironment{acknowledgement}{\relax}{\relax}
\begin{acknowledgement}
\section*{Acknowledgements}
\input{fa_2022-09-28_Opt_C.tex}    %%%%%%% done by webmaster team
\end{acknowledgement}

%%%%%%%% Bibliography (In case of using bibtex generate the bbl requested by arXiv)
%\clearpage
\bibliographystyle{utphys}   % Remember we use title in the biblio
\bibliography{bibliography.bib,alice_papers.bib}

%%%%%%%%% appendix with author list
\newpage
\appendix
\section{The ALICE Collaboration}
\label{app:collab}
\input{2022-09-28-Alice_Authorlist_2022-09-28_Opt_C.tex}  %%%%%%% done by webmaster team
\end{document}

%% file: commands.tex
\usepackage{xspace}
\usepackage{color}
\usepackage{xcolor}
\usepackage{rotating}
\definecolor{light-gray}{gray}{0.8}

\newcommand{\PbPb}{Pb--Pb\xspace}

\newcommand{\mum}{\ensuremath{\mu{\rm m}}\xspace}

\newcommand{\compp}{{\ensuremath{\sqrt{s}} = 5.02 TeV}\xspace}
\newcommand{\comPbPb}{{\ensuremath{\sqrt{s_{\rm NN}}} = 5.02 TeV}\xspace}
\newcommand{\sqrtsNN}{\ensuremath{\sqrt{s_{\rm NN}}}\xspace}

\newcommand{\ptmore}[1]{\ensuremath{\pt > #1 ~\GeVc}\xspace}
\newcommand{\ptless}[1]{\ensuremath{\pt < #1 ~\GeVc}\xspace}
\newcommand{\ptrange}[2]{\ensuremath{#1 < \pt < #2  ~\GeVc}\xspace}

\newcommand{\cmnt}[1]{}

\newcommand{\Raa}{\ensuremath{R_{\rm AA}}\xspace}

\newcommand{\pt}{\ensuremath{p_{\mathrm{T}}}}

\newcommand{\GeVc}{\ensuremath{\mathrm{GeV}/c}}
\newcommand{\ptGeVc}[1][pt]{{#1}\,\mathrm{GeV}/\ensuremath{c}}
\newcommand{\raa}{\ensuremath{R}_{\mathrm{AA}}}

\newcommand{\nsigmaTPCe}{\ensuremath{n_{\sigma,{\rm e}}^{\rm TPC}}}

\newcommand{\nsigmaTOFe}{\ensuremath{n_{\sigma,{\rm e}}^{\rm TOF}}}

%% file: Introduction.tex
\section{Introduction}

The formation of a strongly coupled color-deconfined medium, called quark--gluon plasma (QGP), is predicted by quantum chromodynamics (QCD) calculations on the lattice at high energy density and temperature~\cite{Kheyri:2013sq, Bazavov:2011nk, Borsanyi:2014rza, Bazavov:2018mes, Bazavov:2009zn}.
These extreme conditions can be created in ultrarelativistic heavy-ion collisions, which were used to establish the formation of the QGP and to study its properties~\cite{Braun-Munzinger:2015hba, Shuryak:1978ij} at the SPS~\cite{Csorgo:2000yu, Teaney:2000cw}, RHIC~\cite{Arsene:2004fa, Back:2004je, Adams:2005dq, Adcox:2004mh}, and LHC~\cite{Muller:2012zq, ALICE:2022wpn} accelerators.

Heavy quarks (charm and beauty) are predominantly produced in the initial hard-scattering processes~\cite{Andronic:2006ky}, given that their mass is large compared to the thermal energy scale ($m_{\rm c,b}\gg k_{\rm B} T_{\rm QGP}$, where $k_{\rm B}$ is the Boltzmann constant and $T_{\rm QGP}$ is the temperature of the QGP) and on a timescale shorter than the QGP formation time~\cite{Andronic:2015wma} ($\tau_{\rm QGP}\sim 1 {\rm~fm}/c$~\cite{Liu:2012ax}). Their mass is also large compared to the QCD scale $\Lambda_{\rm QCD}$, allowing perturbative calculations of their production cross section to be applicable down to zero transverse momentum ($\pt$)~\cite{Cacciari:1998it,Cacciari:2001td,Cacciari:2012ny,Bolzoni:2012kx}. Given the early production of heavy quarks during the collision, they  experience all the stages of the system evolution~\cite{Andronic:2015wma, Cao:2018ews}. They interact with the medium constituents via both elastic (collisional) and inelastic (gluon radiation) processes~\cite{Baier:2000mf, Dokshitzer:2001zm, Armesto:2003jh, Wicks:2007am, Zhang:2003wk, Adil:2006ra}, where the relative contribution of the latter increases with \pt. As the heavy quarks from the early collision rarely annihilate or get produced thermally~\cite{Zhang:2007yoa,Andronic:2015wma,Braun-Munzinger:2007fth}, the effect of the interactions with the medium is primarily a change in the momentum distribution of the quarks. Quarks moving faster than the surrounding medium are typically slowed down by the interactions (resulting in in-medium energy loss), while slow quarks may get accelerated and pushed along with the surrounding medium. Measuring the effect of these interactions yields insights into the transport properties of the QGP~\cite{Andronic:2015wma}. The interaction of partons with the medium is expected to depend on the color charge and the mass of the parton~\cite{Wicks:2007am, Armesto:2003jh, Djordjevic:2004nq,Dokshitzer:2001zm}, with a stronger interaction of gluons compared to quarks and of lighter quarks compared to heavier ones for both collisional and radiative processes. The mass dependence of the radiative contribution is connected to the dead-cone effect~\cite{Dokshitzer:2001zm,ALICE:2021aqk}, which predicts gluon radiation to be suppressed for angles $\theta \leq m/E$, where $m$ and $E$ are the mass and energy of the quark. The measurement of hadron species with different quark contents over a large \pt~range is therefore fundamental to shed light on the underlying mechanisms of the in-medium quark energy loss. Measurements of hadrons containing beauty quarks (beauty hadrons) are particularly useful for testing the mass dependence of the parton energy loss up to high transverse momenta.

The effect of the medium is usually quantified using the nuclear modification factor \Raa, defined as the ratio between the $\pt$-differential particle yields in nucleus--nucleus (A--A) collisions (${\rm d}N_{\rm AA}/{\rm d}\pt$) and the corresponding production cross section in pp collisions (${\rm d}\sigma_{\rm pp}/{\rm d}\pt$) at the same energy scaled by the average nuclear overlap function $\left<T_{\rm{AA}}\right>$ for the centrality range under study~\cite{Loizides:2017ack,dEnterria:2020dwq}

\begin{equation}
\Raa = \frac{1}{\left<T_{\rm{AA}}\right>} \frac{ {\rm{d}}N_{\rm{AA}}/{\rm{d}}\pt } { {\rm{d}}\sigma_{\rm{pp}}/{\rm{d}}\pt }.
\end{equation} 

The production cross section of beauty hadrons and their decay products in hadronic collisions has been measured at different energies at RHIC~\cite{Aggarwal:2010xp, Adare:2009ic}, Tevatron ({$\rm{p}\overline{\rm{p}}$})~\cite{Acosta:2004yw}, and at the LHC~\cite{ALICE:2012acz, ALICE:2014ivb, Aaij:2012jd, Aad:2011sp, ALICE:2012vpz, Aad:2012jga, ATLAS:2013cia, Chatrchyan:2012hw, Khachatryan:2010yr, Khachatryan:2011mk, Chatrchyan:2011pw, Chatrchyan:2011vh, Khachatryan:2016csy, ALICE:2021mgk,ALICE:2021edd,ALICE:2018gev, ALICE:2022wpn}.
These measurements are described by perturbative quantum chromodynamics (pQCD) calculations such as Fixed Order plus Next-to-Leading-Log (FONLL)
~\cite{Cacciari:1998it,Cacciari:2001td,Cacciari:2012ny} and General-Mass Variable Flavor-Number Scheme (GM-VFNS)~\cite{Bolzoni:2012kx} within uncertainties.

The interaction of heavy quarks with the medium can be studied via the measurement of charm hadrons, which can be reconstructed using their exclusive hadronic decays. Measurements in Pb--Pb collisions show a significant change in their momentum distribution with respect to pp collision data~\cite{ALICE:2021rxa,Sirunyan:2017xss,ALICE:2021kfc}. Together with measurements of the flow coefficients~\cite{ALICE:2017pbx,ALICE:2020iug,PHENIX:2010xji, STAR:2014yia, STAR:2017kkh}, this can provide significant constraints to models of the transport properties of the QGP medium~\cite{ALICE:2021rxa, Rapp:2018qla, Cao:2018ews, ALICE:2022wpn}. To investigate the quark-mass dependence of the interaction, measurements in the beauty sector are needed as well. Due to the small cross section of beauty-hadron production and their large number of decay channels, the full reconstruction of beauty hadrons is difficult in heavy-ion collisions.  While it is possible to reconstruct beauty hadrons, as for example in the $\rm{B}^{\pm}$ measurement through the decay channel $\rm{B}^{\pm} \rightarrow J/\psi + K^{\pm} \rightarrow \mu^{+} \mu^{-} + K^{\pm}$~\cite{Sirunyan:2017oug, ATLAS:2013cia, ATLAS:2019jpi,Khachatryan:2011mk,Aaij:2012jd} with a BR of $(6.12 \pm 0.19) \times 10^{-5}$~\cite{Zyla:2020zbs}, the small branching ratios make such measurements challenging.
Complementary information can be gained using inclusive measurements of beauty-hadron decay products such as non-prompt $J/\psi$~\cite{ATLAS:2018hqe,CMS:2017uuv,Khachatryan:2010yr}, non-prompt D mesons~\cite{ALICE:2021mgk,ALICE:2022tji,CMS:2018bwt, ALICE:2022xrg}, or through leptons originating from semi-leptonic decays of heavy-flavor hadrons~\cite{ALICE:2016uid, Adare:2015hla,ALICE:2020sjb, STAR:2021uzu, PHENIX:2022wim}.
The nuclear modification factor of leptons from beauty-hadron decays has been measured by the ALICE Collaboration for semielectronic decays in Pb--Pb collisions at $\sqrtsNN = 2.76~{\rm TeV}$~\cite{ALICE:2016uid} and for semimuonic decays at $\sqrtsNN = 5.02~{\rm TeV}$ with $\pt>4~\GeVc$~\cite{ATLAS:2021xtw}, while the elliptic flow coefficient has been measured at $\sqrtsNN = 5.02~{\rm TeV}$ by the ALICE~\cite{ALICE:2020hdw} and ATLAS~\cite{Aad:2020grf} Collaborations. The measurement of leptons from heavy-flavor decays can provide information about a large range of hadron and heavy-quark momenta, which can help to gain an understanding about the interplay of the collisional and radiative processes.

In this article, the ALICE measurements of the $\pt$-differential cross section of electrons from beauty-hadron decays in pp collisions at \compp, the \pt-differential yield in the 10\% most central Pb--Pb collisions at \comPbPb, and the corresponding nuclear modification factor at midrapidity are presented. The results are discussed together with a comparison to theoretical models.

%% file: ALICEDetector.tex
\section{Experimental apparatus and data samples}
The ALICE apparatus consists of a central barrel covering the pseudorapidity region $|\eta| < 0.9$ and a muon spectrometer with $-4 < \eta < -2.5$ coverage. It also contains forward- and backward-pseudorapidity detectors employed for triggering, background rejection, and event characterization. The nominal magnetic field, parallel to the beam axis and provided by the solenoid magnet in which the central barrel detectors are placed, is 0.5 T. A complete description of the detector and an overview of its performance are presented in Refs.~\cite{ALICE:2008ngc, ALICE:2014sbx}.

The central-barrel detectors used in the analyses presented in this article for charged-particle reconstruction and electron identification at midrapidity are the Inner Tracking System (ITS)~\cite{ALICE:2010tia}, the Time Projection Chamber (TPC)~\cite{Alme:2010ke}, the Time-Of-Flight (TOF)~\cite{Akindinov:2010zza} detector, and the Electromagnetic Calorimeter (EMCal)~\cite{Cortese:1121574}.
The ITS consists of six layers of silicon detectors, with the innermost two composed of Silicon Pixel Detectors (SPD),
 two intermediate layers of Silicon Drift Detectors, %(SDD), 
 and the two outermost layers made of double-sided Silicon Strip Detectors. %(SSD). 
 The ITS is used to reconstruct the primary vertex and for tracking charged particles, the latter in combination with the TPC. The SPD crucially also provides very good spatial resolution down to low transverse momentum,  important requirement for the analyses presented in this paper. 
The TPC is the main tracking detector of the central barrel. In addition, it allows for particle identification via the measurement of the particle specific energy loss (${\rm d}E/{\rm d}x$) in the detector gas. Additional information for particle identification is provided by the TOF~\cite{Akindinov:2010zza}, via the measurement of the charged-particle flight time from the interaction point to the detector. The event collision time is determined using the TOF itself or the two T0 arrays~\cite{ALICE:2016ovj}, made of quartz Cherenkov counters and covering the acceptance $4.6 < \eta < 4.9$ and $-3.3 < \eta < -3.0$. 
The EMCal detector ~\cite{Cortese:1121574,ALICE:2022qhn} is a shashlik-type lead and scintillator sampling electromagnetic calorimeter~\cite{Atoian:1992ze} that covers an acceptance of $|\eta| < 0.7$ in pseudorapidity and $\Delta\varphi = 107^{\circ}$ in azimuth. 
The smallest segmentation of the EMCal is a tower, which has a dimension of $6 \times 6~{\rm cm}^{2}$ ($0.0143 \times 0.0143~{\rm rad^2}$) in the $\eta \times \phi$ direction. The EMCal is used for electron identification as well as for triggering on rare events with high momentum particles in its acceptance. 

Two scintillator arrays (V0)~\cite{ALICE:2013axi}, placed on each side of the interaction point (with pseudorapidity coverage $2.8 < \eta < 5.1$ and $-3.7 < \eta < -1.7$), are used to define a minimum-bias trigger, to reject offline beam-induced background events, and for event characterization. The V0 detectors along with the two T0 arrays are employed to measure the cross section corresponding to the minimum-bias trigger condition~\cite{ALICE:2022xir}.
The Zero Degree Calorimeters ~\cite{Cortese:2019nnv} located at 112.5~m on both sides of the interaction point are used to reject electromagnetic interactions and beam-induced background in \PbPb collisions.

The heavy-ion collisions were divided into different centrality classes using the signal of the V0 detectors~\cite{ALICE-PUBLIC-2018-011,ALICE:2013hur}.
The centrality refers to the percentile of the hadronic cross section covered, with lower values corresponding to more central events.

The results presented in this paper were obtained using data recorded by ALICE during the LHC Run\,2 data-taking period in 2017 for pp collisions at center-of-mass energy \compp and in 2015 for \PbPb collisions at center-of-mass energy per nucleon--nucleon collision \comPbPb.
To obtain a uniform acceptance of the detectors, only events with a reconstructed primary vertex position along the beam line located within $\pm10$ cm from the center of the detector were considered for both pp and \PbPb collisions. Additionally, events were selected after standard quality checks on the performance of the detectors used in the analyses.
The analysis in pp collisions was performed using the Minimum Bias (MB) trigger, which requires coincident signals in both scintillator arrays of the V0 detector.
In Pb--Pb collisions, the analysis using TPC and TOF detectors was based on a MB triggered sample, while the TPC--EMCal analysis employed the MB and a high-energy event trigger based on the energy deposited in the EMCal.  The EMCal trigger (EG) is based on the sum of energy in a sliding window of $4\times4$ towers above a given threshold, where the required energy threshold for the Pb--Pb data sample was 10 GeV above the background energy from the underlying event. The background from the underlying event was obtained using the median of $16 \times 16$ tower patches in the opposite side calorimeter referred to as DCAL.
The analyzed number of MB events is $930\times 10^6$ in pp collisions corresponding to an integrated luminosity of 
$L_{\rm{int}} = 18.2 \pm 0.4~{\rm{nb}^{-1}}$ ~\cite{ALICE-PUBLIC-2018-014}, and $6.7\times 10^6$ and $4.8\times 10^6$ events for the 10\% most central Pb--Pb collisions for the TPC--TOF and TPC--EMCal analysis, corresponding to integrated luminosities of $L_{\rm{int}} = 8.9 \pm 0.2~{\rm{\mu b}^{-1}}$ and $6.2 \pm 0.1~{\rm{\mu b}^{-1}}$~\cite{ALICE-PUBLIC-2018-011}, respectively. The number of analyzed EMCal triggered events is $57\times 10^4$ in the 10\% most central Pb--Pb collisions, corresponding to an integrated luminosity of $45.6 \pm 3.6~{\rm{\mu b}^{-1}}$.

%% file: Methodology.tex
\section{Analysis overview}
\label{sec:AnalysisOver}
Throughout this paper, the term ``electron'' is used for indicating both electrons and positrons. In the momentum range $\pt \ge 2~\GeVc$ considered in these analyses, most of the electrons produced near the interaction vertex at midrapidity come from the decays of heavy-flavor hadrons~\cite{ALICE:2012mzy}. They can be produced directly or as part of the decay chain. Other processes producing electrons that constitute a background in the measurement of electrons from heavy flavor decays are Dalitz and dielectron decays of light mesons ($\pi^0$, $\eta$, $\rho$, $\omega$, $\eta^\prime$, $\phi$), photon conversions in the detector material, decays of hadrons containing strange quarks, decays of prompt quarkonia, and decays of vector bosons. For these, the decay chains of heavier particles may also contain lighter electron sources (e.g.~${\rm K}_{\rm S}^0\rightarrow \pi^0 \pi^0$). Generally, such contributions are more significant at low $\pt$.

The measurements of electrons from beauty-hadron decays can be broadly split into four steps: (i) track selection, (ii) electron identification, (iii) signal extraction to estimate the fraction of electrons originating from beauty-hadron decays, and (iv) a correction for selection efficiencies and geometrical acceptance.

Good quality tracks were selected using the criteria detailed in Sec.~\ref{tracking}. The electron identification (eID) was performed in two different ways depending on the particle \pt. At low \pt~($2<\pt<8~\GeVc$) in pp and Pb--Pb collisions, the eID was based on the combination of signals of the TPC and TOF detectors.
In Pb--Pb collisions, a second analysis was performed in the interval $3 < \pt <26~\GeVc$ exploiting the combined eID information from TPC and EMCal. The TPC--EMCal analysis was performed using MB triggered events for the \pt~interval $3 < \pt < 12~\GeVc$ and using EMCal triggered events for $12 < \pt < 26~\GeVc$ to profit from the substantially larger integrated luminosity sampled with this trigger in a momentum interval in which the EG trigger selection is fully efficient. For the final results, the yields obtained from TPC--EMCal analysis were used  for $p_{\rm{T}} > 8~\GeVc$, whereas the overlapping region of $3 < p_{\rm{T}} < 8~\GeVc$~was used to check the consistency between the TPC--TOF and TPC--EMCal measurements. This choice was motivated by the precision of the measurement based on statistical and systematic uncertainties, as will be further discussed in Sec.~\ref{sec_Results}.

The signal of electrons from beauty-hadron decays was separated from the other background electron sources via an analysis of the track impact parameter ($d_0$) distribution, exploiting the comparatively longer lifetime of beauty hadrons ($c\tau \sim 500~{\rm \mu m}$~\cite{ParticleDataGroup:2018ovx}) with respect to other electron sources. The $d_0$ is defined as the distance of closest approach of the electron track to the reconstructed interaction vertex in a plane perpendicular to the beam direction. It has a positive or negative sign depending on whether the track passes on the left or right of the primary vertex.
This value was multiplied by the sign of the charge of the track and by the sign (direction) of the magnetic field along the z-axis.

The signal was extracted through a Monte Carlo (MC) template fit of the $d_0$ distribution, using four templates in the case of the TPC--TOF analysis and two templates in the TPC--EMCal analysis as described in Secs.~\ref{subsec:4Template} and~\ref{sec::TwoTemplate}, respectively.

\subsection{\textit{Track selection and electron identification}}
\label{tracking}
The track selection has two main goals: assuring high quality tracks and reducing the contribution of background electrons. In particular, requiring hits in detectors close to the interaction point removes part of the contribution of electrons from photon conversions that occur in the detector material. For the analyses based on TPC--TOF, all tracks were required to have an associated hit in each of the two innermost layers of the ITS. In the high track-density environment of a central Pb--Pb collision, it is however possible for a track produced at a larger radius by a photon conversion to be associated with the hits of another particle in the inner layers. These are referred to as misassociated conversion electrons. To reduce this contribution, tracks were required to have at most one hit in the ITS which is shared with another track. For the TPC--EMCal analysis, the tracks were required to have at least one hit in one of the two innermost layers of the ITS. This reduces the impact of the inactive channels in the first ITS layer in the acceptance window of the EMCal.

The TPC track quality is ensured by several selections on the clusters associated to the track, reported in Table~\ref{tab:track_sel}. In particular, the tracks are required to have a minimum total number of clusters and a minimum fraction of found clusters relative to the expected maximum considering the track position in the detector geometry (found/findable). Additional requirements on the number of crossed pad rows are applied as in Ref.~\cite{ALICE:2020jsh}.

The resulting resolution of the impact parameter of the selected tracks with \ptmore{2} is better than~60~$\mum$ for the pp measurement, $50~\mum$ for the TPC--EMCal analysis in Pb--Pb, and $40 ~\mu\mathrm{m}$ for TPC--TOF in Pb--Pb. The complete list of selection criteria can be found in Table~\ref{tab:track_sel} and is similar to earlier analyses~\cite{ALICE:2012mzy}.

\begin{table}[t]
\centering
\caption{Selection criteria for electron candidates.}
\begin{tabular}{ c c c c } 
 \hline
  & pp & Pb--Pb & Pb--Pb \\
   & TPC--TOF & TPC--TOF & TPC--EMCal \\\hline
 Rapidity  & $|y| < 0.8$ & $|y| < 0.8$ & $|y| < 0.6$\\
 ${\rm DCA}_z$ & $< 2~\rm{cm}$ & $< 2~\rm{cm}$ & $< 1~\rm{cm}$\\
 TPC clusters for tracking & - & $\geq 100$ & $\geq 80$\\
 TPC crossed rows for tracking & $\geq 70$ & - & $\geq 70$\\
 TPC clusters for ${\rm d}E/{\rm d}x$ & $\geq 80$ & $\geq 80$ & -\\
 found/findable clusters & - & $>0.6$ & -\\
 crossed rows/findable clusters & $>0.8$ & - & $>0.8$\\
 max. $\chi^2$ per cluster in TPC & 4 & 4 & 4\\
 max. $\chi^2$ per cluster in ITS & 36 & 5 & 36\\
 number of ITS clusters & $\geq 4$ & $\geq 4$ & $\geq 3$ \\
 number of SPD hits & 2 & 2 & 1 or 2 \\
 number of shared ITS clusters & - & at most 1 & - \\
 ITS and TPC refit  & yes & yes & yes\\
 Reject kink daughters  & yes & yes & yes\\
 TPC--EMCal matching     & - & - & $|\Delta\eta|< 0.05,|\Delta\varphi|< 0.05$ \\
 \hline
  & & & \\
 TPC eID signal & $-1 < \nsigmaTPCe < 3$ & $-0.16< \nsigmaTPCe <3$ & $-1 < \nsigmaTPCe < 3$ \\
 TOF eID signal & $|\nsigmaTOFe| <$ 3 & $|\nsigmaTOFe| <$ 3 & - \\
 EMCal $E_{\rm{cal}}/p$ & - & - & $0.8 < E_{\rm{cal}}/p < 1.2$ \\
  EMCal shower shape  & - & - & $0.01 < \sigma_{\rm{short}}^{2} < 0.35$\\
 \hline
\end{tabular}
\label{tab:track_sel}
\end{table}

For electron identification with the TPC (TOF), the measured signal was compared to the expected signal for electrons.
The selection was performed on the variables \nsigmaTPCe ~($n^{\rm{TOF}}_{\sigma,\rm{e}}$), defined as the deviation of the signal from the expectation for an electron in units of the expected resolution. For the EMCal, the main variable used for separating electrons from hadrons was $E_{\rm{cal}}/p$: the deposited energy ($E_{\rm{cal}}$) in the calorimeter divided by the reconstructed particle momentum ($p$), together with information about the shape of the electromagnetic shower. The shower shape is characterized by the eigenvalues of the dispersion matrix of the shower shape ellipse defined by the energy distribution within the EMCal cluster~\cite{Awes:1992yp, ALICE:2017nce, ALICE:2022qhn}. In this analysis, it was chosen to require the short axis of the ellipse, $\sigma^{2}_{\rm{short}}$~\cite{ALICE:2005vhb}, within the range $0.01 < \sigma^{2}_{\rm{short}} < 0.35$, to reduce the hadron contamination. The lower threshold of $\sigma_{\rm{short}}^{2}$ was chosen to reduce the contamination caused by neutrons hitting the readout electronics. 
All the eID selection criteria can be found in Table~\ref{tab:track_sel}. The low but finite remaining hadron contamination was explicitly estimated and subtracted in the method employing TPC and EMCal. It was considered as part of the impact parameter fit in the measurements based on TPC and TOF as discussed in Sec.~\ref{FourTemplateSystematicsSection}.

\subsection{\textit{Impact parameter distributions of the different electron sources}}

The electron candidates originate from different sources, as described in Sec.~\ref{sec:AnalysisOver}. As part of the separation is based on the track impact parameter, it is useful to consider the different shapes of the impact parameter distributions. Electrons from beauty-hadron decays have a particularly wide impact parameter distribution due to the large decay length of the hadrons ($c\tau \sim 500~{\rm \mu m}$~\cite{ParticleDataGroup:2018ovx}). For the electrons from charm-hadron decays this distribution is somewhat narrower ($40 < c\tau < 300 ~{\rm \mu m}$~\cite{ParticleDataGroup:2018ovx}) though still wide compared to many of the other background contributions. Light mesons like neutral pions can decay to electrons directly via three-body Dalitz decays, accounting for a significant portion of the light meson background. As these decays, similar to those of quarkonia and vector bosons, essentially occur at the interaction vertex, their impact parameter distribution is narrow and is determined by the reconstructed track resolution. Light mesons can also produce electrons via decays to photons that convert in the detector material. Electrons originating from photon conversions have a very small angle with respect to the photon direction. However, due to the magnetic field, the track acquires a sizable average impact parameter when it is propagated back to the primary vertex with opposite sign for positrons and electrons. Multiplying the impact parameter with the sign of the track charge and magnetic field orientation makes the distribution asymmetric, making it easier to distinguish it from the other sources. Most of the electrons from photon conversions at large radii are removed by the requirement of signals in the inner ITS layers. The few remaining misassociated conversion electrons have a very wide impact parameter distribution. Most of the hadrons left in the sample after electron identification are produced near the interaction vertex. Thus, their impact parameter distribution is similar to that of the Dalitz decay electrons.

The electrons from Dalitz decays and the photons converting in the detector material mostly come from the decays of light-flavor particles, and are produced in electron--positron pairs with low invariant mass. Together, the electrons from these two sources are referred to as ``photonic electrons''.

The analyses presented here extract the beauty and charm contributions with a fit of the inclusive track impact parameter distribution using templates based on event and detector simulations. The approaches using TPC and TOF detectors also include the conversion electrons and remaining sources as templates and will be referred to as the four-template method. The photonic electron contribution, which decreases with \pt, can be subtracted before fitting the $d_0$ distributions utilizing a technique based on electron--positron pairs with low invariant mass.
This approach, referred to as the two-template method, was used in the analysis with the particle identification based on the TPC and EMCal detectors. These methods will be discussed in more detail in the following sections. One effect of the subtraction of the photonic electrons is that it also subtracts a contribution from light mesons produced in the decays of beauty hadrons. The four-template method, instead, includes all electrons produced in the decay chains of beauty hadrons. This was estimated to induce a difference between the results of the two analysis techniques, which is of the order of 2\% in the measured $\pt$ range and decreases with \pt.

The fit templates, as well as the estimation of the tracking and part of the eID efficiencies, are based on MC simulations. The PYTHIA v6.425 event generator with Perugia 2011 tune~\cite{Sjostrand:2006za} was used to simulate pp events, HIJING v1.36~\cite{Wang:1991hta} for Pb--Pb events, while GEANT3~\cite{Brun:1119728} was used to propagate the  generated particles through the ALICE apparatus. The conditions of all the ALICE detectors during the data taking, were taken into account in the simulations. Simulated events were enriched with additional electrons from beauty- and charm-hadron decays as well as decays of $\pi^0$ and $\eta$ mesons to improve the statistical precision for the signal and the main background sources. Any deviation between the MC templates and the data was corrected whenever possible and any uncertainty in the correction was propagated to the systematic uncertainties of the measurement. Corrections were applied for the transverse impact parameter resolution, the momentum distribution of the charm and beauty hadrons, and the relative fractions of the different charm-hadron species (which impacts on the $d_0$ templates because of the significantly different decay lengths of D$^{0}$, D$^{+}$, D$_{\rm s}^{+}$, $\Lambda_{\rm c}^{+}$). Due to the free amplitudes of the contributions in the fit, only effects on the shape of the impact parameter distributions are relevant, not the total number of entries in the templates. Effects of including %like the fraction of 
electrons from strange-hadron decays, the dependence of the conversion electron distribution on the detector occupancy, and the relative contribution of the hadron contamination were considered as systematic uncertainties. These are discussed in more detail in Sec.~\ref{SystematicsSection}.

\subsection{Extraction of electrons from beauty-hadron decays using the four-template method}\label{subsec:4Template}
In this approach, four impact parameter templates for the corresponding electron sources discussed in the previous section are constructed based on MC simulations and are fitted to the measured inclusive electron distribution using a maximum likelihood fit approach. The template corresponding to contributions that are neither from beauty, charm, or photon conversions will be referred to as the Dalitz template. This procedure was applied in the \pt ~interval $2<\pt<8~{\rm GeV}/c$ with an electron identification based on the signals of the TOF and TPC detectors.

\begin{figure}
    \centering
    \includegraphics[width=0.495\columnwidth]{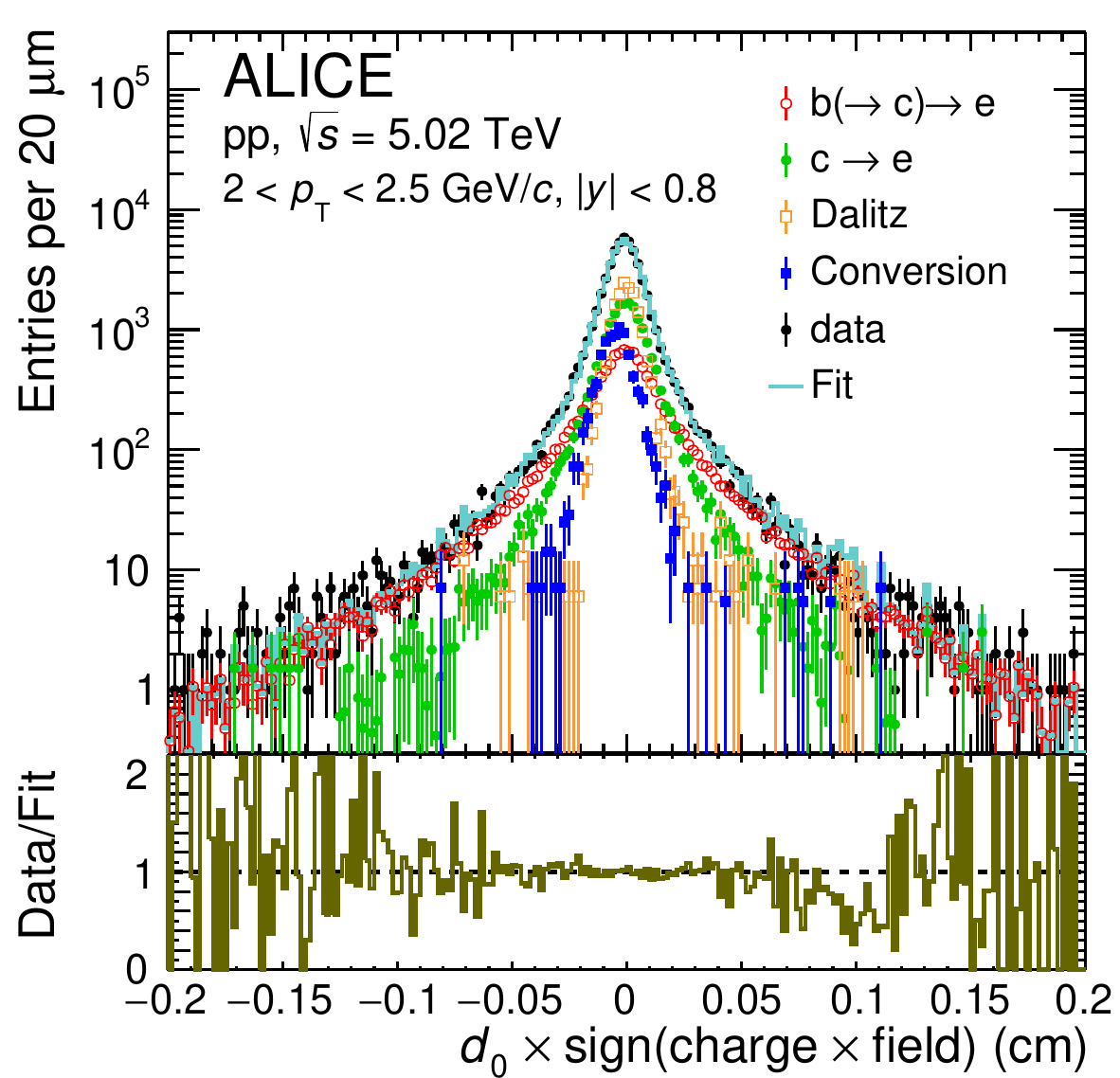}
    \includegraphics[width=0.495\columnwidth]{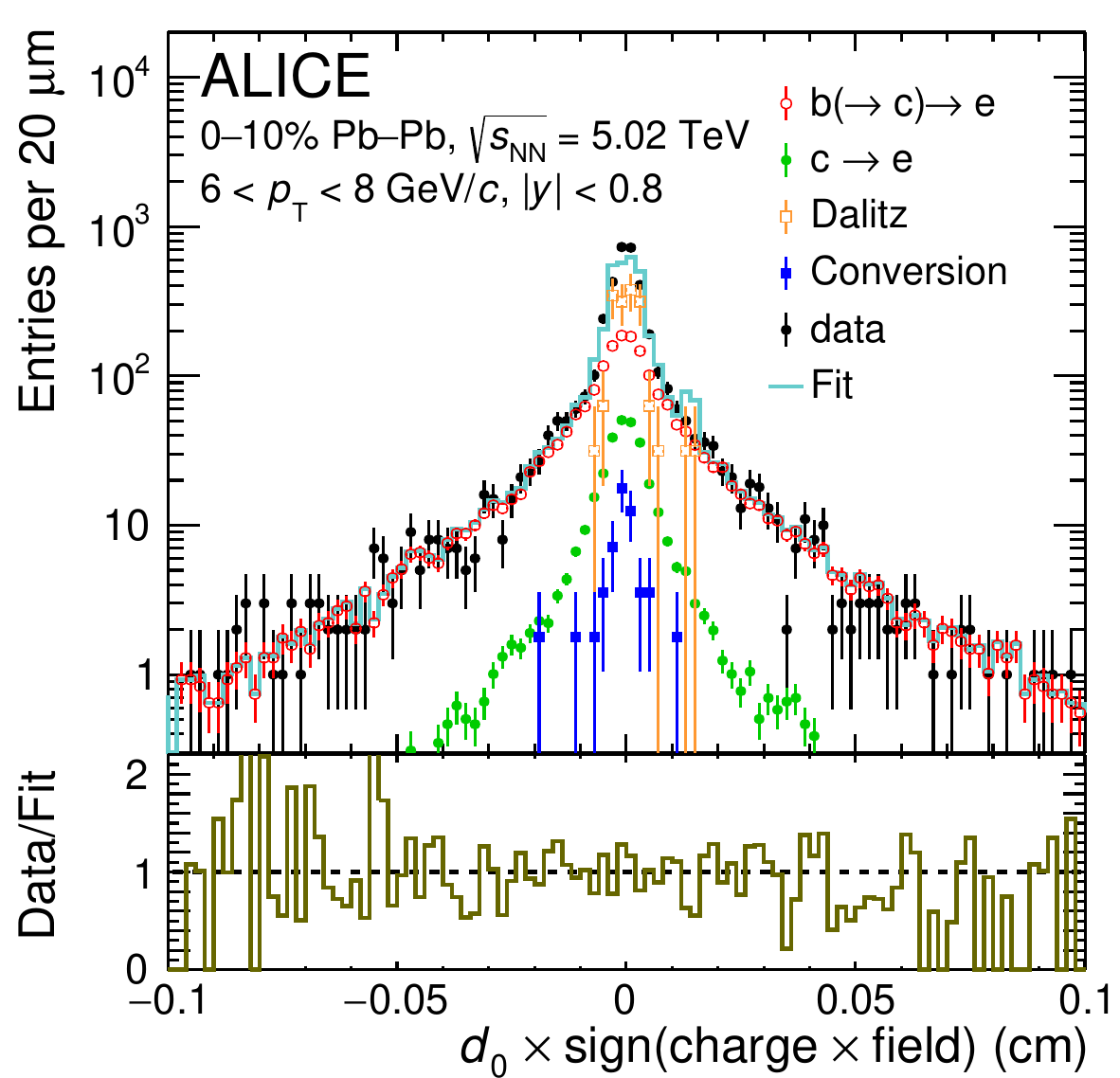}
    \caption{Examples of the impact parameter fits with the four-template fit approach used in the TPC--TOF analyses. The left panel shows a fit with templates for the four classes of electron sources in pp collisions at $\sqrt{s} = 5.02~{\rm TeV}$ in the lowest $\pt$ interval and the right panel shows a fit example in Pb--Pb collisions at $\sqrt{s_{\rm{NN}}} = 5.02~{\rm TeV}$ in the highest $\pt$ range.}
    \label{fig:FitExamples}
\end{figure}

The signal extraction is based on the method of fitting finite-statistics templates proposed in Ref.~\cite{Barlow1993219} and was already used in previous ALICE analyses~\cite{ALICE:2016uid}. The basic idea is to include the fluctuations in the templates by introducing the expectation values $A_{ji}$ of the templates of source $j$ in bin $i$ as free parameters in addition to the overall amplitudes. The contributions from the sources are then estimated from the overall maximum likelihood assuming Poissonian fluctuations. 

Corrections for the fractions of the different charm-hadron species and for the momentum distributions of the charm and beauty hadrons were included via weights of the different entries in the impact parameter histograms. Profiting from the free amplitude parameters, all scaling functions can be changed by an overall multiplicative factor. This was used to keep the weights to values close to unity to not disrupt the Poisson statistics assumed in the maximum likelihood fit. 
For electrons from beauty- and charm-hadron decays, the uncertainties from statistical fluctuations in the templates are much lower than those on the measured distribution (by around a factor of 3).

Examples for the resulting fit together with the scaled template distributions in pp and Pb--Pb collisions are shown in Fig.~\ref{fig:FitExamples}. The ratio of the data to the fit, is also shown to demonstrate the quality of the fit, which is around 1 in the full $d_{0}$ range considered for the fit. Due to the large decay length, the contribution from beauty-hadron decays is most prominent at large absolute impact parameter values and this region thus constrains the magnitude of the beauty contribution the most. The statistical uncertainty of the fit was determined by virtually repeating the measurement using independent samples created from the templates based on measured contributions as input as in Ref.~\cite{ALICE:2016uid}. In addition to the distinct electron sources, there is also a contribution from the remaining hadron contamination. As these hadrons mostly originate near the interaction vertex, the contribution is absorbed into the fit of the Dalitz decay electron contribution.

The slight difference in the average impact parameter and its resolution between data and the simulations used to create the templates was estimated via the measurement of charged hadron tracks and corrected for. The remaining resolution difference is $<4\%$ and was considered as a part of the systematic uncertainties.

Electrons from the decay of heavy-flavor hadrons within a particular $\pt$ interval can originate from different hadron species (e.g. D$^{0}$, D$^{+}$, D$_{\rm s}^{+}$, $\Lambda_{\rm c}^{+}$ for charm hadrons and B$^{0}$, B$^{+}$, B$_{\rm s}^{0}$, $\Lambda_{\rm b}^{0}$ for beauty hadrons, as well as their antiparticles) with different lifetimes and also with a wide range of possible momenta. Thus, the impact parameter distributions depend on the momentum distributions and relative abundances of charm and beauty hadrons. In particular, the charm hadrons have significantly different lifetimes. To correct for this effect, electrons from charm-hadron decays were weighted by the species and momenta of their mothers according to the measured $\rm{D}^0$ momentum distributions and the measured $\Lambda_{\rm{c}}^{+}/\rm{D}^0$~\cite{ALICE:2018hbc,ALICE:2021bib}, $\rm{D}^+/\rm{D}^0$~\cite{ALICE:2018lyv,ALICE:2021rxa} and $\rm{D_s^{+}}/\rm{D}^0$~\cite{ALICE:2018lyv,ALICE:2021kfc} yield ratios, to obtain the MC templates. For beauty, the weights for the \pt-distributions of the beauty hadrons were estimated according to FONLL calculations for the $\pt$ ~distributions using an additional correction for the nuclear modification factor based on the TAMU model~\cite{He:2014cla} in the Pb--Pb analysis. As all beauty hadrons have a similar decay length and thus the corresponding decay electrons are expected to have similar impact parameter distributions, no specific correction was made for the relative contributions of the different beauty hadrons.

The number of measured particles was then taken from the integral of the scaled fit templates after performing the fitting procedure. It was then corrected for the acceptance of the ALICE detector ($\varepsilon^{\rm{geo}}$) and the efficiency of the electron reconstruction ($\epsilon^{\rm reco}$) and identification ($\epsilon^{\rm {eID}}$) and normalized to the number of events according to Eq.~\ref{eg:spectra}. A factor of $\frac{1}{2}$ is included to give the average of electrons and positrons,

\begin{equation}
 \frac{{\rm d}^2 \rm{N}}{{\rm d} p_{\rm T}{\rm d}y} = \frac{1}{2}  \frac{1}{\Delta y \Delta p_{\rm T}} \frac{1}{N_{\rm {events}}} \frac{{N_{\rm{raw}}}} {\epsilon^{\rm{geo}} \times \epsilon^{\rm reco} \times  \epsilon^{\rm {eID}} } {\rm~.}
 \label{eg:spectra}
\end{equation}

The efficiency for the track-quality and TOF eID selections was estimated from MC simulations. For the TPC eID efficiency, a data-driven technique which is based on a fit of the \nsigmaTPCe~distribution after TOF eID, was used for all momentum intervals. The electron peak was fitted using a Gaussian distribution. The resulting mean was around $-0.16$ for all momentum intervals in Pb--Pb collisions. A selection of \nsigmaTPCe~from $-0.16$ to $3$ was applied, resulting in a constant TPC eID efficiency of about 50\%. The small correction for the finite transverse momentum resolution was done via a bin-by-bin correction comparing generated and measured momenta in the simulations.

\subsection{Extraction of electrons from beauty-hadron decays using the two-template method}
\label{sec::TwoTemplate}
In the second procedure, the contributions from hadrons and photonic electrons are subtracted from the $d_{0}$ distribution through data-driven methods similar to the ones used in Ref.~\cite{ALICE:2019nuy}. The contributions of electrons from charm- and beauty-hadron decays are then determined by fitting the $d_{0}$ distribution with two templates, one for charm and one for beauty sources, obtained from MC simulations. This method was used in the Pb--Pb analysis using eID based on the TPC and EMCal detectors in the $\pt$ range from $3$ to $26~\GeVc$.
The inclusive electron sample was obtained by selecting tracks and by applying electron identification criteria in the TPC and the EMCal using \nsigmaTPCe, $E_{\rm{cal}}/p$, and $\sigma_{\rm{short}}^{2}$, as listed in Table~\ref{tab:track_sel}.

\subsubsection{Data-driven background subtraction}
The hadron contamination was estimated by selecting hadron tracks with $\nsigmaTPCe<-4$. The $E_{\rm{cal}}/p$ distribution of these particles was then scaled to match the electron-candidate $E_{\rm{cal}}/p$ distribution in an interval that varies with \pt~ inside the range $0.2 < E_{\rm{cal}}/p < 0.6$.
This hadron $E_{\rm{cal}}/p$ normalization range shifts with increasing \pt~ to account for the shift of the hadron signal towards $E_{\rm{cal}}/p = 1$ with increasing momentum. The electron- and hadron-contamination yields were obtained by integrating the $E_{\rm{cal}}/p$ distributions of electron candidates and the scaled hadron one in $0.8 < E_{\rm{cal}}/p < 1.2$. The hadron contamination is negligible at low $p_{\rm{T}}$, and increases to around $20\%$ at $p_{\rm{T}} = 26~{\rm GeV}/c$. The hadron $d_{0}$ distribution, obtained by selecting particles with $\nsigmaTPCe<-4$, was also scaled to match this estimated hadron contamination. It was then subtracted from the inclusive electron $d_{0}$ distribution (${\rm d}N^{\rm{InclE}}/{\rm d}d_{0}$). 

\begin{figure}
    \centering
    \includegraphics[width=0.495\columnwidth]{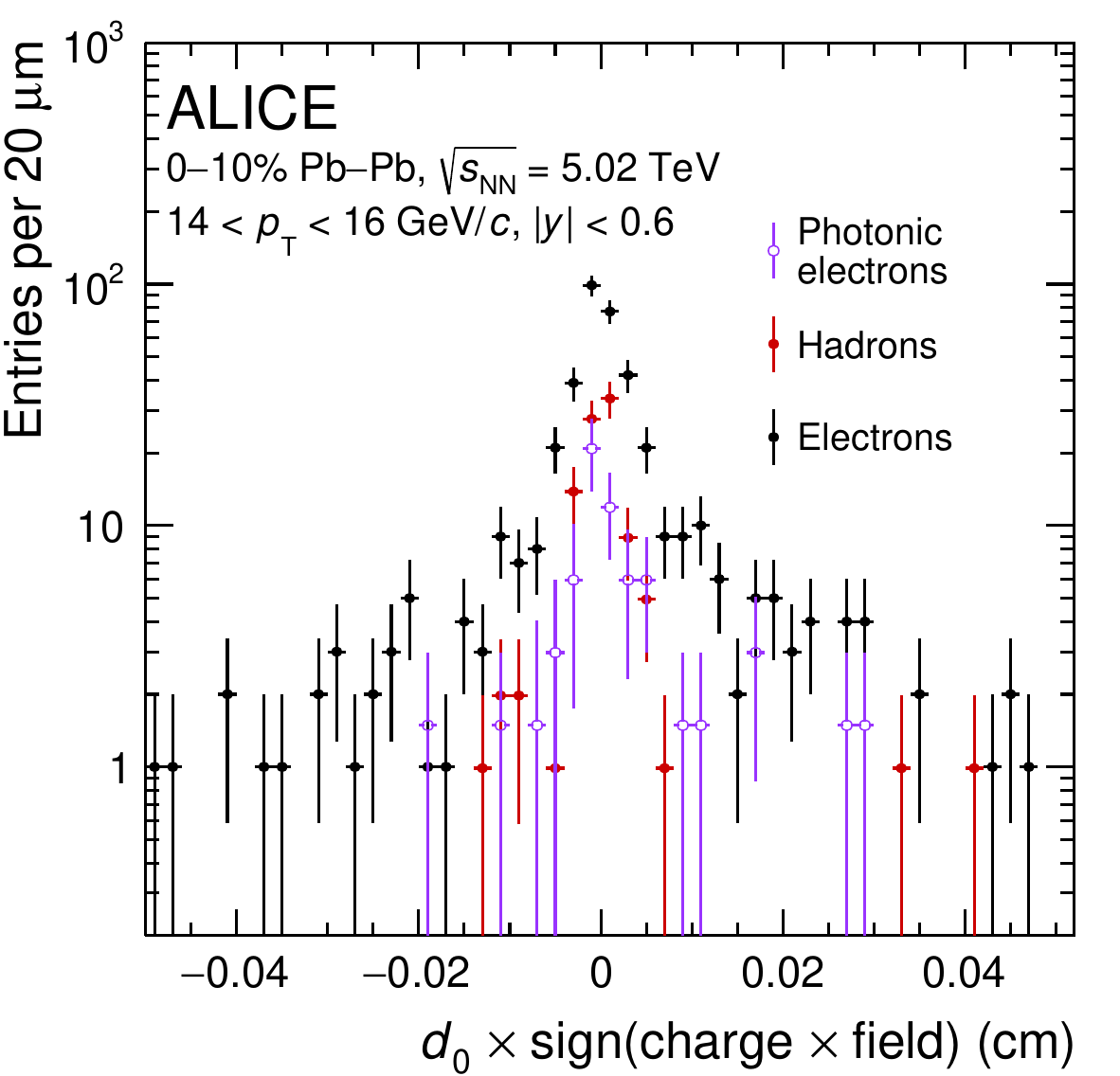}
    \includegraphics[width=0.495\columnwidth]{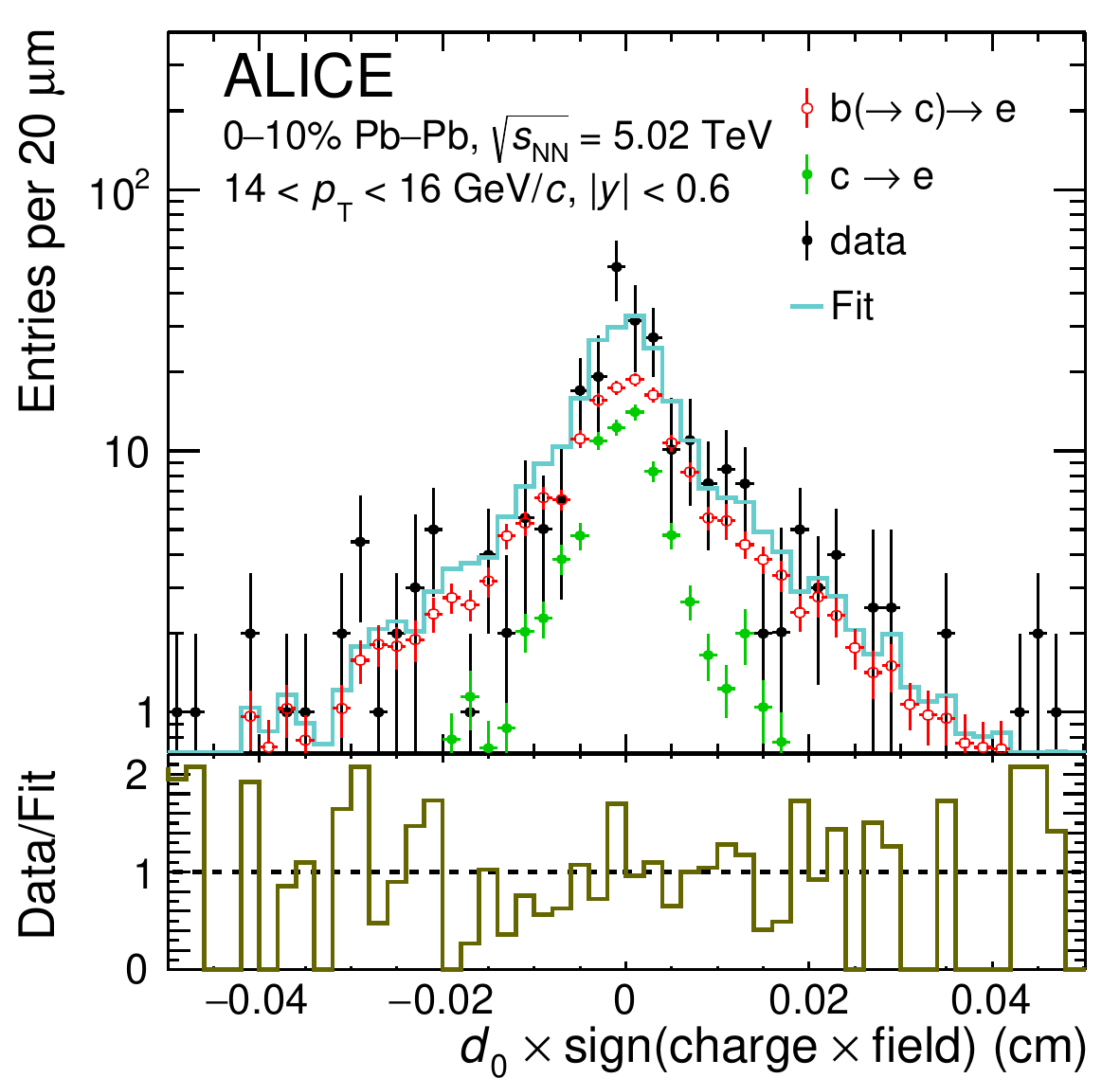}
    \caption{Example of the impact parameter distributions used in the TPC--EMCal analysis in 0--10\% Pb--Pb collisions at $\sqrt{s_{\rm{NN}}} = 5.02~{\rm TeV}$. The hadron contamination (red), photonic electron (purple), and inclusive electron (black) $d_{0}$ distributions are shown in the left panel. The right panel shows an example of a fit with the two template approach.} 
    \label{fig:D02TemplateMethod}
\end{figure}

After the subtraction of the hadron contamination, the contribution of photonic electrons was estimated. Photonic electrons are produced in electron--positron pairs with low invariant mass, peaking close to zero. Thus, they can be identified using an invariant mass analysis of electron pairs. In the invariant mass analysis technique~\cite{ALICE:2015zhm, ALICE:2016mpw},  electron--positron pairs are defined by pairing the selected electrons with opposite-charge electron partners to form unlike-sign (ULS) pairs and calculating their invariant mass ($ m_{\rm e^{+}e^{-}}$). The partner electrons were selected by applying similar but looser track quality and particle identification criteria than those used for selecting signal electrons to increase the efficiency of finding the partner~\cite{ALICE:2016mpw, ALICE:2018yau}. Heavy-flavor decay electrons can form ULS pairs mainly through random combinations with other electrons. The combinatorial contribution was estimated from the invariant mass distribution of like-sign electron (LS) pairs. The $d_{0}$ distributions of electrons forming ULS (${\rm d}N^{\rm{ULS}}/{\rm d}d_{0}$) and LS (${\rm d}N^{\rm{LS}}/{\rm d}d_{0}$) pairs were obtained. The photonic background contribution was then evaluated by subtracting the LS distribution from the ULS distribution in the invariant mass interval $m_{\rm e^{+}e^{-}} < 0.1~{\rm GeV}/c^2$. The efficiency of finding the partner electron, referred to as the tagging efficiency ($\epsilon_{\rm tag}$) from here on, was estimated using the HIJING~\cite{Wang:1991hta} MC simulations with added light flavor signals at high momenta.  The generated particles are propagated through the ALICE apparatus using GEANT3~\cite{Brun:1119728}. The simulated \pt~distributions of $\pi^0$ and $\eta$ mesons were reweighted to match the unbiased \pt~distribution from PYTHIA. The tagging efficiency is $\sim 50\%$ at 3 \GeVc, increasing to $\sim 70\%$ at $\pt > 20 ~\GeVc$. The $d_{0}$ distribution of photonic electrons was corrected for with the tagging efficiency and subtracted from the inclusive electron distribution to obtain the impact parameter distribution of electrons from heavy flavor hadron decays (${\rm d}N^{\rm{HFe}}/{\rm d}d_{0}$) according to

\begin{equation}
    \frac{{\rm d}N^{\rm{HFe}}}{{\rm d}d_{0}} = \frac{{\rm d}N^{\rm{InclE}}}{{\rm d}d_{0}}  -  \frac{1}{\epsilon_{\rm tag}} \left(\frac{{\rm d}N^{\rm{ULS}}}{{\rm d}d_{0}} - \frac{{\rm d}N^{\rm{LS}}}{{\rm d}d_{0}} \right) ~{\rm .}
    \label{eg:hfe}
\end{equation}

An example of the $d_{0}$ distribution of inclusive electrons, estimated hadron contamination, and photonic electrons is shown in the left panel of Fig.~\ref{fig:D02TemplateMethod} for the interval $3 < \pt < 4 ~\GeVc$. 

Electrons from strange-hadron decays are negligible in the \pt ~range considered~\cite{ALICE:2012mzy}. The contribution of electrons from J$/\psi$ decays was estimated to be less than 5\% with a maximum at $2 < \pt < 3~\GeVc$~\cite{ALICE:2016mpw,ALICE:2018yau}, hence not considered in the analysis. The $\rm{W}^{\pm}$ ($\rm{Z}^0$) boson decays have a negligible contribution for $\pt< 15~\GeVc$, increasing to 20\% (10\%) of the heavy-flavor decay electron yield at $\pt = 26~\GeVc$~\cite{ALICE:2019nuy}. These contributions were not subtracted but their small effect on the yield is considered in the systematic uncertainty estimation, as described below.

\subsubsection{Two-template fit procedure}
The electrons from beauty-hadron decays were separated by fitting two MC $d_{0}$ templates to the heavy-flavor electron $d_{0}$ distribution, using a procedure similar to that detailed in Sec.~\ref{subsec:4Template}. The two templates correspond to electrons from charm-hadron decays and beauty-hadron decays. 

Several corrections were applied to the MC $d_{0}$ templates of charm- and beauty-hadron decays to obtain a realistic description of the data, before using them to fit the measured distributions. These corrections include i) the $p_{\rm T}$ shape of the mother charm and beauty hadrons, ii) the relative fraction of the different charm-hadron mother species, and iii) the mean and the resolution of the $d_{0}$ distribution. The corrections were applied using the same procedure described in Sec.~\ref{subsec:4Template}. 

The corrected MC templates of charm and beauty decays were used to fit the heavy-flavor electron $d_{0}$ distribution from data using a weighted log-likelihood fit~\cite{james_statistical_2006}. The statistical uncertainties from the templates were not considered in the fit, as the statistical uncertainties of the data dominate over those of the templates (approximately six times larger).
An example of a fit is shown in the right panel of Fig.~\ref{fig:D02TemplateMethod} for the interval $14<\pt< 16~\GeVc$. The ratio of the data to the fit is also shown to demonstrate the quality of the fit. It is around 1 in the full $d_{0}$ range considered for the fit.

The raw beauty-decay electron yield obtained from the template fit was then corrected for the acceptance of the ALICE detector and the efficiency of the electron reconstruction and identification as described by Eq.~\ref{eg:spectra}. This includes the efficiency corrections for the track quality selections, the procedure to match tracks with EMCal energy-deposition clusters, and the selections used to identify electrons (the $E_{\rm{cal}}/p$, $\sigma_{\rm{short}}^{2}$, and \nsigmaTPCe~ requirements). The efficiencies were estimated using MC simulations, with the exception of the \nsigmaTPCe~ and $\sigma_{\rm{short}}^{2}$ electron selection efficiencies, which was estimated using data-driven procedures.

The \nsigmaTPCe~ efficiency was estimated by parameterizing the \nsigmaTPCe~ distributions by fitting with three Gaussians, for the electron, the pion, and the combined proton and kaon signals. The obtained efficiency varies from about 72\% to 76\% with increasing momentum. The relative efficiency of $\sigma_{\rm{short}}^{2}$ was obtained using $E_{\rm{cal}}/p$ distribution with and without applying the shower shape selection. The resulting efficiency varies from about 80\% to 95\% with increasing momentum. 

The per-event yield obtained with EMCal-triggered events for $\pt > 12~\GeVc$ was normalized using the EMCal trigger rejection factor. The rejection factor expresses the equivalent number of MB events corresponding to a triggered event.
It was estimated with a data-driven method, using the ratio of the EMCal cluster energy distribution in EG-triggered data to the one in minimum-bias data (EG/MB), similar to what is described in Refs.~\cite{ALICE:2016mpw, ALICE:2019nuy}. This ratio forms a stable plateau above $E_{\mathrm{cluster}} > 12~{\rm GeV}/c$. A Fermi function~\cite{Fermi:1934hr, Wilson:1968pwx} was used to fit the ratio and to determine the EMCal trigger rejection factor above the trigger threshold. 
A rejection factor of $61.7 \pm 0.8 (\rm{stat.}) \pm 3.0 (\rm{sys.})$ was obtained. The statistical uncertainty was obtained by varying the fit function within its parameter uncertainties, and the systematic uncertainty was obtained by changing the fit function using a constant above the trigger threshold, and also varying the fit range. The uncertainty on the EMCal trigger rejection factor was propagated to the final measurement. 

\subsection{\textit{Systematic uncertainties}}
\label{SystematicsSection}

\subsubsection{Systematic uncertainty estimation for the four-template method}
\label{FourTemplateSystematicsSection}

\begin{table}[tbp]
\centering
\caption{Systematic uncertainties in the pp and Pb--Pb analyses. Individual sources of systematic uncertainties are $\pt$ dependent.
The values are presented as a range corresponding to the lowest and highest \pt~intervals.}
\begin{tabular}{ c c c c } 
 \hline
 Source & pp & Pb--Pb & Pb--Pb \\
 & TPC--TOF & TPC--TOF & TPC--EMCal \\
  & 2 $< \pt < \ptGeVc[8]$ & 2 $< \pt < \ptGeVc[8]$ & 3 $< \pt < \ptGeVc[26]$ \\
 \hline
 Track selection & 1\% & 3\% & 2\% \\
 ITS--TPC matching & 2\% & 3\% & 3\%\\
 TPC--TOF matching & 2\% & 5\% & -\\
 TPC--EMCal matching & - & - & 1\%\\
 TPC eID & 1\% & 7\% & - \\
 TOF eID & 4\% -- 0\% & 7\% -- 2\% & - \\
 TPC--EMCal eID & - & - & 7\% -- 10\% \\ \hline
 IP resolution & 6$\%$ & 5\% -- 0\% & 2\%\\
 Charm-hadron $\pt$ spectra & 2\% -- 0\% & 5\% -- 2\% & 1\% -- 5\% \\ 
 Charm-hadron species & 4\% -- 0\% & 1\% -- 0\% & 5\% \\ 
 Beauty-hadron $\pt$ spectra & 10\% -- 5\%  & 10\% -- 5\% & 10\% -- 3\% \\
 Hadron contamination & 0\% -- 10\%& 8\% -- 4\% & 1\% -- 3\% \\
 Fit stability & 10\% -- 0\%  & 5\% -- 0\% & 0\% \\
 Strangeness decay contribution & - & 3\% -- 1\% & - \\
 Partner electron selection & - & - & 2\% -- 0\% \\
 Multiplicity effect for conversions & - & 4\% -- 1\% & - \\
 Fit method & 20\% & 15\% & 15\% \\
 Closure test & - & - & 20\% -- 4\% \\
 W/Z contribution & - & - & 0\% -- 6\% \\
 EMCal trigger rejection factor & - & - & 0\% -- 5\% \\
 \hline
 Total & {26\% -- 23\%} & 26 -- 18\% & 28 -- 22\%\\
 Normalization uncertainty & 2.1\% & - & - \\
 \hline
\end{tabular}
\label{tab:SysCombined}
\end{table}

The systematic uncertainties in pp and Pb--Pb collisions with TPC--TOF eID are summarized in Table~\ref{tab:SysCombined}. They can broadly be separated into uncertainties on the estimate of the efficiency of the track selection and eID and uncertainties related to the shape of the fit templates. The different sources of uncertainties were assumed to be uncorrelated and thus added in quadrature. For the calculation of the $\raa$, they were considered as uncorrelated between pp and Pb--Pb. A detailed description of the contributions is given in the following.

The systematic uncertainty for the track quality selection efficiency in pp and Pb--Pb collisions is 1\% and 3\%, respectively, for electrons from beauty- and charm-hadron decays as estimated in Ref.~\cite{ALICE:2019nuy} from variations of the selection criteria. The uncertainty estimated for heavy-flavor decay electrons in Ref.~\cite{ALICE:2019nuy} can be applied in the analysis presented here because, at a given \pt, tracks of electrons from beauty-hadron decays have similar properties as tracks of those from charm-hadron decays and therefore the systematic uncertainty due to track reconstruction and selection is the same. The systematic uncertainty due to the imperfect description in the MC of the efficiency of matching tracks reconstructed in the ITS and TPC, is about 2\% in pp and about 3\% in Pb--Pb collisions~\cite{ALICE:2019nuy}. 

Charged particle tracks reconstructed in the TPC have to be matched with TOF hits, in order to compute the time of flight for particle identification. The detector acceptance does not depend on particle species, thus charged particles can be used to calculate the matching efficiency between TPC and TOF. The systematic uncertainty on the TPC--TOF matching efficiency was obtained by comparing the matching efficiency of charged particles in data and MC and was estimated to be 2\% and 5\% in pp and Pb--Pb collisions, respectively~\cite{ALICE:2019nuy}.

The TPC and TOF eID uncertainties were estimated using a pure sample of electrons produced in photon conversions in the detector material, selected according to the corresponding decay topology (V0 electrons). Since both electrons from beauty-hadron decays and V0 electrons produced in the ITS give the same detector signals in the outer layers, the latter can be used as a proxy to investigate the differences of the eID efficiency between data and simulation. For Pb--Pb collisions, the systematic uncertainty on the TPC eID was estimated to be of 7\% in all \pt ~intervals from the difference between data and MC.
For the analysis in pp collisions, where the TPC signal shape is easier to model than in central Pb--Pb collisions~\cite{ALICE:2016uid}, the model for the extraction of the efficiency was varied, leading to a change of around 1\% in efficiency which was added as a systematic uncertainty. The TOF eID systematic uncertainties reach a maximum of 7\% in the lowest \pt-interval and a minimum of 2\% in the highest \pt-interval for Pb--Pb collisions while the range is 4\% to 0\% in pp collisions.

The correction for differences in the $d_{0}$ resolutions and in the average of the $d_{0}$ distribution between data and simulations extracted for charged-particle tracks was applied to improve the description of the data on average for all particles, independent of the species. Thus, depending on the specific track selection criteria, some deviations can still be present. 
The relative residual deviation was smaller than 4\% for the impact parameter resolution. A systematic uncertainty of 6\% over the entire \pt ~range was estimated for pp collisions by varying the resolution correction accordingly, while for Pb--Pb collisions the effect was 5\% for $\pt < \ptGeVc[5]$ and negligible at larger \pt.

The uncertainties from the correction of the parent hadron \pt ~distribution for electrons from charm-hadron decays originate from the uncertainty of the measurements used as input for the corrections. As the fit has a free amplitude parameter, the correction does not depend on the integrated yield of the measurement, but mostly on its $\pt$ shape. To account for this, the correction was modified by tilting the measured D$^{0}$~\cite{ALICE:2018lyv} spectrum based on its total uncertainty. The uncertainty was then estimated by comparing the fit results for beauty-hadron decay electrons using the different corrections. An uncertainty of 2\% was assigned in pp collisions for electron \pt ~below 6\,\GeVc. In Pb--Pb collisions, a systematic uncertainty of 5\% was assigned below 5~$\GeVc$ and 2\% above.
The uncertainty due to the unknown relative abundances of the different charm-hadron species was estimated by varying the $\Lambda_{\rm{c}}^{+}/\rm{D}^0$~\cite{ALICE:2018hbc,ALICE:2021bib}, $\rm{D}^+/\rm{D}^0$~\cite{ALICE:2018lyv,ALICE:2021rxa} and $\rm{D_s^{+}}/\rm{D}^0$~\cite{ALICE:2018lyv,ALICE:2021kfc} ratios  within the statistical and systematic uncertainties of the measurements,
and assessing the effect of these variations on the yield of electron from beauty-hadron decays. The uncertainty for pp is 4\% for \ptless{2.5}, 2\% for $2.5<p_{\mathrm{T}}<5~\GeVc$, and negligible above. For Pb--Pb it is 1\% for \ptless{3} and negligible above.

For the beauty case, the input \pt~shapes from the considered models were varied. In the Pb--Pb case, the most effective variation on the model is the change of the $\raa$ slope. Two large variations were tested, adding and subtracting half the distance of the $\raa$ to unity to the central values. The resulting beauty-hadron decay electron yields were compared to the central correction to estimate the systematic uncertainty due to the beauty-hadron \pt ~spectrum.
In Pb--Pb collisions, a systematic uncertainty of 10\% was assigned below $3~\GeVc$ and 5\% above. For the pp measurement, the estimation was based on the uncertainties of the FONLL pQCD calculation. To estimate the corresponding uncertainties, the upper and lower limits of the FONLL calculations were used as alternative weights. In this way, the systematic uncertainties on the beauty-hadron decay electron yield reach a maximum of 10\% in the lowest $\pt$ interval and a minimum of 5\% in the highest $\pt$ interval in pp collisions.

Despite the stringent eID selections, some hadron contamination remains in the selected electron sample. Since pions are abundant and the TPC signals for electrons and pions overlap at high $\pt$, the hadron contamination is mostly due to charged pions. In Pb--Pb collisions the hadron contamination contributes over the entire \pt~range, while in pp collisions it is only significant at high \pt, ~similarly as reported in Ref.~\cite{ALICE:2019nuy}. To investigate the effect of the hadron contamination, the template fit was repeated, replacing the Dalitz template by the hadron template obtained from data by requiring $-5<\nsigmaTPCe<-3$. The impact parameter distributions of these two templates are similar because Dalitz electron sources and the charged pions originate mostly from the primary vertex. The differences are mainly due to the resolution of the impact parameter for electrons and pions. From the fit with the hadron templates, the measured yield of the beauty-hadron decay electrons varies by about 8\% (4\%) below (above) $2.5~\GeVc$. These values were assigned as the systematic uncertainty in the Pb--Pb analysis. In the pp case, where the TPC signal shape is easier to model, the hadron contamination was estimated using fits of the TPC signal as described in Ref.~\cite{ALICE:2012mzy}. The hadron impact parameter distribution template was scaled accordingly and subtracted from the total. The difference in the yield of beauty-hadron decay electrons was assigned as a systematic uncertainty yielding no significant change at low \pt~and 10\% in the range $5< \pt < 8~\GeVc$.

The results of the fits to the impact parameter distributions should be mostly independent of the fit range used as well as of the bin width of the templates used for the different sources. For the Pb--Pb case, such variations showed no clear effect on the result, with a possible effect of $<5\%$ found only in the first \pt~interval below 2.5~\GeVc. In pp, an effect is visible below 2~\GeVc, which decreases with \pt. Varying the fit range between $|d_{0}|<0.1$ and  $|d_{0}|<0.2$ and the bin width in the range $5~\mu\rm{m}$ to $20~\mu\rm{m}$ gives an effect on the beauty decay electron yield of the order of $\sim10\%$ in $2{\rm -}2.5~\GeVc$~and $\sim5\%$ up to 5~\GeVc. These values were added as systematic uncertainties.

The electron candidates include contributions from secondary $\pi^{0}$ decays and three body decays of strange hadrons which have a wide impact parameter distribution.
The effect of these secondary tracks on the Dalitz and conversion electron templates was investigated by varying the fraction of this contribution. Based on the results of charged pions and kaons~\cite{ALICE:2019hno}, the secondary contribution in the Dalitz and conversion electron templates was scaled by factors 3 and 0 to estimate the effect on the fitted yield. Considering both variations, the extracted yield of beauty-hadron decay electrons varies from 3\% at $2~\GeVc$ down to 1\% at $8~\GeVc$ in Pb--Pb collisions. The effect of the secondary tracks in pp collisions is negligible.  

The shape of the impact parameter distribution of electrons from photon conversions depends on the influence of the misassociated conversion electrons. The misassociation probability is approximately proportional to the multiplicity of the event. As a result, the misassociated conversion electron contribution is particularly important to consider in central Pb--Pb collisions. Using conversion electron templates from the centrality range 10--30\% instead of the nominal 0--10\% corresponds to a change in multiplicity by a factor of about 1.6. The corresponding extracted beauty contribution changes only by 4\% for $2<\pt<2.5~\GeVc$, decreasing down to 1\% in the highest $\pt$ range ($6<\pt<8~\GeVc$).

The two approaches, TPC--EMCal eID with the two-template fit method and the TPC--TOF eID with the four-template fit method were compared in the overlap region of the analyses and also with a separate test in pp collisions. Comparisons with different eID, centrality, and track selection criteria show a consistent deviation of the order of 15\% in Pb--Pb (20\% in pp). Typically, the approach using the four-template fit gives a higher yield for electrons from beauty-hadron decays than that of two-template fit. Closer inspection of the systematic effects showed that variations in the assumed $d_0$ resolution typically lowered the result of the four-template TPC--TOF approach. On the other hand, the effect of additional background electrons from the primary vertex (e.g.~from ${\rm J}/\psi$ decays) would increase the result for the two-template approach. However, both effects together still do not cover the full difference. To account for this, an additional systematic uncertainty of 15\% in Pb--Pb and 20\% in pp collisions was added.

\subsubsection{Systematic uncertainty estimation for the two-template method}
The systematic uncertainties for the approach using the two-template fits originate in the efficiency correction, the background subtraction, and the signal extraction. A summary of all the sources of systematic uncertainty and the assigned values can be found in Table~\ref{tab:SysCombined}. 

The uncertainty due to track selection was estimated by varying the track selection criteria~\cite{ALICE:2019nuy}, and was found to be about 2\%. The uncertainty from the imperfect description in the MC of the efficiency of matching tracks reconstructed in the ITS and TPC, is about 2\%. An uncertainty of $1\%$ was estimated for matching electron tracks in the TPC to EMCal clusters by varying the matching criteria. The uncertainty on electron identification using the TPC and the EMCal was estimated by varying the selection criteria on \nsigmaTPCe, $E_{\rm{cal}}/p$, and  $\sigma_{\rm{short}}^{2}$. These variations test the procedure of removing the hadron contamination and estimating the efficiency. The chosen variations change the efficiency by a maximum of $\sim20\%$ while still allowing reasonable signal extraction. A total uncertainty from these sources of 7\% for $\pt< 12~\GeVc$~and 10\% for higher \pt~was estimated.
The $E_{\rm{cal}}/p$ interval used to normalize the hadron $E_{\rm{cal}}/p$ distribution to match the electron one in the background region was varied, and the 
effect on the hadron subtraction method was checked. The scale factor was also varied within its statistical uncertainty. The effect of these variations is more pronounced at high \pt ~where the hadron contamination is larger. The uncertainty due to hadron contamination was estimated to be 1\% for $\pt< 12~\GeVc$~and 3\% for higher \pt.

The contribution from photonic electrons was estimated using the invariant mass method. The systematic uncertainty on the procedure, mainly affecting the average correction efficiency,
was obtained by varying the selection criteria of the partner electron tracks, including the minimum \pt~and the invariant-mass window of the electron--positron pairs. While the average tagging efficiency from $\gamma$, $\pi^{0}$, and $\eta$ was used in the analysis, any differences in the efficiency between the different photonic electron sources, estimated using MC simulations, were also considered as a systematic uncertainty.
The resulting systematic uncertainty on the yield of beauty-hadron decay electrons is 2\% for $3 < p_{\mathrm{T}} < 4~\GeVc$ and negligible for $p_{\mathrm{T}} > 4~\GeVc$.

While the uncertainty on the yield of the photonic electrons was obtained using the procedure described above, there can also be an effect from the shape of their $d_0$ distribution. This can be present if there is a difference between the impact parameter distribution of the photonic electron candidates selected via the invariant mass and that of all photonic electrons in the sample of selected tracks. The corresponding effect on the estimated yield of electrons from beauty-hadron decays was evaluated using a MC closure test. Electrons from photon conversions in the detector material have a $d_{0}$ shape that depends on the production vertex, with wider distributions for electrons produced at larger radii. The $d_{0}$ distribution from the invariant mass method gives a combination of contributions from Dalitz decays, produced at the primary vertex, and gamma conversion processes, which occur at different radii in the detector material.
In the MC closure test, the beauty yield was obtained using the data-analysis procedure on simulated data, with realistic fractions of electrons from charm, beauty, and photonic background obtained from previous measurements~\cite{ALICE:2019nuy}.
The photonic electrons were subtracted using the invariant mass method and the resulting $d_{0}$  distribution of candidate heavy-flavor decay electrons was fitted with charm and beauty templates. The beauty yield from the fit was compared to the true input beauty yield in MC simulations, and their difference which was $\sim20\%$ at $3~\GeVc$~decreasing to 4\% at $23~\GeVc$, was taken as a systematic uncertainty.

As described in the previous section, the contribution of electrons from $\rm{W}^{\pm}$ and $\rm{Z}^{0}$ boson decays is non-negligible for $\pt> 15~\GeVc$ and estimated to be $\sim 30\%$ of the yield of heavy-flavor decay electrons at $\pt = 26~\GeVc$~\cite{ALICE:2019nuy}. The effect of this contribution was also studied using a MC closure test, where the $d_{0}$ distribution of Dalitz electrons was used as a proxy for $\rm{W}^{\pm}$ and $\rm{Z}^{0}$ decays, as they both decay close to the primary vertex and the difference is less than the detector resolution of the $d_{0}$ distribution. 
The $d_{0}$ distribution of electrons from $\rm{W}^{\pm}$ decays was added to that of heavy-flavor decay electrons in the simulated sample, which was then fitted using only the charm and beauty templates. The beauty yield obtained from the fit was compared to the true beauty yield to estimate any difference. The maximum difference was $2\%$ for $16 < \pt < 20~\GeVc$ rising to $6\%$ for the highest \pt~interval, and was added as a systematic uncertainty. 

The stability of the weighted log-likelihood fit was studied by changing the histogram bin sizes and the fit ranges. The effect on the beauty yield from the fitting routine was found to be negligible.  

The uncertainty on the impact parameter resolution was estimated as described in Sec.~\ref{FourTemplateSystematicsSection}, and a value of 2\% was assigned. 

As discussed in Sec.~\ref{subsec:4Template}, the $d_0$ templates of charm- and beauty-hadron decays were obtained using MC simulations after several corrections. The systematic effect of these corrections on the yield of electrons from beauty-hadron decays was assessed by varying the weights applied on the D and B hadron \pt~spectra, and varying the ratio of different charm-hadron species, in the same way as for the four-template approach. The uncertainty from the assumed beauty hadron \pt~distribution results in a systematic uncertainty of $10\%$ for $\pt < 6~\GeVc$~and $3\%$ for higher \pt. The uncertainty from the charm-hadron \pt-distributions was obtained by varying the slope of the D-meson \pt~spectra within the statistical and systematic uncertainty of the D-meson measurement~\cite{ALICE:2018lyv} and ranges from$~1\%$ for \pt~up to 20 \GeVc ~to 5\% for the highest \pt~interval. The uncertainty due to the relative abundances of the different charm-hadron species in the MC, obtained by varying the $\Lambda_{\rm{c}}^{+}/\rm{D}^0$~\cite{ALICE:2018hbc,ALICE:2021bib}, $\rm{D}^+/\rm{D}^0$~\cite{ALICE:2018lyv,ALICE:2021rxa}, and $\rm{D_s}^{+}/\rm{D}^0$~\cite{ALICE:2018lyv,ALICE:2021kfc} fractions within the statistical and systematic uncertainty of the measurements, was estimated to be 5\% over the entire \pt~range.

As described for the four-template method, an additional uncertainty of 15\% was assigned to account for differences in the results of the two methods, estimated in their overlap region ($3 < \pt < 8~\GeVc$).

%% file: Results.tex
\section{Results}
\label{sec_Results}

\begin{figure}
    \centering
    \includegraphics[width=0.56\columnwidth]{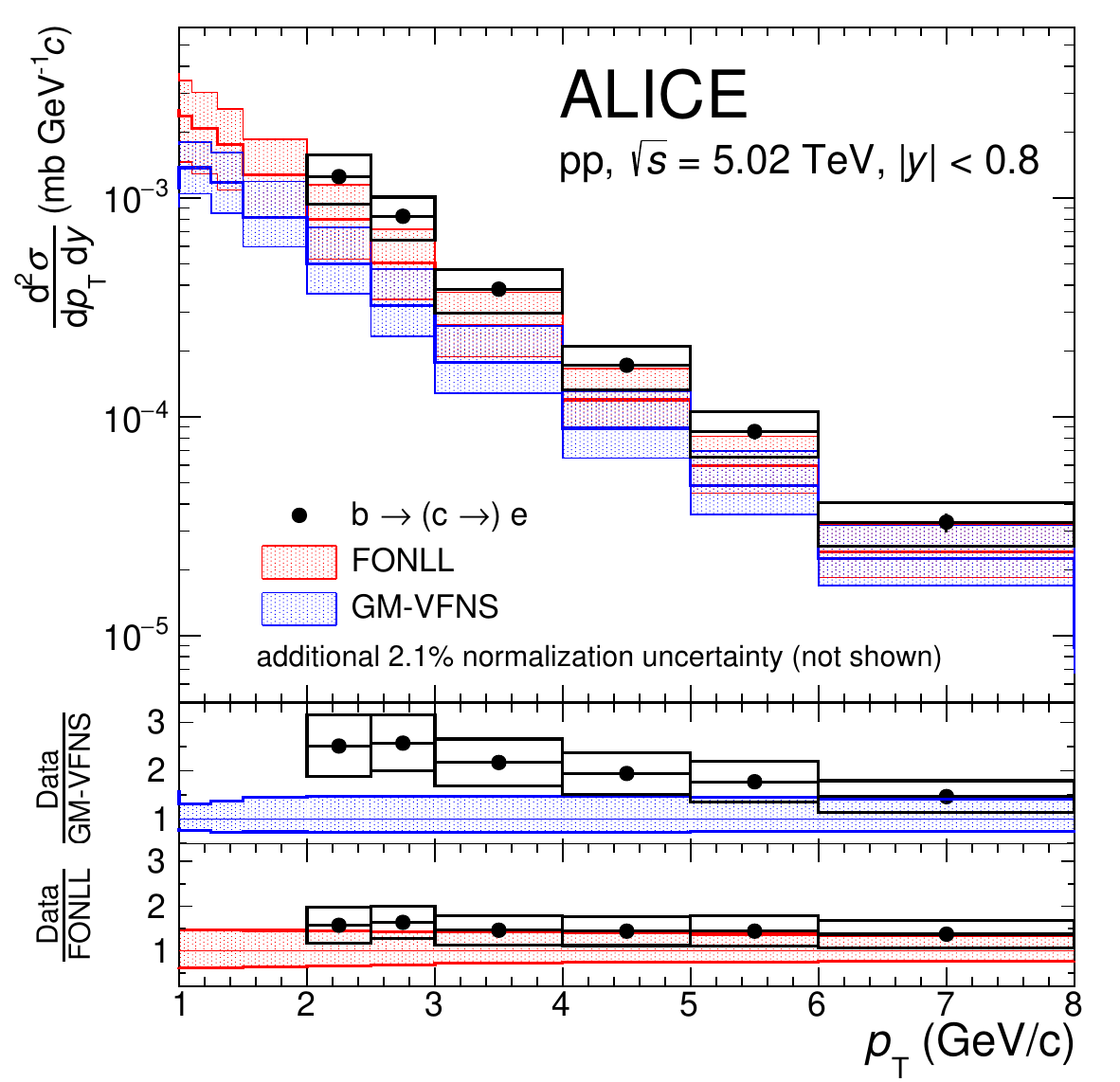}
    \caption{$\pt$-differential cross section of electrons from beauty-hadron decays in pp collisions at $\sqrt{s}=5.02~{\rm TeV}$ using TPC--TOF for electron identification. In the upper panel, the cross section is compared with FONLL~\cite{Cacciari:1998it,Cacciari:2001td,Cacciari:2012ny} and GM-VFNS~\cite{Bolzoni:2012kx} predictions. The ratios of data to these calculations are reported in the bottom panels.
    }
    \label{fig::pTSpectra_pp}
\end{figure}

The $\pt$-differential production cross section of electrons from beauty-hadron decays in pp collisions at $\sqrt{s}=5.02$~TeV, measured at midrapidity in the transverse momentum interval $2<\pt<8~{\rm GeV}/c$, is shown in Fig.~\ref{fig::pTSpectra_pp}. Vertical error bars depict the statistical uncertainties while the systematic uncertainties are shown by rectangular boxes. The cross section is  compared with pQCD calculations such as 
%Fixed-Order-Next-to-Leading-Log (FONLL)
FONLL~\cite{Cacciari:1998it,Cacciari:2001td,Cacciari:2012ny}, and GM-VFNS~\cite{Bolzoni:2012kx}. For the prediction of the \pt ~distribution of leptons from beauty-hadron decays, FONLL calculations use a numerical convolution of a perturbative cross section with a non-perturbative fragmentation function and a decay function for the hadron weak decay into a lepton~\cite{Cacciari:2012ny}. The parameters of the fragmentation function are determined from $\rm{e^{+}} \rm{e}^{-}$ collision data using $\rm{B}^{+}$, $\rm{B}^{-}$, and $\rm{B}^{0}$ mesons.
%assuming that all beauty hadrons fragment in a similar manner.
The weak decay function and the branching ratios are also extracted from experimental data. 
As the fragmentation functions are determined using only B mesons, possible differences in the fragmentation and the decay kinematics of the substantial contribution from beauty baryons~\cite{LHCb:2019fns} are not considered in FONLL calculations. 
The uncertainty bands of the FONLL calculations are the result of different choices for the mass of beauty quarks, and for the factorization and renormalization scales as well as the uncertainty on the set of parton distribution functions (PDF) used in the pQCD calculations.
In the GM-VFNS approach, the contribution of electrons from beauty-hadron decays is calculated from the convolution of the hard-scattering cross section at the partonic level, a non-perturbative fragmentation function, the total beauty-hadron decay width, and the decay spectrum to leptons. The non-perturbative fragmentation functions were obtained based on $\rm{e^{+}} \rm{e}^{-}$ data using all B mesons ($\rm{B}^{+}$, $\rm{B}^{-}$, and $\rm{B}^{0}$) and $\Lambda_{\rm{b}}$~\cite{Bolzoni:2012kx}. The decay width and the spectrum were obtained using the electron energy spectrum in inclusive beauty-meson decays measured by the BABAR experiment~\cite{BaBar:2004bij}. The theoretical uncertainty of the GM-VFNS calculations was obtained by varying the scale parameters related to renormalization and to the factorization of initial- and final-state singularities. The uncertainties due to scale variation are the dominating source and hence PDF-related uncertainties and variations of the bottom mass were not considered.
It should be noted that the results using the four-template method in pp and Pb--Pb collisions include a small contribution ($\sim2\%$) of beauty-hadron decays via light mesons, which is not included in the signal definition of the pQCD calculations.
The data lies on the upper edge of the FONLL uncertainty band, around a factor 1.5 above the central prediction, similarly to earlier measurements in pp collisions at $\sqrt{s}=7~{\rm TeV}$~\cite{ALICE:2012acz}. The measurement using semimuonic decays from the ATLAS Collaboration~\cite{ATLAS:2021xtw} shows a similar behavior in the \pt~interval common between ALICE and ATLAS, though it is close to the FONLL central value for $\pt>10~\GeVc$. The comparison of FONLL with the previous measurement of non-prompt D mesons~\cite{ALICE:2021mgk}, however, shows the center of the theory prediction closer to the data. This difference in the FONLL description of beauty decay electrons and non-prompt D mesons could be due to the contribution from beauty baryons. The beauty baryons produce electrons with a similar branching ratio, and decay kinematics as beauty mesons. However, the branching ratio of beauty baryons into D mesons is smaller than that for beauty mesons. Since all beauty quarks are assumed to fragment into mesons in the FONLL calculations, the non-prompt D meson contribution would be increased relative to the electrons.
Comparison of the results to GM-VFNS predictions shows some tension for $\pt < 4~{\rm GeV}/c$.
In the second \pt ~interval the central point of the measurement is about $1.9\sigma$ above the upper edge of the GM-VFNS uncertainty band when considering the combined statistical and systematic uncertainties of the measurement. At higher \pt, the predictions are in closer agreement between each other and the data. A qualitatively similar result was found previously for non-prompt D mesons~\cite{ALICE:2021mgk}. The models differ in their assumptions concerning the fragmentation functions and the transition to a fixed-flavor number scheme, which should make differences more apparent at low momenta~\cite{Kniehl:2011bk}.

\begin{figure}
    \centering
    \includegraphics[width=0.45\columnwidth]{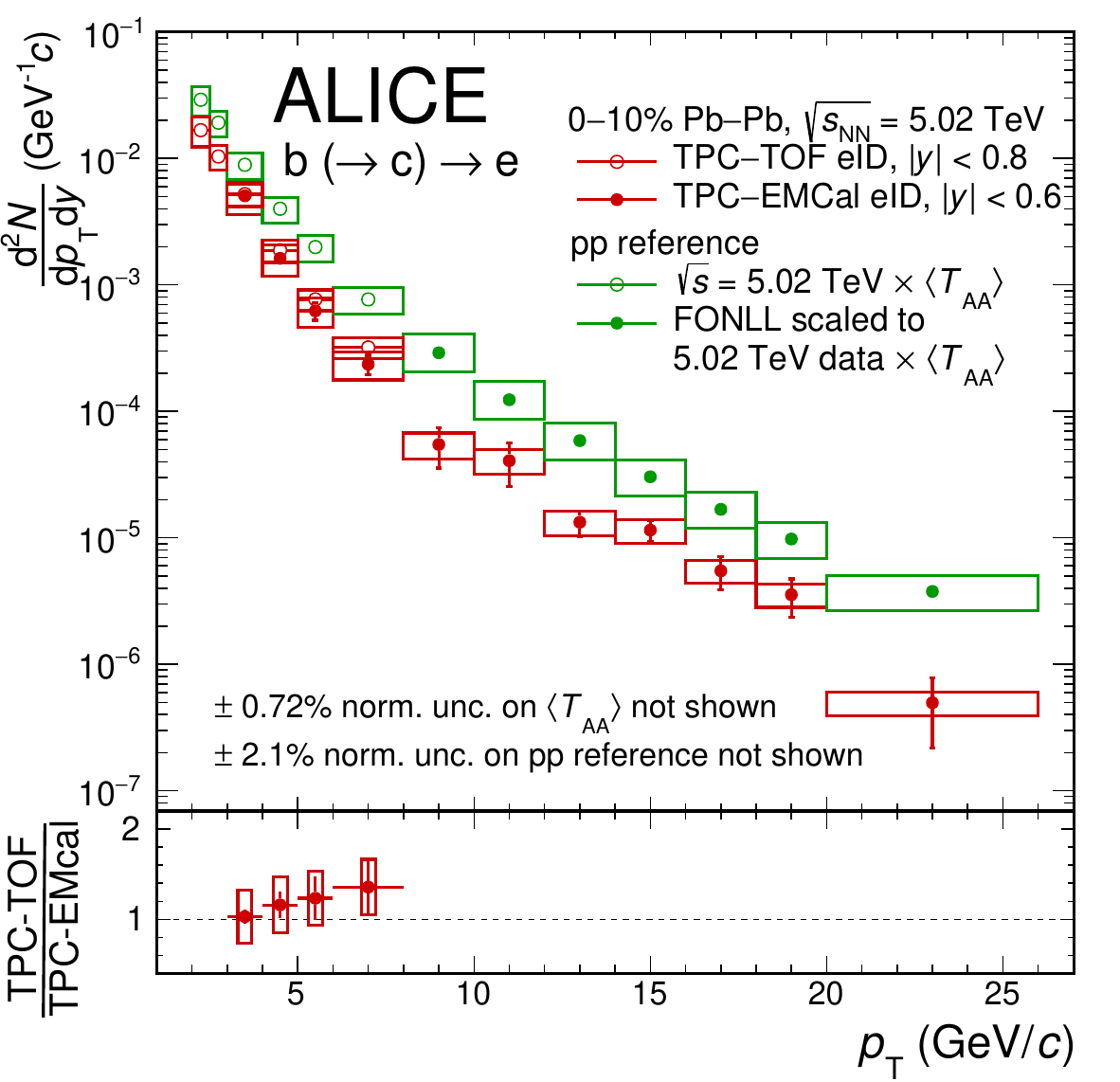}
    \caption{Yield of beauty-hadron decay electrons in 0--10\% central Pb--Pb collisions at $\sqrt{s_{\rm{NN}}} = 5.02~{\rm TeV}$ for the TPC--TOF and TPC--EMCal analyses compared with the pp reference scaled by $\left<T_{\rm{AA}}\right>$, obtained from the measured cross section for \pt~$< 8$ \GeVc, which is extrapolated up to $p_\mathrm{T}= 26~\GeVc$~using FONLL. The ratio of the yields using TPC--TOF and TPC--EMCal analyses in the overlapping interval of $3 < \pt < 8~\GeVc$ in Pb--Pb collisions is shown in the bottom panel.}
    \label{fig::pTSpectra_PbPb}
\end{figure}

The measured cross section in pp collisions at $\sqrt{s} = 5.02~{\rm TeV}$ in Fig.~\ref{fig::pTSpectra_pp} was used as a reference to calculate the $R_{\mathrm{AA}}$ up to $\pt = 8~{\rm GeV}/c$. For $\pt > 8~{\rm GeV}/c$, the FONLL prediction for electrons from beauty-hadron decays in pp collisions at $\sqrt{s} = 5.02~{\rm TeV}$ was used. Since the FONLL central prediction is lower than the data at low momentum, the FONLL reference was scaled to match the data.
The scaling factor was determined by taking the ratio between the measured cross section and the FONLL prediction for $\pt < 8~{\rm GeV}/c$. This ratio reaches a plateau for $\pt > 4~{\rm GeV}/c$, and was assumed to be constant at higher momenta where the data points are unavailable. The scale factor of $1.40 \pm 0.08$ was determined by fitting the ratio with a constant for $\pt> 4~\GeVc$. The statistical and systematic uncertainties of the measurement were propagated to obtain the systematic uncertainty associated with the scaling factor, assuming no correlation between \pt ~intervals for the statistical uncertainties and full correlation for the systematic ones.
The total uncertainty assigned to the scaled FONLL reference is calculated by taking the sum in quadrature of the uncertainty from the original FONLL prediction and the systematic uncertainty of the scaling, and is approximately 30\% for $p_{\mathrm{T}} > 8~{\rm GeV}/c$. The result of the extrapolation, scaled with the nuclear overlap function, is shown in Fig.~\ref{fig::pTSpectra_PbPb}.

%\subsection{$p_\mathrm{T}$-differential invariant yield in Pb--Pb collisions}

The $\pt$-differential yields for electrons from beauty-hadron decays in the 10\% most central Pb--Pb collisions are shown in Fig.~\ref{fig::pTSpectra_PbPb}. The yields are obtained using the four-template method with TPC--TOF detectors in the interval $2 < \pt < 8~\GeVc$, and using the two-template method with TPC--EMCal detectors in the interval $3 < \pt < 26~\GeVc$. The ratio of the yields from the two methods in the overlapping interval of $3 < \pt < 8~\GeVc$ is shown in the bottom panel of Fig.~\ref{fig::pTSpectra_PbPb}. The systematic uncertainties are propagated as uncorrelated from all sources, except for the 15\% uncertainty assigned for the differences in the results of the two methods, which is not considered in the ratio. Some of the remaining systematic uncertainties have common sources and are thus correlated to some degree, which is difficult to estimate. The ratio is consistent with unity within statistical and systematic uncertainties.
For the final yield, TPC--TOF results were used in the overlapping \pt~range because of their smaller statistical and systematic uncertainties. The Pb--Pb results are shown together with the pp results, scaled by the estimated nuclear overlap function~\cite{ALICE-PUBLIC-2018-011}, which is proportional to the number of binary collisions. 

\begin{figure}
    \centering
    \includegraphics[width=0.495\columnwidth]{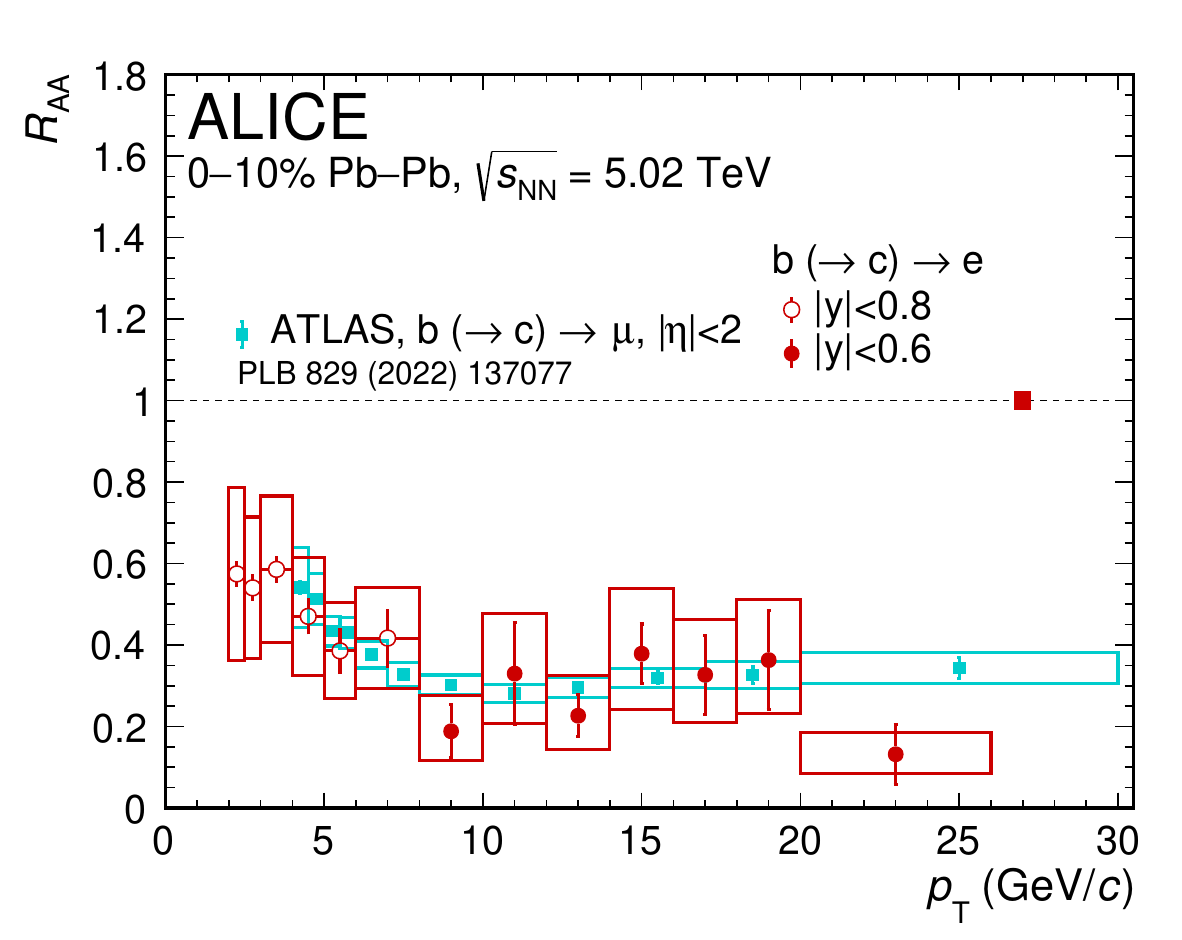}
    \includegraphics[width=0.495\columnwidth]{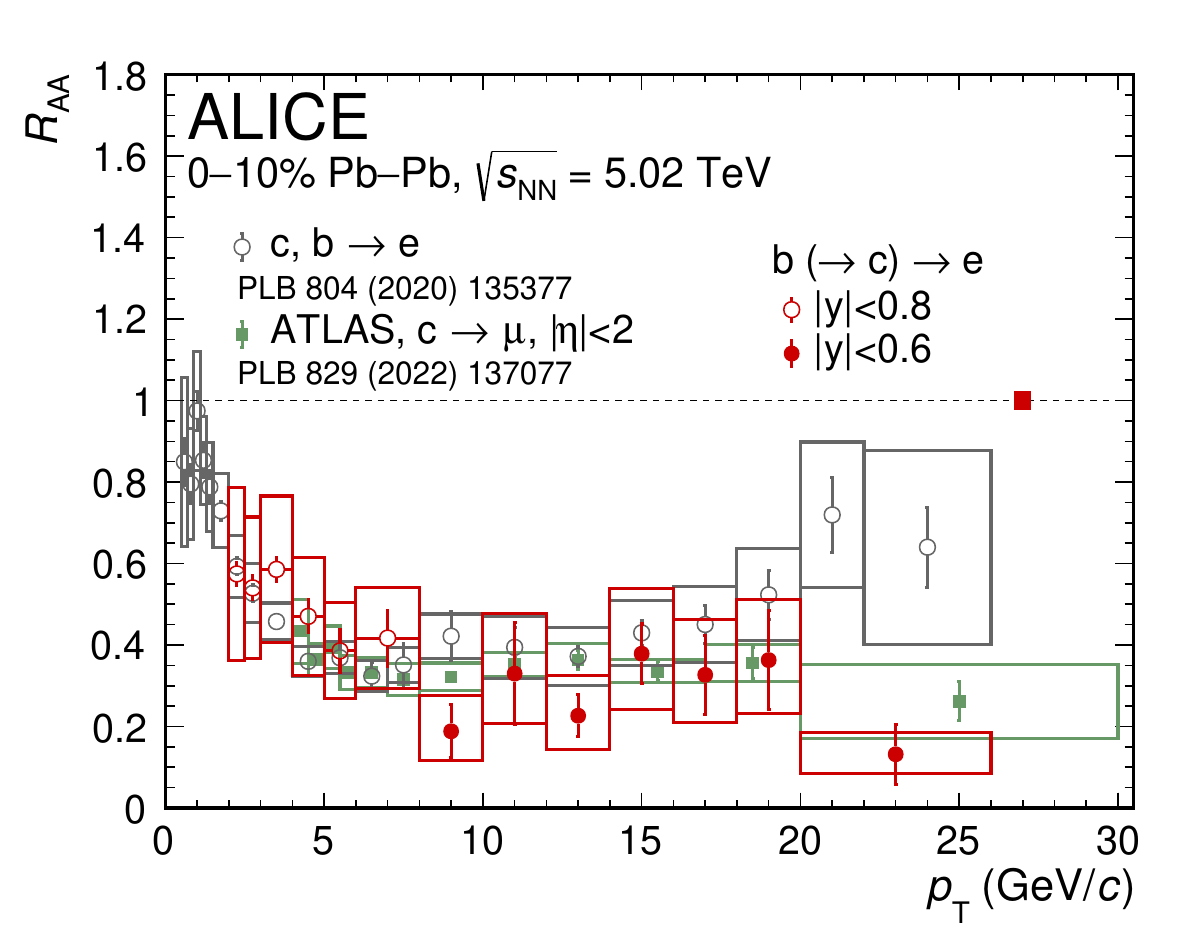}
    \caption{Nuclear modification factor of electrons from beauty-hadron decays in the $10\%$ most central Pb--Pb collisions at $\sqrt{s_{\rm{NN}}} = 5.02~{\rm TeV}$. Left: Comparison with \Raa of muons from beauty-hadron decays measured by the ATLAS Collaboration~\cite{ATLAS:2021xtw}. Right:  comparison with the measurements of \Raa of electrons from heavy-flavor hadron (beauty plus charm) decays~\cite{ALICE:2019nuy}, and with the \Raa of muons from charm-hadron decays measured by the ATLAS Collaboration~\cite{ATLAS:2021xtw}. }
    \label{fig:RAA_CompData}
\end{figure}

The nuclear modification factor for electrons from beauty-hadron decays in Pb--Pb collisions as a function of \pt ~in the  0--10\% centrality interval is shown in Fig.~\ref{fig:RAA_CompData}.  The nuclear modification factor in the measured \pt ~range of $2 < \pt < 26 ~\GeVc$ is lower than unity, consistent with the expectation of suppression of the yield in this \pt ~interval due to in-medium parton energy loss. Considering that the systematic uncertainties are mostly correlated across \pt ~intervals, the measured \Raa suggests a broadly increasing suppression with increasing \pt ~up to  $\pt \sim 5~\GeVc$. For \pt ~above 8 \GeVc,~the \Raa does not change significantly with values between 0.2 and 0.4, and a maximum suppression observed for \pt ~around 8--10~\GeVc.  
The measured \Raa of electrons from beauty-hadron decays is compared with the \Raa of muons from beauty-hadron decays, measured by the ATLAS Collaboration~\cite{ATLAS:2021xtw}, in the 10\% most central Pb--Pb collisions at $\sqrt{s_{\rm{NN}}} = 5.02$ TeV in the interval $4 < \pt < 30~\GeVc$ and $|y| < 2$. As mentioned above, the pp reference for the ATLAS measurement decreases faster with \pt~than the central FONLL extrapolation used here. However, the two measurements are consistent with each other within statistical and systematic uncertainties.  
The non-prompt $\rm{D}^{0}$ meson~\cite{ALICE:2022tji} and the non-prompt $\rm{D}^{+}_{\rm{s}}$ meson~\cite{ALICE:2022xrg} \Raa measured by the ALICE Collaboration in the 10\% most central Pb--Pb collisions also show a similar \pt ~dependence, with the minimum value at around 0.35 for \pt ~$\sim 10$~\GeVc. Possible differences may arise between the \Raa of leptons and non-prompt $\rm{D}$ mesons due to different decay kinematics of B mesons in the two decay channels.
For $\pt > 4$ \GeVc, the \Raa of electrons from beauty-hadron decays shows a similar suppression as observed in similar measurements at lower collision energies, namely at $\sqrt{s_{\rm{NN}}} = 2.76$~TeV by the ALICE Collaboration~\cite{ALICE:2016uid} at the LHC, and at $\sqrt{s_{\rm{NN}}} = 0.2$~TeV by the PHENIX~\cite{PHENIX:2022wim} and the STAR~\cite{STAR:2021uzu} Collaborations at RHIC. Within large uncertainties, the results at lower collision energies show somewhat higher \Raa values at lower \pt. The similar \Raa values at high \pt ~could be caused by the the interplay between the medium temperature and density, and the \pt ~distribution of beauty quarks at the two collision energies~\cite{Djordjevic:2015hra}. A similar trend was observed for prompt D mesons when comparing results at the two collisions energies~\cite{ALICE:2021rxa}.

To understand the mass ordering effects on the energy loss, the \Raa of electrons from beauty-hadron decays is compared with the \Raa of electrons from heavy-flavor hadron (beauty and charm) decays~\cite{ALICE:2019nuy}, and with the \Raa of muons from charm-hadron decays measured by the ATLAS Collaboration~\cite{ATLAS:2021xtw}, as a function of \pt ~in the 10\% most central Pb--Pb collisions, in the right panel of Fig.~\ref{fig:RAA_CompData}. The electrons from heavy-flavor hadron decays originate mostly from charm-hadron decays at low \pt, more than 70\% for \pt~$< 4$ \GeVc, with the contribution from beauty-hadron decays that increases with increasing \pt~\cite{ALICE:2014aev}, and becomes the dominant source ($> 70\%$) for \pt~$> 8$~\GeVc. 
Within uncertainties, the \Raa of b($\rightarrow$~c)$\rightarrow$~e in the 10\% most central Pb--Pb collisions shows similar values to that of c$\rightarrow \mu$ and c,b$\rightarrow$~e. While the central points of the \Raa values  might be slightly higher for b($\rightarrow$ c)$\rightarrow$~e compared to c$\rightarrow \mu$ at low~\pt~($< 5  ~\GeVc$), the values are very similar and consistent within the uncertainties, and thus precise measurements of leptons from beauty-hadron decays would be required to see a potential difference.

\begin{figure}
    \centering
    \includegraphics[width=0.495\columnwidth]{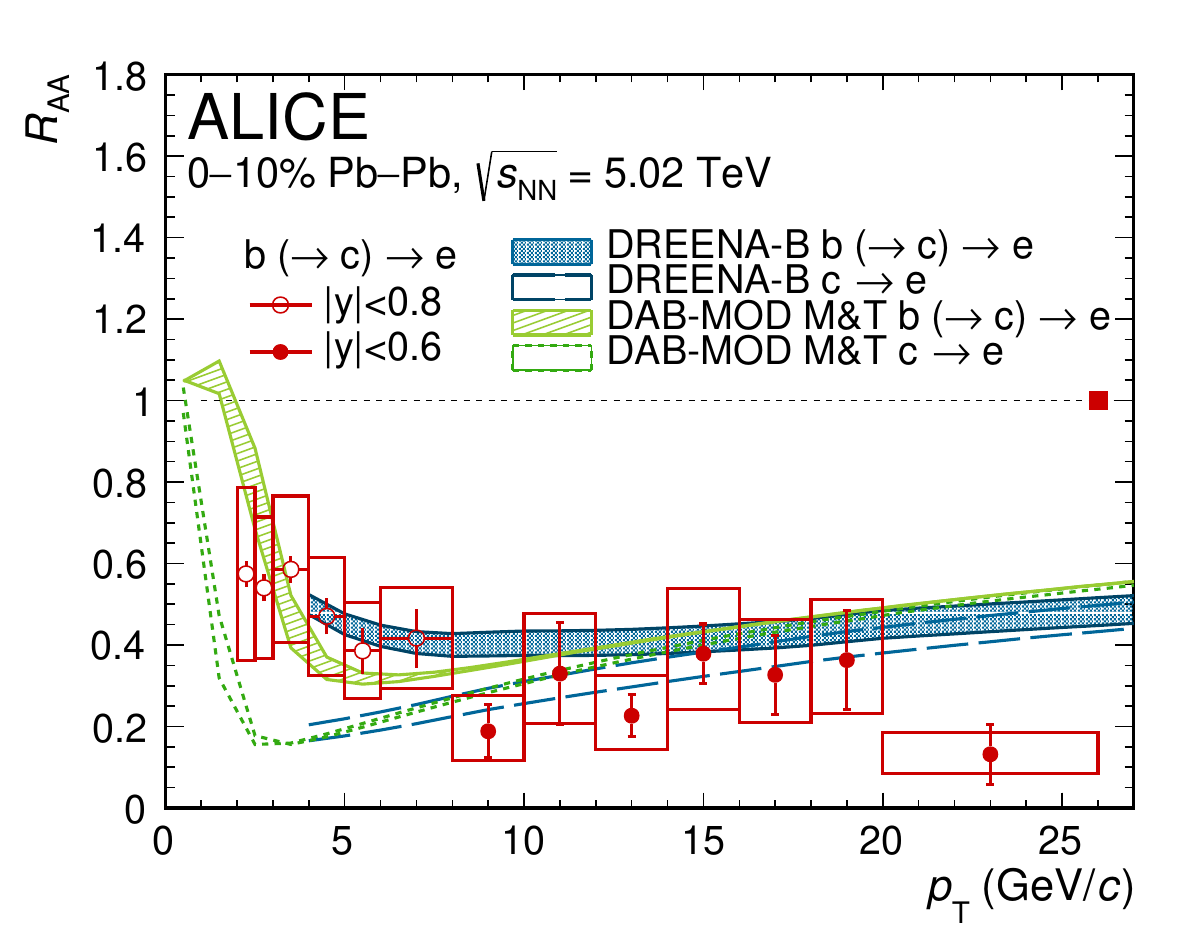}
    \includegraphics[width=0.495\columnwidth]{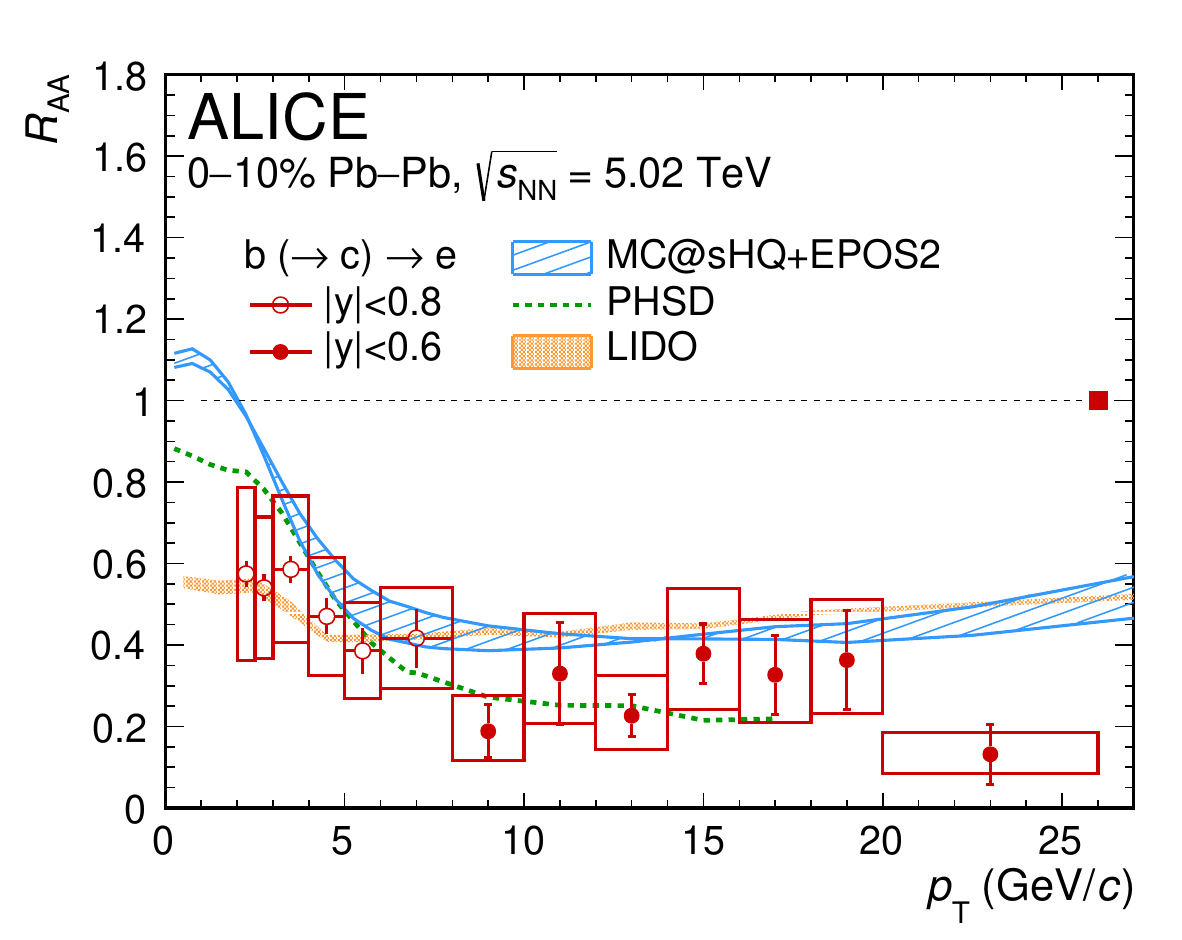}
    \caption{Nuclear modification factor of electrons from beauty-hadron decays in the $10\%$ most central Pb--Pb collisions at $\sqrt{s_{\rm{NN}}} = 5.02~{\rm TeV}$ compared with predictions from several theoretical calculations. Left: Comparison with predictions from DREENA-B~\cite{zigic_dreena-b_2019} and DAB-MOD M\&T~\cite{Prado:2016szr} models for electrons from beauty and charm decays. Right: Comparison with predictions from MC@sHQ~\cite{Nahrgang:2013xaa}, PHSD~\cite{Song:2015sfa}, and LIDO~\cite{Ke:2018tsh} calculations for electrons from beauty decays.}
    \label{fig::RAA_models}
\end{figure}

The nuclear modification factor of electrons from beauty-hadron decays is compared with DREENA-B~\cite{zigic_dreena-b_2019} and DAB-MOD M\&T~\cite{Prado:2016szr} model predictions in the left panel of Fig.~\ref{fig::RAA_models}, and with MC@sHQ~\cite{Nahrgang:2013xaa}, PHSD~\cite{Song:2015sfa}, and LIDO~\cite{Ke:2018tsh} models in the right panel of Fig.~\ref{fig::RAA_models}. All models include the assumption of a dynamically expanding QGP. Each model makes different hypotheses about the mass dependence of the energy loss within the QGP, transport dynamics, and hadronization of the beauty quarks. All the models include collisional and radiative energy-loss processes with the exception of PHSD, which only includes collisional energy loss. The models MC@sHQ, PHSD, and LIDO include hadronization via coalescence~\cite{Prino:2016cni} at low and intermediate momentum, and via fragmentation at high momentum, while the DREENA-B and DAB-MOD M\&T models use fragmentation in the full momentum range considered. Initial-state effects are included by using nuclear PDFs in the calculation of the initial \pt ~distributions of heavy quarks in all models, except for DAB-MOD. All the models give a fair description of the data within the uncertainties of the measurement. For $\pt>5$ \GeVc, all models except for PHSD predict similar \Raa, with values lying on the upper edge of the data uncertainties. The PHSD model gives lower \Raa values possibly due to the higher probability of large momentum transfers in the in-medium interactions as compared to the other models~\cite{Song:2015sfa}. Predictions for all models, but DREENA-B, are available down to very low \pt, where they show significant differences among each other.
However, measurements with improved precision will be needed to discriminate among these models.
The left panel of Fig.~\ref{fig::RAA_models} also shows the predictions for $\rm{c}\rightarrow \rm{e}$ \Raa from DREENA-B~\cite{zigic_dreena-b_2019} and DAB-MOD M\&T~\cite{Prado:2016szr} models. The difference in the \Raa of electrons from charm- and beauty-hadron decays is larger at low \pt~and reduces at high \pt~where the mass effects become negligible. However, the current measurements (see right panel of Fig.~\ref{fig:RAA_CompData}) do not have enough precision to confirm this prediction.

As previously observed with charm measurements~\cite{ALICE:2021rxa, ALICE:2017pbx}, it is challenging for models to simultaneously describe \Raa and $v_2$ of heavy-flavor particles, allowing data to provide constraints to the model ingredients and parameters to describe the interaction of heavy quarks with the QGP medium. Among the models presented in this article, the MC@sHQ~\cite{Nahrgang:2013xaa} and LIDO~\cite{Ke:2018tsh} models best describe the \Raa, $v_2$, and $v_3$ measurements of D mesons~\cite{ALICE:2021rxa}. In the beauty sector, the $v_2$ of leptons from beauty-hadron decays was measured by the ALICE~\cite{ALICE:2020hdw} and ATLAS~\cite{ATLAS:2021xtw} Collaborations. The ALICE measurement of electrons from beauty-hadron decays was performed in the interval $1.3 < \pt < 6~\GeVc$, and compared with predictions from  MC@sHQ~\cite{Nahrgang:2013xaa}, PHSD~\cite{Song:2015sfa}, and LIDO~\cite{Ke:2018tsh} models. These models predict similar \pt-dependent $v_2$ values in the full \pt~interval, and all models describe the data within uncertainties above 2~\GeVc. Some parameters of the LIDO model were calibrated to reproduce previous D meson and B meson measurements by the ALICE and CMS Collaborations~\cite{Ke:2018tsh}. Extension of $v_2$ measurements to higher \pt~would be beneficial for comparing the models at high \pt. The $v_2$ of muons from beauty-hadron decays from the ATLAS Collaboration~\cite{ATLAS:2021xtw}, measured in the interval $4 < \pt < 20~\GeVc$, was compared with  DREENA-B~\cite{zigic_dreena-b_2019} and DAB-MOD M\&T~\cite{Prado:2016szr} models. While the two models provide similar $\rm{b} \rightarrow \rm{e}$ \Raa predictions in the available \pt~range, they significantly vary in $v_2$ predictions below 10~\GeVc, where DREENA-B model is qualitatively in better agreement with the data. In this context, measurements of beauty decay electrons can provide additional and important constraints for modeling the heavy quark in-medium interactions and hadronization.

%% file: Summary.tex
\section{Summary}

The $\pt$-differential production of beauty-hadron decay electrons was measured in pp collisions and in the 10\% most central Pb--Pb collisions at \comPbPb. The measurements are based on electron identification together with a fit to the track impact parameter distributions to extract the beauty contribution. The measured \pt-differential cross section in pp collisions lies at the upper edge of the FONLL uncertainty band, and shows some tension with GM-VFNS calculations at low \pt. The measured nuclear modification factor in central Pb--Pb collisions shows an increasing suppression with increasing \pt ~up to $\sim5$ \GeVc, and is almost constant at higher \pt. The maximum suppression of about a factor 3 is observed around 8--10~\GeVc. The measured \Raa of $\rm{b}(\rightarrow\rm{c})\rightarrow\rm{e}$ is consistent with the measurement of $\rm{b}(\rightarrow\rm{c})\rightarrow\mu$ by the ATLAS Collaboration. The \Raa of leptons from beauty-hadron decays shows a  similar suppression and shape compared to charm-hadron decays within uncertainties. The results are consistent with several transport models implementing interactions of heavy quarks with a QGP formed in Pb--Pb collisions. While models implementing radiative and collisional energy loss processes predict similar \Raa values at high \pt, significant differences among models exist at low \pt, but more precise measurements are needed to constrain the model parameters further.

%% file: fa_2022-09-28_Opt_C.tex
% Version: 2022-09-28

The ALICE Collaboration would like to thank all its engineers and technicians for their invaluable contributions to the construction of the experiment and the CERN accelerator teams for the outstanding performance of the LHC complex.
The ALICE Collaboration gratefully acknowledges the resources and support provided by all Grid centres and the Worldwide LHC Computing Grid (WLCG) collaboration.
The ALICE Collaboration acknowledges the following funding agencies for their support in building and running the ALICE detector:
A. I. Alikhanyan National Science Laboratory (Yerevan Physics Institute) Foundation (ANSL), State Committee of Science and World Federation of Scientists (WFS), Armenia;
Austrian Academy of Sciences, Austrian Science Fund (FWF): [M 2467-N36] and Nationalstiftung f\"{u}r Forschung, Technologie und Entwicklung, Austria;
Ministry of Communications and High Technologies, National Nuclear Research Center, Azerbaijan;
Conselho Nacional de Desenvolvimento Cient\'{\i}fico e Tecnol\'{o}gico (CNPq), Financiadora de Estudos e Projetos (Finep), Funda\c{c}\~{a}o de Amparo \`{a} Pesquisa do Estado de S\~{a}o Paulo (FAPESP) and Universidade Federal do Rio Grande do Sul (UFRGS), Brazil;
Bulgarian Ministry of Education and Science, within the National Roadmap for Research Infrastructures 2020¿2027 (object CERN), Bulgaria;
Ministry of Education of China (MOEC) , Ministry of Science \& Technology of China (MSTC) and National Natural Science Foundation of China (NSFC), China;
Ministry of Science and Education and Croatian Science Foundation, Croatia;
Centro de Aplicaciones Tecnol\'{o}gicas y Desarrollo Nuclear (CEADEN), Cubaenerg\'{\i}a, Cuba;
Ministry of Education, Youth and Sports of the Czech Republic, Czech Republic;
The Danish Council for Independent Research | Natural Sciences, the VILLUM FONDEN and Danish National Research Foundation (DNRF), Denmark;
Helsinki Institute of Physics (HIP), Finland;
Commissariat \`{a} l'Energie Atomique (CEA) and Institut National de Physique Nucl\'{e}aire et de Physique des Particules (IN2P3) and Centre National de la Recherche Scientifique (CNRS), France;
Bundesministerium f\"{u}r Bildung und Forschung (BMBF) and GSI Helmholtzzentrum f\"{u}r Schwerionenforschung GmbH, Germany;
General Secretariat for Research and Technology, Ministry of Education, Research and Religions, Greece;
National Research, Development and Innovation Office, Hungary;
Department of Atomic Energy Government of India (DAE), Department of Science and Technology, Government of India (DST), University Grants Commission, Government of India (UGC) and Council of Scientific and Industrial Research (CSIR), India;
National Research and Innovation Agency - BRIN, Indonesia;
Istituto Nazionale di Fisica Nucleare (INFN), Italy;
Japanese Ministry of Education, Culture, Sports, Science and Technology (MEXT) and Japan Society for the Promotion of Science (JSPS) KAKENHI, Japan;
Consejo Nacional de Ciencia (CONACYT) y Tecnolog\'{i}a, through Fondo de Cooperaci\'{o}n Internacional en Ciencia y Tecnolog\'{i}a (FONCICYT) and Direcci\'{o}n General de Asuntos del Personal Academico (DGAPA), Mexico;
Nederlandse Organisatie voor Wetenschappelijk Onderzoek (NWO), Netherlands;
The Research Council of Norway, Norway;
Commission on Science and Technology for Sustainable Development in the South (COMSATS), Pakistan;
Pontificia Universidad Cat\'{o}lica del Per\'{u}, Peru;
Ministry of Education and Science, National Science Centre and WUT ID-UB, Poland;
Korea Institute of Science and Technology Information and National Research Foundation of Korea (NRF), Republic of Korea;
Ministry of Education and Scientific Research, Institute of Atomic Physics, Ministry of Research and Innovation and Institute of Atomic Physics and University Politehnica of Bucharest, Romania;
Ministry of Education, Science, Research and Sport of the Slovak Republic, Slovakia;
National Research Foundation of South Africa, South Africa;
Swedish Research Council (VR) and Knut \& Alice Wallenberg Foundation (KAW), Sweden;
European Organization for Nuclear Research, Switzerland;
Suranaree University of Technology (SUT), National Science and Technology Development Agency (NSTDA), Thailand Science Research and Innovation (TSRI) and National Science, Research and Innovation Fund (NSRF), Thailand;
Turkish Energy, Nuclear and Mineral Research Agency (TENMAK), Turkey;
National Academy of  Sciences of Ukraine, Ukraine;
Science and Technology Facilities Council (STFC), United Kingdom;
National Science Foundation of the United States of America (NSF) and United States Department of Energy, Office of Nuclear Physics (DOE NP), United States of America.
In addition, individual groups or members have received support from:
Marie Sk\l{}odowska Curie, European Research Council, Strong 2020 - Horizon 2020 (grant nos. 950692, 824093, 896850), European Union;
Academy of Finland (Center of Excellence in Quark Matter) (grant nos. 346327, 346328), Finland;
Programa de Apoyos para la Superaci\'{o}n del Personal Acad\'{e}mico, UNAM, Mexico;
the DST-DAAD Project-based Personnel Exchange Programme, India.

%% file: 2022-09-28-Alice_Authorlist_2022-09-28_Opt_C.tex
% ALICE Collaboration author list for 2022-09-28
\begin{flushleft} 
\small

S.~Acharya\,\orcidlink{0000-0002-9213-5329}\,$^{\rm 125}$, 
D.~Adamov\'{a}\,\orcidlink{0000-0002-0504-7428}\,$^{\rm 86}$, 
A.~Adler$^{\rm 69}$, 
G.~Aglieri Rinella\,\orcidlink{0000-0002-9611-3696}\,$^{\rm 32}$, 
M.~Agnello\,\orcidlink{0000-0002-0760-5075}\,$^{\rm 29}$, 
N.~Agrawal\,\orcidlink{0000-0003-0348-9836}\,$^{\rm 50}$, 
Z.~Ahammed\,\orcidlink{0000-0001-5241-7412}\,$^{\rm 132}$, 
S.~Ahmad\,\orcidlink{0000-0003-0497-5705}\,$^{\rm 15}$, 
S.U.~Ahn\,\orcidlink{0000-0001-8847-489X}\,$^{\rm 70}$, 
I.~Ahuja\,\orcidlink{0000-0002-4417-1392}\,$^{\rm 37}$
% ...

\section*{Affiliation Notes}

$^{\rm I}$ Deceased\\
$^{\rm II}$ Also at: Max-Planck-Institut f\"{u}r Physik, Munich, Germany\\
$^{\rm III}$ Also at: Italian National Agency for New Technologies, Energy and Sustainable Economic Development (ENEA), Bologna, Italy\\
$^{\rm IV}$ Also at: Dipartimento DET del Politecnico di Torino, Turin, Italy\\
$^{\rm V}$ Also at: Department of Applied Physics, Aligarh Muslim University, Aligarh, India\\
$^{\rm VI}$ Also at: Institute of Theoretical Physics, University of Wroclaw, Poland\\
$^{\rm VII}$ Also at: An institution covered by a cooperation agreement with CERN\\

\section*{Collaboration Institutes}

$^{1}$ A.I. Alikhanyan National Science Laboratory (Yerevan Physics Institute) Foundation, Yerevan, Armenia\\
$^{2}$ AGH University of Science and Technology, Cracow, Poland\\
$^{3}$ Bogolyubov Institute for Theoretical Physics, National Academy of Sciences of Ukraine, Kiev, Ukraine\\
$^{4}$ Bose Institute, Department of Physics  and Centre for Astroparticle Physics and Space Science (CAPSS), Kolkata, India\\
$^{5}$ California Polytechnic State University, San Luis Obispo, California, United States\\
$^{6}$ Central China Normal University, Wuhan, China\\
$^{7}$ Centro de Aplicaciones Tecnol\'{o}gicas y Desarrollo Nuclear (CEADEN), Havana, Cuba\\
$^{8}$ Centro de Investigaci\'{o}n y de Estudios Avanzados (CINVESTAV), Mexico City and M\'{e}rida, Mexico\\
$^{9}$ Chicago State University, Chicago, Illinois, United States\\
$^{10}$ China Institute of Atomic Energy, Beijing, China\\
$^{11}$ Chungbuk National University, Cheongju, Republic of Korea\\
$^{12}$ Comenius University Bratislava, Faculty of Mathematics, Physics and Informatics, Bratislava, Slovak Republic\\
$^{13}$ COMSATS University Islamabad, Islamabad, Pakistan\\
$^{14}$ Creighton University, Omaha, Nebraska, United States\\
$^{15}$ Department of Physics, Aligarh Muslim University, Aligarh, India\\
$^{16}$ Department of Physics, Pusan National University, Pusan, Republic of Korea\\
$^{17}$ Department of Physics, Sejong University, Seoul, Republic of Korea\\
$^{18}$ Department of Physics, University of California, Berkeley, California, United States\\
$^{19}$ Department of Physics, University of Oslo, Oslo, Norway\\
$^{20}$ Department of Physics and Technology, University of Bergen, Bergen, Norway\\
$^{21}$ Dipartimento di Fisica, Universit\`{a} di Pavia, Pavia, Italy\\
$^{22}$ Dipartimento di Fisica dell'Universit\`{a} and Sezione INFN, Cagliari, Italy\\
$^{23}$ Dipartimento di Fisica dell'Universit\`{a} and Sezione INFN, Trieste, Italy\\
$^{24}$ Dipartimento di Fisica dell'Universit\`{a} and Sezione INFN, Turin, Italy\\
$^{25}$ Dipartimento di Fisica e Astronomia dell'Universit\`{a} and Sezione INFN, Bologna, Italy\\
$^{26}$ Dipartimento di Fisica e Astronomia dell'Universit\`{a} and Sezione INFN, Catania, Italy\\
$^{27}$ Dipartimento di Fisica e Astronomia dell'Universit\`{a} and Sezione INFN, Padova, Italy\\
$^{28}$ Dipartimento di Fisica `E.R.~Caianiello' dell'Universit\`{a} and Gruppo Collegato INFN, Salerno, Italy\\
$^{29}$ Dipartimento DISAT del Politecnico and Sezione INFN, Turin, Italy\\
$^{30}$ Dipartimento di Scienze MIFT, Universit\`{a} di Messina, Messina, Italy\\
$^{31}$ Dipartimento Interateneo di Fisica `M.~Merlin' and Sezione INFN, Bari, Italy\\
$^{32}$ European Organization for Nuclear Research (CERN), Geneva, Switzerland\\
$^{33}$ Faculty of Electrical Engineering, Mechanical Engineering and Naval Architecture, University of Split, Split, Croatia\\
$^{34}$ Faculty of Engineering and Science, Western Norway University of Applied Sciences, Bergen, Norway\\
$^{35}$ Faculty of Nuclear Sciences and Physical Engineering, Czech Technical University in Prague, Prague, Czech Republic\\
$^{36}$ Faculty of Physics, Sofia University, Sofia, Bulgaria\\
$^{37}$ Faculty of Science, P.J.~\v{S}af\'{a}rik University, Ko\v{s}ice, Slovak Republic\\
$^{38}$ Frankfurt Institute for Advanced Studies, Johann Wolfgang Goethe-Universit\"{a}t Frankfurt, Frankfurt, Germany\\
$^{39}$ Fudan University, Shanghai, China\\
$^{40}$ Gangneung-Wonju National University, Gangneung, Republic of Korea\\
$^{41}$ Gauhati University, Department of Physics, Guwahati, India\\
$^{42}$ Helmholtz-Institut f\"{u}r Strahlen- und Kernphysik, Rheinische Friedrich-Wilhelms-Universit\"{a}t Bonn, Bonn, Germany\\
$^{43}$ Helsinki Institute of Physics (HIP), Helsinki, Finland\\
$^{44}$ High Energy Physics Group,  Universidad Aut\'{o}noma de Puebla, Puebla, Mexico\\
$^{45}$ Horia Hulubei National Institute of Physics and Nuclear Engineering, Bucharest, Romania\\
$^{46}$ Indian Institute of Technology Bombay (IIT), Mumbai, India\\
$^{47}$ Indian Institute of Technology Indore, Indore, India\\
$^{48}$ INFN, Laboratori Nazionali di Frascati, Frascati, Italy\\
$^{49}$ INFN, Sezione di Bari, Bari, Italy\\
$^{50}$ INFN, Sezione di Bologna, Bologna, Italy\\
$^{51}$ INFN, Sezione di Cagliari, Cagliari, Italy\\
$^{52}$ INFN, Sezione di Catania, Catania, Italy\\
$^{53}$ INFN, Sezione di Padova, Padova, Italy\\
$^{54}$ INFN, Sezione di Pavia, Pavia, Italy\\
$^{55}$ INFN, Sezione di Torino, Turin, Italy\\
$^{56}$ INFN, Sezione di Trieste, Trieste, Italy\\
$^{57}$ Inha University, Incheon, Republic of Korea\\
$^{58}$ Institute for Gravitational and Subatomic Physics (GRASP), Utrecht University/Nikhef, Utrecht, Netherlands\\
$^{59}$ Institute of Experimental Physics, Slovak Academy of Sciences, Ko\v{s}ice, Slovak Republic\\
$^{60}$ Institute of Physics, Homi Bhabha National Institute, Bhubaneswar, India\\
$^{61}$ Institute of Physics of the Czech Academy of Sciences, Prague, Czech Republic\\
$^{62}$ Institute of Space Science (ISS), Bucharest, Romania\\
$^{63}$ Institut f\"{u}r Kernphysik, Johann Wolfgang Goethe-Universit\"{a}t Frankfurt, Frankfurt, Germany\\
$^{64}$ Instituto de Ciencias Nucleares, Universidad Nacional Aut\'{o}noma de M\'{e}xico, Mexico City, Mexico\\
$^{65}$ Instituto de F\'{i}sica, Universidade Federal do Rio Grande do Sul (UFRGS), Porto Alegre, Brazil\\
$^{66}$ Instituto de F\'{\i}sica, Universidad Nacional Aut\'{o}noma de M\'{e}xico, Mexico City, Mexico\\
$^{67}$ iThemba LABS, National Research Foundation, Somerset West, South Africa\\
$^{68}$ Jeonbuk National University, Jeonju, Republic of Korea\\
$^{69}$ Johann-Wolfgang-Goethe Universit\"{a}t Frankfurt Institut f\"{u}r Informatik, Fachbereich Informatik und Mathematik, Frankfurt, Germany\\
$^{70}$ Korea Institute of Science and Technology Information, Daejeon, Republic of Korea\\
$^{71}$ KTO Karatay University, Konya, Turkey\\
$^{72}$ Laboratoire de Physique des 2 Infinis, Ir\`{e}ne Joliot-Curie, Orsay, France\\
$^{73}$ Laboratoire de Physique Subatomique et de Cosmologie, Universit\'{e} Grenoble-Alpes, CNRS-IN2P3, Grenoble, France\\
$^{74}$ Lawrence Berkeley National Laboratory, Berkeley, California, United States\\
$^{75}$ Lund University Department of Physics, Division of Particle Physics, Lund, Sweden\\
$^{76}$ Nagasaki Institute of Applied Science, Nagasaki, Japan\\
$^{77}$ Nara Women{'}s University (NWU), Nara, Japan\\
$^{78}$ National and Kapodistrian University of Athens, School of Science, Department of Physics , Athens, Greece\\
$^{79}$ National Centre for Nuclear Research, Warsaw, Poland\\
$^{80}$ National Institute of Science Education and Research, Homi Bhabha National Institute, Jatni, India\\
$^{81}$ National Nuclear Research Center, Baku, Azerbaijan\\
$^{82}$ National Research and Innovation Agency - BRIN, Jakarta, Indonesia\\
$^{83}$ Niels Bohr Institute, University of Copenhagen, Copenhagen, Denmark\\
$^{84}$ Nikhef, National institute for subatomic physics, Amsterdam, Netherlands\\
$^{85}$ Nuclear Physics Group, STFC Daresbury Laboratory, Daresbury, United Kingdom\\
$^{86}$ Nuclear Physics Institute of the Czech Academy of Sciences, Husinec-\v{R}e\v{z}, Czech Republic\\
$^{87}$ Oak Ridge National Laboratory, Oak Ridge, Tennessee, United States\\
$^{88}$ Ohio State University, Columbus, Ohio, United States\\
$^{89}$ Physics department, Faculty of science, University of Zagreb, Zagreb, Croatia\\
$^{90}$ Physics Department, Panjab University, Chandigarh, India\\
$^{91}$ Physics Department, University of Jammu, Jammu, India\\
$^{92}$ Physics Program and International Institute for Sustainability with Knotted Chiral Meta Matter (SKCM2), Hiroshima University, Hiroshima, Japan\\
$^{93}$ Physikalisches Institut, Eberhard-Karls-Universit\"{a}t T\"{u}bingen, T\"{u}bingen, Germany\\
$^{94}$ Physikalisches Institut, Ruprecht-Karls-Universit\"{a}t Heidelberg, Heidelberg, Germany\\
$^{95}$ Physik Department, Technische Universit\"{a}t M\"{u}nchen, Munich, Germany\\
$^{96}$ Politecnico di Bari and Sezione INFN, Bari, Italy\\
$^{97}$ Research Division and ExtreMe Matter Institute EMMI, GSI Helmholtzzentrum f\"ur Schwerionenforschung GmbH, Darmstadt, Germany\\
$^{98}$ Saga University, Saga, Japan\\
$^{99}$ Saha Institute of Nuclear Physics, Homi Bhabha National Institute, Kolkata, India\\
$^{100}$ School of Physics and Astronomy, University of Birmingham, Birmingham, United Kingdom\\
$^{101}$ Secci\'{o}n F\'{\i}sica, Departamento de Ciencias, Pontificia Universidad Cat\'{o}lica del Per\'{u}, Lima, Peru\\
$^{102}$ Stefan Meyer Institut f\"{u}r Subatomare Physik (SMI), Vienna, Austria\\
$^{103}$ SUBATECH, IMT Atlantique, Nantes Universit\'{e}, CNRS-IN2P3, Nantes, France\\
$^{104}$ Sungkyunkwan University, Suwon City, Republic of Korea\\
$^{105}$ Suranaree University of Technology, Nakhon Ratchasima, Thailand\\
$^{106}$ Technical University of Ko\v{s}ice, Ko\v{s}ice, Slovak Republic\\
$^{107}$ The Henryk Niewodniczanski Institute of Nuclear Physics, Polish Academy of Sciences, Cracow, Poland\\
$^{108}$ The University of Texas at Austin, Austin, Texas, United States\\
$^{109}$ Universidad Aut\'{o}noma de Sinaloa, Culiac\'{a}n, Mexico\\
$^{110}$ Universidade de S\~{a}o Paulo (USP), S\~{a}o Paulo, Brazil\\
$^{111}$ Universidade Estadual de Campinas (UNICAMP), Campinas, Brazil\\
$^{112}$ Universidade Federal do ABC, Santo Andre, Brazil\\
$^{113}$ University of Cape Town, Cape Town, South Africa\\
$^{114}$ University of Houston, Houston, Texas, United States\\
$^{115}$ University of Jyv\"{a}skyl\"{a}, Jyv\"{a}skyl\"{a}, Finland\\
$^{116}$ University of Kansas, Lawrence, Kansas, United States\\
$^{117}$ University of Liverpool, Liverpool, United Kingdom\\
$^{118}$ University of Science and Technology of China, Hefei, China\\
$^{119}$ University of South-Eastern Norway, Kongsberg, Norway\\
$^{120}$ University of Tennessee, Knoxville, Tennessee, United States\\
$^{121}$ University of the Witwatersrand, Johannesburg, South Africa\\
$^{122}$ University of Tokyo, Tokyo, Japan\\
$^{123}$ University of Tsukuba, Tsukuba, Japan\\
$^{124}$ University Politehnica of Bucharest, Bucharest, Romania\\
$^{125}$ Universit\'{e} Clermont Auvergne, CNRS/IN2P3, LPC, Clermont-Ferrand, France\\
$^{126}$ Universit\'{e} de Lyon, CNRS/IN2P3, Institut de Physique des 2 Infinis de Lyon, Lyon, France\\
$^{127}$ Universit\'{e} de Strasbourg, CNRS, IPHC UMR 7178, F-67000 Strasbourg, France, Strasbourg, France\\
$^{128}$ Universit\'{e} Paris-Saclay Centre d'Etudes de Saclay (CEA), IRFU, D\'{e}partment de Physique Nucl\'{e}aire (DPhN), Saclay, France\\
$^{129}$ Universit\`{a} degli Studi di Foggia, Foggia, Italy\\
$^{130}$ Universit\`{a} del Piemonte Orientale, Vercelli, Italy\\
$^{131}$ Universit\`{a} di Brescia, Brescia, Italy\\
$^{132}$ Variable Energy Cyclotron Centre, Homi Bhabha National Institute, Kolkata, India\\
$^{133}$ Warsaw University of Technology, Warsaw, Poland\\
$^{134}$ Wayne State University, Detroit, Michigan, United States\\
$^{135}$ Westf\"{a}lische Wilhelms-Universit\"{a}t M\"{u}nster, Institut f\"{u}r Kernphysik, M\"{u}nster, Germany\\
$^{136}$ Wigner Research Centre for Physics, Budapest, Hungary\\
$^{137}$ Yale University, New Haven, Connecticut, United States\\
$^{138}$ Yonsei University, Seoul, Republic of Korea\\
$^{139}$  Zentrum  f\"{u}r Technologie und Transfer (ZTT), Worms, Germany\\
$^{140}$ Affiliated with an institute covered by a cooperation agreement with CERN\\
$^{141}$ Affiliated with an international laboratory covered by a cooperation agreement with CERN.\\

\end{flushleft} 

%% file: main.bbl
\providecommand{\href}[2]{#2}\begingroup\raggedright\begin{thebibliography}{100}

\bibitem{Kheyri:2013sq}
F.~Kheyri, M.~Khodadi, and H.~R. Sepangi, ``{Horava-Lifshitz early universe
  phase transition beyond detailed balance}'',
  \href{http://dx.doi.org/10.1140/epjc/s10052-013-2286-0}{{\em Eur. Phys. J. C}
  {\bfseries 73} (2013) 2286}, \href{http://arxiv.org/abs/1301.5460}{{\ttfamily
  arXiv:1301.5460 [gr-qc]}}.

\bibitem{Bazavov:2011nk}
A.~Bazavov {\em et~al.}, ``{The chiral and deconfinement aspects of the QCD
  transition}'', \href{http://dx.doi.org/10.1103/PhysRevD.85.054503}{{\em Phys.
  Rev. D} {\bfseries 85} (2012) 054503},
  \href{http://arxiv.org/abs/1111.1710}{{\ttfamily arXiv:1111.1710 [hep-lat]}}.

\bibitem{Borsanyi:2014rza}
S.~Bors\'anyi, Z.~Fodor, C.~Hoelbling, S.~D. Katz, S.~Krieg, C.~Ratti, and
  K.~K. Szabo, ``{Recent results on the Equation of State of QCD}'',
  \href{http://dx.doi.org/10.22323/1.214.0224}{{\em PoS} {\bfseries
  LATTICE2014} (2015) 224}, \href{http://arxiv.org/abs/1410.7917}{{\ttfamily
  arXiv:1410.7917 [hep-lat]}}.

\bibitem{Bazavov:2018mes}
{\bfseries HotQCD} Collaboration, A.~Bazavov {\em et~al.}, ``{Chiral crossover
  in QCD at zero and non-zero chemical potentials}'',
  \href{http://dx.doi.org/10.1016/j.physletb.2019.05.013}{{\em Phys. Lett. B}
  {\bfseries 795} (2019) 15--21},
  \href{http://arxiv.org/abs/1812.08235}{{\ttfamily arXiv:1812.08235
  [hep-lat]}}.

\bibitem{Bazavov:2009zn}
A.~Bazavov {\em et~al.}, ``{Equation of state and QCD transition at finite
  temperature}'', \href{http://dx.doi.org/10.1103/PhysRevD.80.014504}{{\em
  Phys. Rev. D} {\bfseries 80} (2009) 014504},
  \href{http://arxiv.org/abs/0903.4379}{{\ttfamily arXiv:0903.4379 [hep-lat]}}.

\bibitem{Braun-Munzinger:2015hba}
P.~Braun-Munzinger, V.~Koch, T.~Sch\"afer, and J.~Stachel, ``{Properties of hot
  and dense matter from relativistic heavy ion collisions}'',
  \href{http://dx.doi.org/10.1016/j.physrep.2015.12.003}{{\em Phys. Rept.}
  {\bfseries 621} (2016) 76--126},
  \href{http://arxiv.org/abs/1510.00442}{{\ttfamily arXiv:1510.00442
  [nucl-th]}}.

\bibitem{Shuryak:1978ij}
E.~V. Shuryak, ``{Quark-Gluon Plasma and Hadronic Production of Leptons,
  Photons and Psions}'',
  \href{http://dx.doi.org/10.1016/0370-2693(78)90370-2}{{\em Phys. Lett. B}
  {\bfseries 78} (1978) 150}.

\bibitem{Csorgo:2000yu}
T.~Csorgo, ``{New form of matter at CERN SPS: Quark matter but not quark gluon
  plasma}'', \href{http://dx.doi.org/10.1016/S0920-5632(00)01020-3}{{\em Nucl.
  Phys. B Proc. Suppl.} {\bfseries 92} (2001) 62--74},
  \href{http://arxiv.org/abs/hep-ph/0011339}{{\ttfamily arXiv:hep-ph/0011339}}.

\bibitem{Teaney:2000cw}
D.~Teaney, J.~Lauret, and E.~V. Shuryak, ``{Flow at the SPS and RHIC as a quark
  gluon plasma signature}'',
  \href{http://dx.doi.org/10.1103/PhysRevLett.86.4783}{{\em Phys. Rev. Lett.}
  {\bfseries 86} (2001) 4783--4786},
  \href{http://arxiv.org/abs/nucl-th/0011058}{{\ttfamily
  arXiv:nucl-th/0011058}}.

\bibitem{Arsene:2004fa}
{\bfseries BRAHMS} Collaboration, I.~Arsene {\em et~al.}, ``{Quark gluon plasma
  and color glass condensate at RHIC? The Perspective from the BRAHMS
  experiment}'', \href{http://dx.doi.org/10.1016/j.nuclphysa.2005.02.130}{{\em
  Nucl. Phys. A} {\bfseries 757} (2005) 1--27},
  \href{http://arxiv.org/abs/nucl-ex/0410020}{{\ttfamily
  arXiv:nucl-ex/0410020}}.

\bibitem{Back:2004je}
{\bfseries PHOBOS} Collaboration, B.~B. Back {\em et~al.}, ``{The PHOBOS
  perspective on discoveries at RHIC}'',
  \href{http://dx.doi.org/10.1016/j.nuclphysa.2005.03.084}{{\em Nucl. Phys. A}
  {\bfseries 757} (2005) 28--101},
  \href{http://arxiv.org/abs/nucl-ex/0410022}{{\ttfamily
  arXiv:nucl-ex/0410022}}.

\bibitem{Adams:2005dq}
{\bfseries STAR} Collaboration, J.~Adams {\em et~al.}, ``{Experimental and
  theoretical challenges in the search for the quark gluon plasma: The STAR
  Collaboration's critical assessment of the evidence from RHIC collisions}'',
  \href{http://dx.doi.org/10.1016/j.nuclphysa.2005.03.085}{{\em Nucl. Phys. A}
  {\bfseries 757} (2005) 102--183},
  \href{http://arxiv.org/abs/nucl-ex/0501009}{{\ttfamily
  arXiv:nucl-ex/0501009}}.

\bibitem{Adcox:2004mh}
{\bfseries PHENIX} Collaboration, K.~Adcox {\em et~al.}, ``{Formation of dense
  partonic matter in relativistic nucleus-nucleus collisions at RHIC:
  Experimental evaluation by the PHENIX collaboration}'',
  \href{http://dx.doi.org/10.1016/j.nuclphysa.2005.03.086}{{\em Nucl. Phys. A}
  {\bfseries 757} (2005) 184--283},
  \href{http://arxiv.org/abs/nucl-ex/0410003}{{\ttfamily
  arXiv:nucl-ex/0410003}}.

\bibitem{Muller:2012zq}
B.~Muller, J.~Schukraft, and B.~Wyslouch, ``{First Results from Pb+Pb
  collisions at the LHC}'',
  \href{http://dx.doi.org/10.1146/annurev-nucl-102711-094910}{{\em Ann. Rev.
  Nucl. Part. Sci.} {\bfseries 62} (2012) 361--386},
  \href{http://arxiv.org/abs/1202.3233}{{\ttfamily arXiv:1202.3233 [hep-ex]}}.

\bibitem{ALICE:2022wpn}
{\bfseries ALICE} Collaboration, ``{The ALICE experiment -- A journey through
  QCD}'', \href{http://arxiv.org/abs/2211.04384}{{\ttfamily arXiv:2211.04384
  [nucl-ex]}}.

\bibitem{Andronic:2006ky}
A.~Andronic, P.~Braun-Munzinger, K.~Redlich, and J.~Stachel, ``{Statistical
  hadronization of heavy quarks in ultra-relativistic nucleus-nucleus
  collisions}'', \href{http://dx.doi.org/10.1016/j.nuclphysa.2007.02.013}{{\em
  Nucl. Phys. A} {\bfseries 789} (2007) 334--356},
  \href{http://arxiv.org/abs/nucl-th/0611023}{{\ttfamily
  arXiv:nucl-th/0611023}}.

\bibitem{Andronic:2015wma}
A.~Andronic {\em et~al.}, ``{Heavy-flavour and quarkonium production in the LHC
  era: from proton\textendash{}proton to heavy-ion collisions}'',
  \href{http://dx.doi.org/10.1140/epjc/s10052-015-3819-5}{{\em Eur. Phys. J. C}
  {\bfseries 76} (2016) 107}, \href{http://arxiv.org/abs/1506.03981}{{\ttfamily
  arXiv:1506.03981 [nucl-ex]}}.

\bibitem{Liu:2012ax}
F.-M. Liu and S.-X. Liu, ``{Quark-gluon plasma formation time and direct
  photons from heavy ion collisions}'',
  \href{http://dx.doi.org/10.1103/PhysRevC.89.034906}{{\em Phys. Rev. C}
  {\bfseries 89} (2014) 034906},
  \href{http://arxiv.org/abs/1212.6587}{{\ttfamily arXiv:1212.6587 [nucl-th]}}.

\bibitem{Cacciari:1998it}
M.~Cacciari, M.~Greco, and P.~Nason, ``{The {$p_{\rm{T}}$} spectrum in heavy
  flavor hadroproduction}'',
  \href{http://dx.doi.org/10.1088/1126-6708/1998/05/007}{{\em JHEP} {\bfseries
  05} (1998) 007}, \href{http://arxiv.org/abs/hep-ph/9803400}{{\ttfamily
  arXiv:hep-ph/9803400}}.

\bibitem{Cacciari:2001td}
M.~Cacciari, S.~Frixione, and P.~Nason, ``{The {$p_{\rm{T}}$} spectrum in heavy
  flavor photoproduction}'',
  \href{http://dx.doi.org/10.1088/1126-6708/2001/03/006}{{\em JHEP} {\bfseries
  03} (2001) 006}, \href{http://arxiv.org/abs/hep-ph/0102134}{{\ttfamily
  arXiv:hep-ph/0102134}}.

\bibitem{Cacciari:2012ny}
M.~Cacciari, S.~Frixione, N.~Houdeau, M.~L. Mangano, P.~Nason, and G.~Ridolfi,
  ``{Theoretical predictions for charm and bottom production at the LHC}'',
  \href{http://dx.doi.org/10.1007/JHEP10(2012)137}{{\em JHEP} {\bfseries 10}
  (2012) 137}, \href{http://arxiv.org/abs/1205.6344}{{\ttfamily arXiv:1205.6344
  [hep-ph]}}.

\bibitem{Bolzoni:2012kx}
P.~Bolzoni and G.~Kramer, ``{Inclusive lepton production from heavy-hadron
  decay in pp collisions at the LHC}'',
  \href{http://dx.doi.org/10.1016/j.nuclphysb.2013.04.002}{{\em Nucl. Phys. B}
  {\bfseries 872} (2013) 253--264},
  \href{http://arxiv.org/abs/1212.4356}{{\ttfamily arXiv:1212.4356 [hep-ph]}}.
  [Erratum: Nucl. Phys. B 876 (2013) 334--337].

\bibitem{Cao:2018ews}
S.~Cao {\em et~al.}, ``{Toward the determination of heavy-quark transport
  coefficients in quark-gluon plasma}'',
  \href{http://dx.doi.org/10.1103/PhysRevC.99.054907}{{\em Phys. Rev. C}
  {\bfseries 99} (2019) 054907},
  \href{http://arxiv.org/abs/1809.07894}{{\ttfamily arXiv:1809.07894
  [nucl-th]}}.

\bibitem{Baier:2000mf}
R.~Baier, D.~Schiff, and B.~G. Zakharov, ``{Energy loss in perturbative QCD}'',
  \href{http://dx.doi.org/10.1146/annurev.nucl.50.1.37}{{\em Ann. Rev. Nucl.
  Part. Sci.} {\bfseries 50} (2000) 37--69},
  \href{http://arxiv.org/abs/hep-ph/0002198}{{\ttfamily arXiv:hep-ph/0002198}}.

\bibitem{Dokshitzer:2001zm}
Y.~L. Dokshitzer and D.~E. Kharzeev, ``{Heavy quark colorimetry of QCD
  matter}'', \href{http://dx.doi.org/10.1016/S0370-2693(01)01130-3}{{\em Phys.
  Lett. B} {\bfseries 519} (2001) 199--206},
  \href{http://arxiv.org/abs/hep-ph/0106202}{{\ttfamily arXiv:hep-ph/0106202}}.

\bibitem{Armesto:2003jh}
N.~Armesto, C.~A. Salgado, and U.~A. Wiedemann, ``{Medium induced gluon
  radiation off massive quarks fills the dead cone}'',
  \href{http://dx.doi.org/10.1103/PhysRevD.69.114003}{{\em Phys. Rev. D}
  {\bfseries 69} (2004) 114003},
  \href{http://arxiv.org/abs/hep-ph/0312106}{{\ttfamily arXiv:hep-ph/0312106}}.

\bibitem{Wicks:2007am}
S.~Wicks, W.~Horowitz, M.~Djordjevic, and M.~Gyulassy, ``{Heavy quark jet
  quenching with collisional plus radiative energy loss and path length
  fluctuations}'',
  \href{http://dx.doi.org/10.1016/j.nuclphysa.2006.11.102}{{\em Nucl. Phys. A}
  {\bfseries 783} (2007) 493--496},
  \href{http://arxiv.org/abs/nucl-th/0701063}{{\ttfamily
  arXiv:nucl-th/0701063}}.

\bibitem{Zhang:2003wk}
B.-W. Zhang, E.~Wang, and X.-N. Wang, ``{Heavy quark energy loss in nuclear
  medium}'', \href{http://dx.doi.org/10.1103/PhysRevLett.93.072301}{{\em Phys.
  Rev. Lett.} {\bfseries 93} (2004) 072301},
  \href{http://arxiv.org/abs/nucl-th/0309040}{{\ttfamily
  arXiv:nucl-th/0309040}}.

\bibitem{Adil:2006ra}
A.~Adil and I.~Vitev, ``{Collisional dissociation of heavy mesons in dense QCD
  matter}'', \href{http://dx.doi.org/10.1016/j.physletb.2007.03.050}{{\em Phys.
  Lett. B} {\bfseries 649} (2007) 139--146},
  \href{http://arxiv.org/abs/hep-ph/0611109}{{\ttfamily arXiv:hep-ph/0611109}}.

\bibitem{Zhang:2007yoa}
B.-W. Zhang, C.-M. Ko, and W.~Liu, ``{Thermal charm production in a quark-gluon
  plasma in Pb-Pb collisions at $\sqrt{s_{\rm{NN}}}$= 5.5 TeV}'',
  \href{http://dx.doi.org/10.1103/PhysRevC.77.024901}{{\em Phys. Rev. C}
  {\bfseries 77} (2008) 024901},
  \href{http://arxiv.org/abs/0709.1684}{{\ttfamily arXiv:0709.1684 [nucl-th]}}.

\bibitem{Braun-Munzinger:2007fth}
P.~Braun-Munzinger, ``{Quarkonium production in ultra-relativistic nuclear
  collisions: Suppression versus enhancement}'',
  \href{http://dx.doi.org/10.1088/0954-3899/34/8/S36}{{\em J. Phys. G}
  {\bfseries 34} (2007) S471--478},
  \href{http://arxiv.org/abs/nucl-th/0701093}{{\ttfamily
  arXiv:nucl-th/0701093}}.

\bibitem{Djordjevic:2004nq}
M.~Djordjevic, M.~Gyulassy, and S.~Wicks, ``{{}Open Charm and Beauty at
  Ultrarelativistic Heavy Ion Colliders}'',
  \href{http://dx.doi.org/10.1103/PhysRevLett.94.112301}{{\em Phys. Rev. Lett.}
  {\bfseries 94} (2005) 112301},
  \href{http://arxiv.org/abs/hep-ph/0410372}{{\ttfamily arXiv:hep-ph/0410372}}.

\bibitem{ALICE:2021aqk}
{\bfseries ALICE} Collaboration, S.~Acharya {\em et~al.}, ``{Direct observation
  of the dead-cone effect in quantum chromodynamics}'',
  \href{http://dx.doi.org/10.1038/s41586-022-04572-w}{{\em Nature} {\bfseries
  605} (2022) 440--446}, \href{http://arxiv.org/abs/2106.05713}{{\ttfamily
  arXiv:2106.05713 [nucl-ex]}}.

\bibitem{Loizides:2017ack}
C.~Loizides, J.~Kamin, and D.~d'Enterria, ``{Improved Monte Carlo Glauber
  predictions at present and future nuclear colliders}'',
  \href{http://dx.doi.org/10.1103/PhysRevC.97.054910}{{\em Phys. Rev. C}
  {\bfseries 97} (2018) 054910},
  \href{http://arxiv.org/abs/1710.07098}{{\ttfamily arXiv:1710.07098
  [nucl-ex]}}. [Erratum: Phys.Rev.C 99, 019901 (2019)].

\bibitem{dEnterria:2020dwq}
D.~d'Enterria and C.~Loizides, ``{Progress in the Glauber Model at Collider
  Energies}'', \href{http://dx.doi.org/10.1146/annurev-nucl-102419-060007}{{\em
  Ann. Rev. Nucl. Part. Sci.} {\bfseries 71} (2021) 315--344},
  \href{http://arxiv.org/abs/2011.14909}{{\ttfamily arXiv:2011.14909
  [hep-ph]}}.

\bibitem{Aggarwal:2010xp}
{\bfseries STAR} Collaboration, M.~M. Aggarwal {\em et~al.}, ``{Measurement of
  the Bottom contribution to non-photonic electron production in $p+p$
  collisions at $\sqrt{s} $=200 GeV}'',
  \href{http://dx.doi.org/10.1103/PhysRevLett.105.202301}{{\em Phys. Rev.
  Lett.} {\bfseries 105} (2010) 202301},
  \href{http://arxiv.org/abs/1007.1200}{{\ttfamily arXiv:1007.1200 [nucl-ex]}}.

\bibitem{Adare:2009ic}
{\bfseries PHENIX} Collaboration, A.~Adare {\em et~al.}, ``{Measurement of
  Bottom versus Charm as a Function of Transverse Momentum with Electron-Hadron
  Correlations in $p+ p$ Collisions at $\sqrt{s}=200$ GeV}'',
  \href{http://dx.doi.org/10.1103/PhysRevLett.103.082002}{{\em Phys. Rev.
  Lett.} {\bfseries 103} (2009) 082002},
  \href{http://arxiv.org/abs/0903.4851}{{\ttfamily arXiv:0903.4851 [hep-ex]}}.

\bibitem{Acosta:2004yw}
{\bfseries CDF} Collaboration, D.~Acosta {\em et~al.}, ``{Measurement of the
  $J/\psi$ meson and $b-$hadron production cross sections in $p\bar{p}$
  collisions at $\sqrt{s} = 1960$ GeV}'',
  \href{http://dx.doi.org/10.1103/PhysRevD.71.032001}{{\em Phys. Rev. D}
  {\bfseries 71} (2005) 032001},
  \href{http://arxiv.org/abs/hep-ex/0412071}{{\ttfamily arXiv:hep-ex/0412071}}.

\bibitem{ALICE:2012acz}
{\bfseries ALICE} Collaboration, B.~Abelev {\em et~al.}, ``{Measurement of
  electrons from beauty hadron decays in pp collisions at $\sqrt{s}=7$ TeV}'',
  \href{http://dx.doi.org/10.1016/j.physletb.2013.01.069}{{\em Phys. Lett. B}
  {\bfseries 721} (2013) 13--23},
  \href{http://arxiv.org/abs/1208.1902}{{\ttfamily arXiv:1208.1902 [hep-ex]}}.
  [Erratum: Phys. Lett. B 763 (2016) 507--509].

\bibitem{ALICE:2014ivb}
{\bfseries ALICE} Collaboration, B.~Abelev {\em et~al.}, ``{Measurement of
  electrons from semileptonic heavy-flavor hadron decays in pp collisions at
  $\sqrt{s} = 2.76$ TeV}'',
  \href{http://dx.doi.org/10.1103/PhysRevD.91.012001}{{\em Phys. Rev. D}
  {\bfseries 91} (2015) 012001},
  \href{http://arxiv.org/abs/1405.4117}{{\ttfamily arXiv:1405.4117 [nucl-ex]}}.

\bibitem{Aaij:2012jd}
{\bfseries LHCb} Collaboration, R.~Aaij {\em et~al.}, ``{Measurement of the
  $B^\pm$ production cross-section in pp collisions at $\sqrt{s}=7$ TeV}'',
  \href{http://dx.doi.org/10.1007/JHEP04(2012)093}{{\em JHEP} {\bfseries 04}
  (2012) 093}, \href{http://arxiv.org/abs/1202.4812}{{\ttfamily arXiv:1202.4812
  [hep-ex]}}.

\bibitem{Aad:2011sp}
{\bfseries ATLAS} Collaboration, G.~Aad {\em et~al.}, ``{Measurement of the
  differential cross-sections of inclusive, prompt and non-prompt $J/\psi$
  production in proton-proton collisions at $\sqrt{s}=7$ TeV}'',
  \href{http://dx.doi.org/10.1016/j.nuclphysb.2011.05.015}{{\em Nucl. Phys. B}
  {\bfseries 850} (2011) 387--444},
  \href{http://arxiv.org/abs/1104.3038}{{\ttfamily arXiv:1104.3038 [hep-ex]}}.

\bibitem{ALICE:2012vpz}
{\bfseries ALICE} Collaboration, B.~Abelev {\em et~al.}, ``{Measurement of
  prompt $J/\psi$ and beauty hadron production cross sections at mid-rapidity
  in pp collisions at $\sqrt{s} = 7$ TeV}'',
  \href{http://dx.doi.org/10.1007/JHEP11(2012)065}{{\em JHEP} {\bfseries 11}
  (2012) 065}, \href{http://arxiv.org/abs/1205.5880}{{\ttfamily arXiv:1205.5880
  [hep-ex]}}.

\bibitem{Aad:2012jga}
{\bfseries ATLAS} Collaboration, G.~Aad {\em et~al.}, ``{Measurement of the
  b-hadron production cross section using decays to $D^{*}\mu^-X$ final states
  in pp collisions at $\sqrt{s}$ = 7 TeV with the ATLAS detector}'',
  \href{http://dx.doi.org/10.1016/j.nuclphysb.2012.07.009}{{\em Nucl. Phys. B}
  {\bfseries 864} (2012) 341--381},
  \href{http://arxiv.org/abs/1206.3122}{{\ttfamily arXiv:1206.3122 [hep-ex]}}.

\bibitem{ATLAS:2013cia}
{\bfseries ATLAS} Collaboration, G.~Aad {\em et~al.}, ``{Measurement of the
  differential cross-section of $B^{+}$ meson production in pp collisions at
  $\sqrt{s}$ = 7 TeV at ATLAS}'',
  \href{http://dx.doi.org/10.1007/JHEP10(2013)042}{{\em JHEP} {\bfseries 10}
  (2013) 042}, \href{http://arxiv.org/abs/1307.0126}{{\ttfamily arXiv:1307.0126
  [hep-ex]}}.

\bibitem{Chatrchyan:2012hw}
{\bfseries CMS} Collaboration, S.~Chatrchyan {\em et~al.}, ``{Measurement of
  the cross section for production of $b b^-$ bar $X$, decaying to muons in pp
  collisions at $\sqrt{s}=7$ TeV}'',
  \href{http://dx.doi.org/10.1007/JHEP06(2012)110}{{\em JHEP} {\bfseries 06}
  (2012) 110}, \href{http://arxiv.org/abs/1203.3458}{{\ttfamily arXiv:1203.3458
  [hep-ex]}}.

\bibitem{Khachatryan:2010yr}
{\bfseries CMS} Collaboration, V.~Khachatryan {\em et~al.}, ``{Prompt and
  Non-Prompt $J/\psi$ Production in pp Collisions at $\sqrt{s}=7$ TeV}'',
  \href{http://dx.doi.org/10.1140/epjc/s10052-011-1575-8}{{\em Eur. Phys. J. C}
  {\bfseries 71} (2011) 1575}, \href{http://arxiv.org/abs/1011.4193}{{\ttfamily
  arXiv:1011.4193 [hep-ex]}}.

\bibitem{Khachatryan:2011mk}
{\bfseries CMS} Collaboration, V.~Khachatryan {\em et~al.}, ``{Measurement of
  the $B^+$ Production Cross Section in pp Collisions at $\sqrt{s} = 7$ TeV}'',
  \href{http://dx.doi.org/10.1103/PhysRevLett.106.112001}{{\em Phys. Rev.
  Lett.} {\bfseries 106} (2011) 112001},
  \href{http://arxiv.org/abs/1101.0131}{{\ttfamily arXiv:1101.0131 [hep-ex]}}.

\bibitem{Chatrchyan:2011pw}
{\bfseries CMS} Collaboration, S.~Chatrchyan {\em et~al.}, ``{Measurement of
  the $B^0$ production cross section in pp Collisions at $\sqrt{s}=7$ TeV}'',
  \href{http://dx.doi.org/10.1103/PhysRevLett.106.252001}{{\em Phys. Rev.
  Lett.} {\bfseries 106} (2011) 252001},
  \href{http://arxiv.org/abs/1104.2892}{{\ttfamily arXiv:1104.2892 [hep-ex]}}.

\bibitem{Chatrchyan:2011vh}
{\bfseries CMS} Collaboration, S.~Chatrchyan {\em et~al.}, ``{Measurement of
  the ${B}_{s}^{0}$ Production Cross Section with
  ${B}_{s}^{0}\ensuremath{\rightarrow}J/\ensuremath{\psi}\ensuremath{\phi}$
  Decays in pp Collisions at $\sqrt{s}\mathbf{=}7\text{ }\text{
  }\mathrm{TeV}$}'', \href{http://dx.doi.org/10.1103/PhysRevD.84.052008}{{\em
  Phys. Rev. D} {\bfseries 84} (2011) 052008},
  \href{http://arxiv.org/abs/1106.4048}{{\ttfamily arXiv:1106.4048 [hep-ex]}}.

\bibitem{Khachatryan:2016csy}
{\bfseries CMS} Collaboration, V.~Khachatryan {\em et~al.}, ``{Measurement of
  the total and differential inclusive $B^+$ hadron cross sections in pp
  collisions at $\sqrt{s}$ = 13 TeV}'',
  \href{http://dx.doi.org/10.1016/j.physletb.2017.05.074}{{\em Phys. Lett. B}
  {\bfseries 771} (2017) 435--456},
  \href{http://arxiv.org/abs/1609.00873}{{\ttfamily arXiv:1609.00873
  [hep-ex]}}.

\bibitem{ALICE:2021mgk}
{\bfseries ALICE} Collaboration, S.~Acharya {\em et~al.}, ``{Measurement of
  beauty and charm production in pp collisions at $ \sqrt{s} $ = 5.02 TeV via
  non-prompt and prompt D mesons}'',
  \href{http://dx.doi.org/10.1007/JHEP05(2021)220}{{\em JHEP} {\bfseries 05}
  (2021) 220}, \href{http://arxiv.org/abs/2102.13601}{{\ttfamily
  arXiv:2102.13601 [nucl-ex]}}.

\bibitem{ALICE:2021edd}
{\bfseries ALICE} Collaboration, S.~Acharya {\em et~al.}, ``{Prompt and
  non-prompt J/\ensuremath{\psi} production cross sections at midrapidity in
  proton-proton collisions at $ \sqrt{\mathrm{s}} $ = 5.02 and 13 TeV}'',
  \href{http://dx.doi.org/10.1007/JHEP03(2022)190}{{\em JHEP} {\bfseries 03}
  (2022) 190}, \href{http://arxiv.org/abs/2108.02523}{{\ttfamily
  arXiv:2108.02523 [nucl-ex]}}.

\bibitem{ALICE:2018gev}
{\bfseries ALICE} Collaboration, S.~Acharya {\em et~al.}, ``{Dielectron and
  heavy-quark production in inelastic and high-multiplicity
  proton\textendash{}proton collisions at $\sqrt {s_{\rm{NN}}}=$ 13 TeV}'',
  \href{http://dx.doi.org/10.1016/j.physletb.2018.11.009}{{\em Phys. Lett. B}
  {\bfseries 788} (2019) 505--518},
  \href{http://arxiv.org/abs/1805.04407}{{\ttfamily arXiv:1805.04407
  [hep-ex]}}.

\bibitem{ALICE:2021rxa}
{\bfseries ALICE} Collaboration, S.~Acharya {\em et~al.}, ``{Prompt D$^{0}$,
  D$^{+}$, and D$^{*+}$ production in Pb\textendash{}Pb collisions at $
  \sqrt{s_{\mathrm{NN}}} $ = 5.02 TeV}'',
  \href{http://dx.doi.org/10.1007/JHEP01(2022)174}{{\em JHEP} {\bfseries 01}
  (2022) 174}, \href{http://arxiv.org/abs/2110.09420}{{\ttfamily
  arXiv:2110.09420 [nucl-ex]}}.

\bibitem{Sirunyan:2017xss}
{\bfseries CMS} Collaboration, A.~M. Sirunyan {\em et~al.}, ``{Nuclear
  modification factor of D$^0$ mesons in PbPb collisions at
  $\sqrt{s_\mathrm{\rm{NN}}} = 5.02$ TeV}'',
  \href{http://dx.doi.org/10.1016/j.physletb.2018.05.074}{{\em Phys. Lett. B}
  {\bfseries 782} (2018) 474--496},
  \href{http://arxiv.org/abs/1708.04962}{{\ttfamily arXiv:1708.04962
  [nucl-ex]}}.

\bibitem{ALICE:2021kfc}
{\bfseries ALICE} Collaboration, S.~Acharya {\em et~al.}, ``{Measurement of
  prompt $D_{\rm{s}}^+$-meson production and azimuthal anisotropy in Pb--Pb
  collisions at $\sqrt {s_{\rm{NN}}}$ = 5.02TeV}'',
  \href{http://dx.doi.org/10.1016/j.physletb.2022.136986}{{\em Phys. Lett. B}
  {\bfseries 827} (2022) 136986},
  \href{http://arxiv.org/abs/2110.10006}{{\ttfamily arXiv:2110.10006
  [nucl-ex]}}.

\bibitem{ALICE:2017pbx}
{\bfseries ALICE} Collaboration, S.~Acharya {\em et~al.}, ``{$D$-meson
  azimuthal anisotropy in midcentral Pb-Pb collisions at $\sqrt{s_{\rm
  NN}}=5.02$ TeV }'',
  \href{http://dx.doi.org/10.1103/PhysRevLett.120.102301}{{\em Phys. Rev.
  Lett.} {\bfseries 120} (2018) 102301},
  \href{http://arxiv.org/abs/1707.01005}{{\ttfamily arXiv:1707.01005
  [nucl-ex]}}.

\bibitem{ALICE:2020iug}
{\bfseries ALICE} Collaboration, S.~Acharya {\em et~al.},
  ``{Transverse-momentum and event-shape dependence of D-meson flow harmonics
  in Pb--Pb collisions at $\sqrt {s_{\rm{NN}}}$ = 5.02 TeV}'',
  \href{http://dx.doi.org/10.1016/j.physletb.2020.136054}{{\em Phys. Lett. B}
  {\bfseries 813} (2021) 136054},
  \href{http://arxiv.org/abs/2005.11131}{{\ttfamily arXiv:2005.11131
  [nucl-ex]}}.

\bibitem{PHENIX:2010xji}
{\bfseries PHENIX} Collaboration, A.~Adare {\em et~al.}, ``{Heavy Quark
  Production in $p+p$ and Energy Loss and Flow of Heavy Quarks in Au+Au
  Collisions at $\sqrt{s_{\rm{NN}}}=200$ GeV}'',
  \href{http://dx.doi.org/10.1103/PhysRevC.84.044905}{{\em Phys. Rev. C}
  {\bfseries 84} (2011) 044905},
  \href{http://arxiv.org/abs/1005.1627}{{\ttfamily arXiv:1005.1627 [nucl-ex]}}.

\bibitem{STAR:2014yia}
{\bfseries STAR} Collaboration, L.~Adamczyk {\em et~al.}, ``{Elliptic flow of
  electrons from heavy-flavor hadron decays in Au + Au collisions at
  $\sqrt{s_{\rm NN}} = $ 200, 62.4, and 39 GeV}'',
  \href{http://dx.doi.org/10.1103/PhysRevC.95.034907}{{\em Phys. Rev. C}
  {\bfseries 95} (2017) 034907},
  \href{http://arxiv.org/abs/1405.6348}{{\ttfamily arXiv:1405.6348 [hep-ex]}}.

\bibitem{STAR:2017kkh}
{\bfseries STAR} Collaboration, L.~Adamczyk {\em et~al.}, ``{Measurement of
  $D^0$ Azimuthal Anisotropy at Midrapidity in Au+Au Collisions at
  $\sqrt{s_{NN}}$=200 GeV}'',
  \href{http://dx.doi.org/10.1103/PhysRevLett.118.212301}{{\em Phys. Rev.
  Lett.} {\bfseries 118} (2017) 212301},
  \href{http://arxiv.org/abs/1701.06060}{{\ttfamily arXiv:1701.06060
  [nucl-ex]}}.

\bibitem{Rapp:2018qla}
A.~Beraudo {\em et~al.}, ``{Extraction of Heavy-Flavor Transport Coefficients
  in QCD Matter}'',
  \href{http://dx.doi.org/10.1016/j.nuclphysa.2018.09.002}{{\em Nucl. Phys. A}
  {\bfseries 979} (2018) 21--86},
  \href{http://arxiv.org/abs/1803.03824}{{\ttfamily arXiv:1803.03824
  [nucl-th]}}.

\bibitem{Sirunyan:2017oug}
{\bfseries CMS} Collaboration, A.~M. Sirunyan {\em et~al.}, ``{Measurement of
  the ${B}^{\pm}$ Meson Nuclear Modification Factor in Pb-Pb Collisions at
  $\sqrt{{s}_{\rm{NN}}}=5.02\text{ }\text{ }\mathrm{TeV}$}'',
  \href{http://dx.doi.org/10.1103/PhysRevLett.119.152301}{{\em Phys. Rev.
  Lett.} {\bfseries 119} (2017) 152301},
  \href{http://arxiv.org/abs/1705.04727}{{\ttfamily arXiv:1705.04727
  [hep-ex]}}.

\bibitem{ATLAS:2019jpi}
{\bfseries ATLAS} Collaboration, M.~Aaboud {\em et~al.}, ``{Measurement of the
  relative $B^{\pm}_{c}/B^{\pm}$ production cross section with the ATLAS
  detector at $\sqrt{s}=8$ TeV}'',
  \href{http://dx.doi.org/10.1103/PhysRevD.104.012010}{{\em Phys. Rev. D}
  {\bfseries 104} (2021) 012010},
  \href{http://arxiv.org/abs/1912.02672}{{\ttfamily arXiv:1912.02672
  [hep-ex]}}.

\bibitem{Zyla:2020zbs}
{\bfseries Particle Data Group} Collaboration, P.~Zyla {\em et~al.}, ``{Review
  of Particle Physics}'', \href{http://dx.doi.org/10.1093/ptep/ptaa104}{{\em
  PTEP} {\bfseries 2020} (2020) 083C01}.

\bibitem{ATLAS:2018hqe}
{\bfseries ATLAS} Collaboration, M.~Aaboud {\em et~al.}, ``{Prompt and
  non-prompt $J/\psi $ and $\psi (2\mathrm {S})$ suppression at high transverse
  momentum in $5.02~\mathrm {TeV}$ Pb+Pb collisions with the ATLAS
  experiment}'', \href{http://dx.doi.org/10.1140/epjc/s10052-018-6219-9}{{\em
  Eur. Phys. J. C} {\bfseries 78} (2018) 762},
  \href{http://arxiv.org/abs/1805.04077}{{\ttfamily arXiv:1805.04077
  [nucl-ex]}}.

\bibitem{CMS:2017uuv}
{\bfseries CMS} Collaboration, A.~M. Sirunyan {\em et~al.}, ``{Measurement of
  prompt and non-prompt charmonium suppression in $\text {PbPb}$ collisions at
  5.02 $\,\text {Te}\text {V}$}'',
  \href{http://dx.doi.org/10.1140/epjc/s10052-018-5950-6}{{\em Eur. Phys. J. C}
  {\bfseries 78} (2018) 509}, \href{http://arxiv.org/abs/1712.08959}{{\ttfamily
  arXiv:1712.08959 [nucl-ex]}}.

\bibitem{ALICE:2022tji}
{\bfseries ALICE} Collaboration, S.~Acharya {\em et~al.}, ``{Measurement of
  beauty production via non-prompt ${\rm D}^{0}$ mesons in Pb-Pb collisions at
  $\sqrt{s_{\rm NN}}$ = 5.02 TeV}'',
  \href{http://dx.doi.org/10.1007/JHEP12(2022)126}{{\em JHEP} {\bfseries 12}
  (2022) 126}, \href{http://arxiv.org/abs/2202.00815}{{\ttfamily
  arXiv:2202.00815 [nucl-ex]}}.

\bibitem{CMS:2018bwt}
{\bfseries CMS} Collaboration, A.~M. Sirunyan {\em et~al.}, ``{Studies of
  Beauty Suppression via Non-prompt $D^0$ Mesons in Pb-Pb Collisions at $Q^2 =
  4$ $\rm GeV^2$}'',
  \href{http://dx.doi.org/10.1103/PhysRevLett.123.022001}{{\em Phys. Rev.
  Lett.} {\bfseries 123} (2019) 022001},
  \href{http://arxiv.org/abs/1810.11102}{{\ttfamily arXiv:1810.11102
  [hep-ex]}}.

\bibitem{ALICE:2022xrg}
{\bfseries ALICE} Collaboration, ``{Measurement of beauty-strange meson
  production in Pb$-$Pb collisions at $\sqrt{s_{\rm NN}} = 5.02$ TeV via
  non-prompt $\mathrm{D_s}^{+}$ mesons}'',
  \href{http://dx.doi.org/10.1016/j.physletb.2022.137561}{{\em Phys. Lett. B}
  {\bfseries 846} (2023) 137561},
  \href{http://arxiv.org/abs/2204.10386}{{\ttfamily arXiv:2204.10386
  [nucl-ex]}}.

\bibitem{ALICE:2016uid}
{\bfseries ALICE} Collaboration, J.~Adam {\em et~al.}, ``{Measurement of
  electrons from beauty-hadron decays in p-Pb collisions at $
  \sqrt{s_{\mathrm{NN}}}=5.02 $ TeV and Pb-Pb collisions at $
  \sqrt{s_{\mathrm{NN}}}=2.76 $ TeV}'',
  \href{http://dx.doi.org/10.1007/JHEP07(2017)052}{{\em JHEP} {\bfseries 07}
  (2017) 052}, \href{http://arxiv.org/abs/1609.03898}{{\ttfamily
  arXiv:1609.03898 [nucl-ex]}}.

\bibitem{Adare:2015hla}
{\bfseries PHENIX} Collaboration, A.~Adare {\em et~al.}, ``{Single electron
  yields from semileptonic charm and bottom hadron decays in Au$+$Au collisions
  at $\sqrt{s_{\rm{NN}}}=200$ GeV}'',
  \href{http://dx.doi.org/10.1103/PhysRevC.93.034904}{{\em Phys. Rev. C}
  {\bfseries 93} (2016) 034904},
  \href{http://arxiv.org/abs/1509.04662}{{\ttfamily arXiv:1509.04662
  [nucl-ex]}}.

\bibitem{ALICE:2020sjb}
{\bfseries ALICE} Collaboration, S.~Acharya {\em et~al.}, ``{Production of
  muons from heavy-flavour hadron decays at high transverse momentum in Pb--Pb
  collisions at $\sqrt{s_{\rm{NN}}} = 5.02$ and 2.76 TeV}'',
  \href{http://dx.doi.org/10.1016/j.physletb.2021.136558}{{\em Phys. Lett. B}
  {\bfseries 820} (2021) 136558},
  \href{http://arxiv.org/abs/2011.05718}{{\ttfamily arXiv:2011.05718
  [nucl-ex]}}.

\bibitem{STAR:2021uzu}
{\bfseries STAR} Collaboration, M.~S. Abdallah {\em et~al.}, ``{Evidence of
  Mass Ordering of Charm and Bottom Quark Energy Loss in Au+Au Collisions at
  RHIC}'', \href{http://dx.doi.org/10.1140/epjc/s10052-022-11003-7}{{\em Eur.
  Phys. J. C} {\bfseries 82} (2022) 1150},
  \href{http://arxiv.org/abs/2111.14615}{{\ttfamily arXiv:2111.14615
  [nucl-ex]}}. [Erratum: Eur.Phys.J.C 83, 455 (2023)].

\bibitem{PHENIX:2022wim}
{\bfseries PHENIX} Collaboration, U.~A. Acharya {\em et~al.}, ``{Charm- and
  Bottom-Quark Production in Au$+$Au Collisions at $\sqrt{s_{_{NN}}}$ = 200
  GeV}'', \href{http://arxiv.org/abs/2203.17058}{{\ttfamily arXiv:2203.17058
  [nucl-ex]}}.

\bibitem{ATLAS:2021xtw}
{\bfseries ATLAS} Collaboration, G.~Aad {\em et~al.}, ``{Measurement of the
  nuclear modification factor for muons from charm and bottom hadrons in Pb+Pb
  collisions at 5.02 TeV with the ATLAS detector}'',
  \href{http://dx.doi.org/10.1016/j.physletb.2022.137077}{{\em Phys. Lett. B}
  {\bfseries 829} (2022) 137077},
  \href{http://arxiv.org/abs/2109.00411}{{\ttfamily arXiv:2109.00411
  [nucl-ex]}}.

\bibitem{ALICE:2020hdw}
{\bfseries ALICE} Collaboration, S.~Acharya {\em et~al.}, ``{Elliptic Flow of
  Electrons from Beauty-Hadron Decays in Pb-Pb Collisions at $\sqrt
  {s_{\rm{NN}}}$ = 5.02 TeV}'',
  \href{http://dx.doi.org/10.1103/PhysRevLett.126.162001}{{\em Phys. Rev.
  Lett.} {\bfseries 126} (2021) 162001},
  \href{http://arxiv.org/abs/2005.11130}{{\ttfamily arXiv:2005.11130
  [nucl-ex]}}.

\bibitem{Aad:2020grf}
{\bfseries ATLAS} Collaboration, G.~Aad {\em et~al.}, ``{Measurement of
  azimuthal anisotropy of muons from charm and bottom hadrons in Pb+Pb
  collisions at $\sqrt {s_{\rm{NN}}}$ = 5.02 TeV with the ATLAS detector}'',
  \href{http://dx.doi.org/10.1016/j.physletb.2020.135595}{{\em Phys. Lett. B}
  {\bfseries 807} (2020) 135595},
  \href{http://arxiv.org/abs/2003.03565}{{\ttfamily arXiv:2003.03565
  [nucl-ex]}}.

\bibitem{ALICE:2008ngc}
{\bfseries ALICE} Collaboration, K.~Aamodt {\em et~al.}, ``{The ALICE
  experiment at the CERN LHC}'',
  \href{http://dx.doi.org/10.1088/1748-0221/3/08/S08002}{{\em JINST} {\bfseries
  3} (2008) S08002}.

\bibitem{ALICE:2014sbx}
{\bfseries ALICE} Collaboration, B.~Abelev {\em et~al.}, ``{Performance of the
  ALICE Experiment at the CERN LHC}'',
  \href{http://dx.doi.org/10.1142/S0217751X14300440}{{\em Int. J. Mod. Phys. A}
  {\bfseries 29} (2014) 1430044},
  \href{http://arxiv.org/abs/1402.4476}{{\ttfamily arXiv:1402.4476 [nucl-ex]}}.

\bibitem{ALICE:2010tia}
{\bfseries ALICE} Collaboration, K.~Aamodt {\em et~al.}, ``{Alignment of the
  ALICE Inner Tracking System with cosmic-ray tracks}'',
  \href{http://dx.doi.org/10.1088/1748-0221/5/03/P03003}{{\em JINST} {\bfseries
  5} (2010) P03003}, \href{http://arxiv.org/abs/1001.0502}{{\ttfamily
  arXiv:1001.0502 [physics.ins-det]}}.

\bibitem{Alme:2010ke}
J.~Alme {\em et~al.}, ``{The ALICE TPC, a large 3-dimensional tracking device
  with fast readout for ultra-high multiplicity events}'',
  \href{http://dx.doi.org/10.1016/j.nima.2010.04.042}{{\em Nucl. Instrum. Meth.
  A} {\bfseries 622} (2010) 316--367},
  \href{http://arxiv.org/abs/1001.1950}{{\ttfamily arXiv:1001.1950
  [physics.ins-det]}}.

\bibitem{Akindinov:2010zza}
A.~Akindinov {\em et~al.}, ``{The commissioning of the ALICE time-of-flight
  detector and results from the 2008 cosmic-ray data taking}'',
  \href{http://dx.doi.org/10.1016/j.nima.2010.01.004}{{\em Nucl. Instrum. Meth.
  A} {\bfseries 615} (2010) 37--41}.

\bibitem{Cortese:1121574}
{\bfseries ALICE} Collaboration, P.~Cortese {\em et~al.}, ``{ALICE
  Electromagnetic Calorimeter Technical Design Report}'', Tech. Rep.
  CERN-LHCC-2008-014. ALICE-TDR-14, Aug, 2008.
\newblock \url{http://cds.cern.ch/record/1121574}.

\bibitem{ALICE:2016ovj}
{\bfseries ALICE} Collaboration, J.~Adam {\em et~al.}, ``{Determination of the
  event collision time with the ALICE detector at the LHC}'',
  \href{http://dx.doi.org/10.1140/epjp/i2017-11279-1}{{\em Eur. Phys. J. Plus}
  {\bfseries 132} (2017) 99}, \href{http://arxiv.org/abs/1610.03055}{{\ttfamily
  arXiv:1610.03055 [physics.ins-det]}}.

\bibitem{ALICE:2022qhn}
{\bfseries ALICE} Collaboration, ``{Performance of the ALICE Electromagnetic
  Calorimeter}'', \href{http://dx.doi.org/10.1088/1748-0221/18/08/P08007}{{\em
  JINST} {\bfseries 18} (2023) P08007},
  \href{http://arxiv.org/abs/2209.04216}{{\ttfamily arXiv:2209.04216
  [physics.ins-det]}}.

\bibitem{Atoian:1992ze}
G.~S. Atoian {\em et~al.}, ``{Lead scintillator electromagnetic calorimeter
  with wavelength shifting fiber readout}'',
  \href{http://dx.doi.org/10.1016/0168-9002(92)90773-W}{{\em Nucl. Instrum.
  Meth. A} {\bfseries 320} (1992) 144--154}.

\bibitem{ALICE:2013axi}
{\bfseries ALICE} Collaboration, E.~Abbas {\em et~al.}, ``{Performance of the
  ALICE VZERO system}'',
  \href{http://dx.doi.org/10.1088/1748-0221/8/10/P10016}{{\em JINST} {\bfseries
  8} (2013) P10016}, \href{http://arxiv.org/abs/1306.3130}{{\ttfamily
  arXiv:1306.3130 [nucl-ex]}}.

\bibitem{ALICE:2022xir}
{\bfseries ALICE} Collaboration, ``{ALICE luminosity determination for Pb$-$Pb
  collisions at $\sqrt{s_{\mathrm{NN}}} = 5.02$ TeV}'',
  \href{http://arxiv.org/abs/2204.10148}{{\ttfamily arXiv:2204.10148
  [nucl-ex]}}.

\bibitem{Cortese:2019nnv}
{\bfseries ALICE} Collaboration, P.~Cortese, ``{Performance of the ALICE Zero
  Degree Calorimeters and upgrade strategy}'',
  \href{http://dx.doi.org/10.1088/1742-6596/1162/1/012006}{{\em J. Phys. Conf.
  Ser.} {\bfseries 1162} (2019) 012006}.

\bibitem{ALICE-PUBLIC-2018-011}
{\bfseries ALICE} Collaboration, ``{Centrality determination in heavy ion
  collisions}'', Aug, 2018.
\newblock \url{http://cds.cern.ch/record/2636623}. ALICE-PUBLIC-2018-011.

\bibitem{ALICE:2013hur}
{\bfseries ALICE} Collaboration, B.~Abelev {\em et~al.}, ``{Centrality
  determination of Pb-Pb collisions at $\sqrt{s_{\rm{NN}}} = 2.76$ TeV with
  ALICE}'', \href{http://dx.doi.org/10.1103/PhysRevC.88.044909}{{\em Phys. Rev.
  C} {\bfseries 88} (2013) 044909},
  \href{http://arxiv.org/abs/1301.4361}{{\ttfamily arXiv:1301.4361 [nucl-ex]}}.

\bibitem{ALICE-PUBLIC-2018-014}
{\bfseries ALICE} Collaboration, ``{ALICE 2017 luminosity determination for pp
  collisions at $\sqrt{s}$ = 5 TeV}'', Nov, 2018.
\newblock \url{http://cds.cern.ch/record/2648933}. ALICE-PUBLIC-2018-014.

\bibitem{ALICE:2012mzy}
{\bfseries ALICE} Collaboration, B.~Abelev {\em et~al.}, ``{Measurement of
  electrons from semileptonic heavy-flavour hadron decays in pp collisions at
  $\sqrt{s}=7$ TeV}'', \href{http://dx.doi.org/10.1103/PhysRevD.86.112007}{{\em
  Phys. Rev. D} {\bfseries 86} (2012) 112007},
  \href{http://arxiv.org/abs/1205.5423}{{\ttfamily arXiv:1205.5423 [hep-ex]}}.

\bibitem{ParticleDataGroup:2018ovx}
{\bfseries Particle Data Group} Collaboration, M.~Tanabashi {\em et~al.},
  ``{Review of Particle Physics}'',
  \href{http://dx.doi.org/10.1103/PhysRevD.98.030001}{{\em Phys. Rev. D}
  {\bfseries 98} (2018) 030001}.

\bibitem{ALICE:2020jsh}
{\bfseries ALICE} Collaboration, S.~Acharya {\em et~al.}, ``{Production of
  light-flavor hadrons in pp collisions at $\sqrt{s}~=~7$ and $\sqrt{s} = 13$
  TeV }'', \href{http://dx.doi.org/10.1140/epjc/s10052-020-08690-5}{{\em Eur.
  Phys. J. C} {\bfseries 81} (2021) 256},
  \href{http://arxiv.org/abs/2005.11120}{{\ttfamily arXiv:2005.11120
  [nucl-ex]}}.

\bibitem{Awes:1992yp}
T.~C. Awes, F.~E. Obenshain, F.~Plasil, S.~Saini, S.~P. Sorensen, and G.~R.
  Young, ``{A Simple method of shower localization and identification in
  laterally segmented calorimeters}'',
  \href{http://dx.doi.org/10.1016/0168-9002(92)90858-2}{{\em Nucl. Instrum.
  Meth. A} {\bfseries 311} (1992) 130--138}.

\bibitem{ALICE:2017nce}
{\bfseries ALICE} Collaboration, S.~Acharya {\em et~al.}, ``{Production of
  ${\pi ^0}$ and $\eta $ mesons up to high transverse momentum in pp collisions
  at 2.76 TeV}'', \href{http://dx.doi.org/10.1140/epjc/s10052-017-4890-x}{{\em
  Eur. Phys. J. C} {\bfseries 77} (2017) 339},
  \href{http://arxiv.org/abs/1702.00917}{{\ttfamily arXiv:1702.00917
  [hep-ex]}}.

\bibitem{ALICE:2005vhb}
{\bfseries ALICE} Collaboration, C.~W. Fabjan {\em et~al.}, ``{ALICE: Physics
  Performance Report}'',
  \href{http://dx.doi.org/10.1088/0954-3899/32/10/001}{{\em J. Phys. G}
  {\bfseries 32} (2006) 1295--2040}.

\bibitem{Sjostrand:2006za}
T.~Sjostrand, S.~Mrenna, and P.~Z. Skands, ``{PYTHIA 6.4 Physics and Manual}'',
  \href{http://dx.doi.org/10.1088/1126-6708/2006/05/026}{{\em JHEP} {\bfseries
  05} (2006) 026}, \href{http://arxiv.org/abs/hep-ph/0603175}{{\ttfamily
  arXiv:hep-ph/0603175}}.

\bibitem{Wang:1991hta}
X.-N. Wang and M.~Gyulassy, ``{HIJING: A Monte Carlo model for multiple jet
  production in p p, p A and A A collisions}'',
  \href{http://dx.doi.org/10.1103/PhysRevD.44.3501}{{\em Phys. Rev. D}
  {\bfseries 44} (1991) 3501--3516}.

\bibitem{Brun:1119728}
R.~Brun, F.~Bruyant, M.~Maire, A.~C. McPherson, and P.~Zanarini, {\em {GEANT 3:
  user's guide Geant 3.10, Geant 3.11; rev. version}}.
\newblock CERN, Geneva, 1987.
\newblock \url{https://cds.cern.ch/record/1119728}.

\bibitem{Barlow1993219}
R.~Barlow and C.~Beeston, ``Fitting using finite monte carlo samples'', {\em
  Comp. Phys. Comm.} {\bfseries 77} (1993) 219.

\bibitem{ALICE:2018hbc}
{\bfseries ALICE} Collaboration, S.~Acharya {\em et~al.},
  ``{$\Lambda_\mathrm{c}^+$ production in Pb-Pb collisions at $\sqrt{s_{\rm
  NN}} = 5.02$ TeV}'',
  \href{http://dx.doi.org/10.1016/j.physletb.2019.04.046}{{\em Phys. Lett. B}
  {\bfseries 793} (2019) 212--223},
  \href{http://arxiv.org/abs/1809.10922}{{\ttfamily arXiv:1809.10922
  [nucl-ex]}}.

\bibitem{ALICE:2021bib}
{\bfseries ALICE} Collaboration, S.~Acharya {\em et~al.}, ``{Constraining
  hadronization mechanisms with $\rm \Lambda_{\rm c}^{+}$/D$^0$ production
  ratios in Pb-Pb collisions at $\sqrt{s_{\rm NN}} = 5.02$ TeV}'',
  \href{http://dx.doi.org/10.1016/j.physletb.2023.137796}{{\em Phys. Lett. B}
  {\bfseries 839} (2023) 137796},
  \href{http://arxiv.org/abs/2112.08156}{{\ttfamily arXiv:2112.08156
  [nucl-ex]}}.

\bibitem{ALICE:2018lyv}
{\bfseries ALICE} Collaboration, S.~Acharya {\em et~al.}, ``{Measurement of
  D$^{0}$, D$^{+}$, D$^{*+}$ and D$_{s}^{+}$ production in Pb-Pb collisions at
  $ \sqrt{{\mathrm{s}}_{\mathrm{NN}}}=5.02 $ TeV}'',
  \href{http://dx.doi.org/10.1007/JHEP10(2018)174}{{\em JHEP} {\bfseries 10}
  (2018) 174}, \href{http://arxiv.org/abs/1804.09083}{{\ttfamily
  arXiv:1804.09083 [nucl-ex]}}.

\bibitem{He:2014cla}
M.~He, R.~J. Fries, and R.~Rapp, ``{Heavy Flavor at the Large Hadron Collider
  in a Strong Coupling Approach}'',
  \href{http://dx.doi.org/10.1016/j.physletb.2014.05.050}{{\em Phys. Lett. B}
  {\bfseries 735} (2014) 445--450},
  \href{http://arxiv.org/abs/1401.3817}{{\ttfamily arXiv:1401.3817 [nucl-th]}}.

\bibitem{ALICE:2019nuy}
{\bfseries ALICE} Collaboration, S.~Acharya {\em et~al.}, ``{Measurement of
  electrons from semileptonic heavy-flavour hadron decays at midrapidity in pp
  and Pb-Pb collisions at $\sqrt{s_{\rm{NN}}}$ = 5.02 TeV}'',
  \href{http://dx.doi.org/10.1016/j.physletb.2020.135377}{{\em Phys. Lett. B}
  {\bfseries 804} (2020) 135377},
  \href{http://arxiv.org/abs/1910.09110}{{\ttfamily arXiv:1910.09110
  [nucl-ex]}}.

\bibitem{ALICE:2015zhm}
{\bfseries ALICE} Collaboration, J.~Adam {\em et~al.}, ``{Measurement of
  electrons from heavy-flavour hadron decays in p-Pb collisions at
  $\sqrt{s_{\rm NN}} =$ 5.02 TeV}'',
  \href{http://dx.doi.org/10.1016/j.physletb.2015.12.067}{{\em Phys. Lett. B}
  {\bfseries 754} (2016) 81--93},
  \href{http://arxiv.org/abs/1509.07491}{{\ttfamily arXiv:1509.07491
  [nucl-ex]}}.

\bibitem{ALICE:2016mpw}
{\bfseries ALICE} Collaboration, J.~Adam {\em et~al.}, ``{Measurement of the
  production of high-$p_{\rm T}$ electrons from heavy-flavour hadron decays in
  Pb-Pb collisions at $\mathbf{\sqrt{\it s_{\rm{NN}}}}$ = 2.76 TeV}'',
  \href{http://dx.doi.org/10.1016/j.physletb.2017.05.060}{{\em Phys. Lett. B}
  {\bfseries 771} (2017) 467--481},
  \href{http://arxiv.org/abs/1609.07104}{{\ttfamily arXiv:1609.07104
  [nucl-ex]}}.

\bibitem{ALICE:2018yau}
{\bfseries ALICE} Collaboration, S.~Acharya {\em et~al.}, ``{Measurements of
  low-$p_{\rm T}$ electrons from semileptonic heavy-flavour hadron decays at
  mid-rapidity in pp and Pb-Pb collisions at $ \sqrt{s_{\mathrm{NN}}}=2.76 $
  TeV}'', \href{http://dx.doi.org/10.1007/JHEP10(2018)061}{{\em JHEP}
  {\bfseries 10} (2018) 061}, \href{http://arxiv.org/abs/1805.04379}{{\ttfamily
  arXiv:1805.04379 [nucl-ex]}}.

\bibitem{james_statistical_2006}
F.~James, \href{http://dx.doi.org/10.1142/6096}{{\em Statistical {Methods} in
  {Experimental} {Physics}}}.
\newblock WORLD SCIENTIFIC, 2~ed., Nov., 2006.
\newblock \url{https://www.worldscientific.com/worldscibooks/10.1142/6096}.

\bibitem{Fermi:1934hr}
E.~Fermi, ``{An attempt of a theory of beta radiation. 1.}'',
  \href{http://dx.doi.org/10.1007/BF01351864}{{\em Z. Phys.} {\bfseries 88}
  (1934) 161--177}.

\bibitem{Wilson:1968pwx}
F.~L. Wilson, ``{Fermi's Theory of Beta Decay}'',
  \href{http://dx.doi.org/10.1119/1.1974382}{{\em Am. J. Phys.} {\bfseries 36}
  (1968) 1150--1160}.

\bibitem{ALICE:2019hno}
{\bfseries ALICE} Collaboration, S.~Acharya {\em et~al.}, ``{Production of
  charged pions, kaons, and (anti-)protons in Pb-Pb and inelastic pp collisions
  at $\sqrt {s_{\rm{NN}}}$ = 5.02 TeV}'',
  \href{http://dx.doi.org/10.1103/PhysRevC.101.044907}{{\em Phys. Rev. C}
  {\bfseries 101} (2020) 044907},
  \href{http://arxiv.org/abs/1910.07678}{{\ttfamily arXiv:1910.07678
  [nucl-ex]}}.

\bibitem{LHCb:2019fns}
{\bfseries LHCb} Collaboration, R.~Aaij {\em et~al.}, ``{Measurement of $b$
  hadron fractions in 13 TeV pp collisions}'',
  \href{http://dx.doi.org/10.1103/PhysRevD.100.031102}{{\em Phys. Rev. D}
  {\bfseries 100} (2019) 031102},
  \href{http://arxiv.org/abs/1902.06794}{{\ttfamily arXiv:1902.06794
  [hep-ex]}}.

\bibitem{BaBar:2004bij}
{\bfseries BaBar} Collaboration, B.~Aubert {\em et~al.}, ``{Measurement of the
  electron energy spectrum and its moments in inclusive $B \to X e \nu$
  decays}'', \href{http://dx.doi.org/10.1103/PhysRevD.69.111104}{{\em Phys.
  Rev. D} {\bfseries 69} (2004) 111104},
  \href{http://arxiv.org/abs/hep-ex/0403030}{{\ttfamily arXiv:hep-ex/0403030}}.

\bibitem{Kniehl:2011bk}
B.~A. Kniehl, G.~Kramer, I.~Schienbein, and H.~Spiesberger, ``{Inclusive
  B-Meson Production at the LHC in the GM-VFN Scheme}'',
  \href{http://dx.doi.org/10.1103/PhysRevD.84.094026}{{\em Phys. Rev. D}
  {\bfseries 84} (2011) 094026},
  \href{http://arxiv.org/abs/1109.2472}{{\ttfamily arXiv:1109.2472 [hep-ph]}}.

\bibitem{Djordjevic:2015hra}
M.~Djordjevic and M.~Djordjevic, ``{Predictions of heavy-flavor suppression at
  5.1 TeV Pb + Pb collisions at the CERN Large Hadron Collider}'',
  \href{http://dx.doi.org/10.1103/PhysRevC.92.024918}{{\em Phys. Rev. C}
  {\bfseries 92} (2015) 024918},
  \href{http://arxiv.org/abs/1505.04316}{{\ttfamily arXiv:1505.04316
  [nucl-th]}}.

\bibitem{ALICE:2014aev}
{\bfseries ALICE} Collaboration, B.~Abelev {\em et~al.}, ``{Beauty production
  in pp collisions at $\sqrt{s}$ = 2.76 TeV measured via semi-electronic
  decays}'', \href{http://dx.doi.org/10.1016/j.physletb.2014.09.026}{{\em Phys.
  Lett. B} {\bfseries 738} (2014) 97--108},
  \href{http://arxiv.org/abs/1405.4144}{{\ttfamily arXiv:1405.4144 [nucl-ex]}}.

\bibitem{zigic_dreena-b_2019}
D.~Zigic, I.~Salom, J.~Auvinen, M.~Djordjevic, and M.~Djordjevic, ``{DREENA-B
  framework: first predictions of $R_{AA}$ and $v_2$ within dynamical energy
  loss formalism in evolving QCD medium}'',
  \href{http://dx.doi.org/10.1016/j.physletb.2019.02.020}{{\em Phys. Lett. B}
  {\bfseries 791} (2019) 236--241},
  \href{http://arxiv.org/abs/1805.04786}{{\ttfamily arXiv:1805.04786
  [nucl-th]}}.

\bibitem{Prado:2016szr}
C.~A.~G. Prado, J.~Noronha-Hostler, R.~Katz, A.~A.~P. Suaide, J.~Noronha, M.~G.
  Munhoz, and M.~R. Cosentino, ``{Event-by-event correlations between soft
  hadrons and $D^0$ mesons in 5.02 TeV PbPb collisions at the CERN Large Hadron
  Collider}'', \href{http://dx.doi.org/10.1103/PhysRevC.96.064903}{{\em Phys.
  Rev. C} {\bfseries 96} (2017) 064903},
  \href{http://arxiv.org/abs/1611.02965}{{\ttfamily arXiv:1611.02965
  [nucl-th]}}.

\bibitem{Nahrgang:2013xaa}
M.~Nahrgang, J.~Aichelin, P.~B. Gossiaux, and K.~Werner, ``{Influence of
  hadronic bound states above $T_c$ on heavy-quark observables in Pb + Pb
  collisions at at the CERN Large Hadron Collider}'',
  \href{http://dx.doi.org/10.1103/PhysRevC.89.014905}{{\em Phys. Rev. C}
  {\bfseries 89} (2014) 014905},
  \href{http://arxiv.org/abs/1305.6544}{{\ttfamily arXiv:1305.6544 [hep-ph]}}.

\bibitem{Song:2015sfa}
T.~Song, H.~Berrehrah, D.~Cabrera, J.~M. Torres-Rincon, L.~Tolos, W.~Cassing,
  and E.~Bratkovskaya, ``{Tomography of the Quark-Gluon-Plasma by Charm
  Quarks}'', \href{http://dx.doi.org/10.1103/PhysRevC.92.014910}{{\em Phys.
  Rev. C} {\bfseries 92} (2015) 014910},
  \href{http://arxiv.org/abs/1503.03039}{{\ttfamily arXiv:1503.03039
  [nucl-th]}}.

\bibitem{Ke:2018tsh}
W.~Ke, Y.~Xu, and S.~A. Bass, ``{{Linearized Boltzmann-Langevin model for heavy
  quark transport in hot and dense QCD matter}}'',
  \href{http://dx.doi.org/10.1103/PhysRevC.98.064901}{{\em Phys. Rev. C}
  {\bfseries 98} (2018) 064901},
  \href{http://arxiv.org/abs/1806.08848}{{\ttfamily arXiv:1806.08848
  [nucl-th]}}.

\bibitem{Prino:2016cni}
F.~Prino and R.~Rapp, ``{Open Heavy Flavor in QCD Matter and in Nuclear
  Collisions}'', \href{http://dx.doi.org/10.1088/0954-3899/43/9/093002}{{\em J.
  Phys. G} {\bfseries 43} (2016) 093002},
  \href{http://arxiv.org/abs/1603.00529}{{\ttfamily arXiv:1603.00529
  [nucl-ex]}}.

\end{thebibliography}\endgroup
